# Poverty Mapping: Data, Models and Applications

Suoyi Tan [a,+], Mengning Wang [a,+], Yixiu Kong [b], Huimin Bai [c], Jianguo Liu [d], Dirk Brockmann [e], Yicheng Zhang [f], Xin Lu [a,*]

[a] *College of Systems Engineering, National University of Defense Technology, Changsha, 410073, China.*

[b] *International Research Center for Complexity Sciences, Hangzhou International Innovation Institute, Beihang University, Hangzhou, 311115, China*

[c] *Department of Mathematics, North University of China, Taiyuan 030051, Shanxi, China.*

[d] *Institute of Accounting and Finance, Shanghai University of Finance & Economics, Shanghai, 200433, China.*

[e] *Center Synergy of Systems, Dresden University of Technology, Dresden, 01069, Germany.*

[f] *Department of Physics, University of Fribourg, CH-1700 Fribourg, Switzerland.*

*+ These authors contributed equally to this work.*
** Corresponding authors: xin.lu.lab@outlook.com (XL)*

**Abstract**

Poverty mapping is increasingly important for monitoring Sustainable Development Goal 1 (SDG 1) of the United Nations 2030 Agenda, which aims to end poverty in all its forms everywhere. Yet timely and fine-resolution poverty estimation remains difficult because conventional census- and survey-based approaches are costly, infrequent, and often sparse precisely where deprivation is most severe. As poverty emerges from complex socioeconomic systems shaped by human mobility, social interactions, infrastructure, and economic activities, emerging computational methods and nontraditional data sources have created new opportunities for poverty estimation and mapping. At the intersection of statistical physics, complex systems science, and data science, these approaches enable poverty estimation at finer spatial and temporal resolutions. This review summarizes the main concepts of poverty and the principal frameworks used to measure it, and examines recent advances on poverty estimation and mapping using satellite imagery, mobile phone data, social media data, and multisource data fusion. The review also discusses persistent challenges related to representativeness, transferability across regions, interpretability, and uncertainty quantification. Finally, the review clarifies both the analytical promise and the practical limits of contemporary poverty mapping.

## Contents

## 1. Introduction

Poverty mapping is fundamentally a problem of measurement under spatial, temporal, and institutional constraints. Poverty is not only unevenly distributed across countries, but also highly heterogeneous within regions, cities, and neighborhoods. Poverty monitoring therefore cannot rely only on national aggregates, but requires estimates that are timely, spatially disaggregated, and interpretable for policy intervention. This measurement problem is closely tied to the international development agenda. The first Millennium Development Goal (MDG 1), adopted as part of the United Nations Millennium Development Goals, focused on reducing extreme poverty and hunger [1, 2]. It was followed by the 17 Sustainable Development Goals (SDGs, 2015–2030), adopted under the United Nations 2030 Agenda, with SDG 1 aiming to end poverty in all its forms everywhere [3]. This transition broadened the monitoring task from tracking aggregate poverty reduction to identifying where deprivation persists, which dimensions of poverty are involved, and which populations remain poorly represented in official statistics [4].

Poverty estimation has traditionally relied on household surveys and population censuses. Demographic and Health Surveys (DHS) [5], one of the most widely used sources of household-level socioeconomic information, have conducted more than 400 surveys in more than 90 low- and middle-income countries (LMICs) since 1984 [6]. In practice, survey and census records also provide the empirical basis for small-area estimation, through which local observations can be extrapolated to broader administrative or geographic units [7, 8]. These sources remain difficult to replace, since they provide direct information on income, consumption, assets, health, education, and living conditions. However, they are costly, infrequently updated, and often sparse in contexts where deprivation is severe or difficult to observe. Conducting an annual DHS in a single country typically incurs expenses ranging from US$1.5 to 2 million [9], while official poverty statistics are often released with delays of months or years. In many LMICs, particularly in sub-Saharan Africa, economic statistics remain rudimentary and may not be updated for decades [10]. A World Bank assessment in 2015 found that censuses and surveys were inadequate for monitoring poverty in more than one third of LMICs [11]. These limitations become especially consequential when poverty is geographically uneven. Economic inequality, racial segregation, forced displacement, and informal urbanization can produce sharp differences in deprivation within the same country or even within the same city. Identifying poverty hotspots below the national scale is therefore necessary for targeted responses [12], including the detection of slum areas [13] and the separate mapping of urban and rural poverty [14].

The demand for timely and spatially resolved poverty estimates has become still more urgent under recent global shocks. The three-year COVID-19 pandemic interrupted several decades of progress in global poverty reduction and exposed the fragility of gains achieved in many developing economies [15, 16]. Climate-related disruptions, inflation, a substantial rise in trade barriers, and geopolitical conflicts, including the Israeli–Palestinian conflict and Russian–Ukrainian war, have further increased the volatility of poverty dynamics [17-21]. Under these conditions, poverty conditions may change faster than survey cycles can observe, and its spatial distribution may shift across urban, rural, and conflict-affected settings. This mismatch between changing poverty conditions and slow data collection cycles calls for poverty estimates that are more frequently updated and more spatially detailed. Such estimates require both granular, up-to-date data and computational methods that can translate heterogeneous observations into policy-relevant poverty indicators. Recent advances in computational capacity, geospatial data, and artificial intelligence have made it possible to infer poverty-related conditions from large-scale observational data. As a result, nontraditional data sources and computational models into poverty mapping as complements to survey-based statistics [22-24].

From a complex-systems perspective, poverty-related conditions can be viewed as macroscopic outcomes of coupled spatial, behavioral, infrastructural, and networked processes. Nontraditional data sources capture observable traces of these processes. Poverty mapping is therefore not only a predictive task but also a problem of representing and interpreting poverty-related socioeconomic conditions across spatial scales.

Satellite imagery offers spatially extensive observations of nighttime illumination, land use, infrastructure, settlement morphology, and the built environment. Nighttime light (NTL) imagery has been widely used as a proxy for economic activity [25]. Daytime imagery, high-resolution imagery, very-high-resolution imagery, and OpenStreetMap (OSM) data have extended poverty mapping to finer spatial scales and richer visual features. When combined with machine learning techniques, these data have been used to infer the distribution of deprived populations [26-30]. The widespread use of mobile phones, even in remote areas, provides another rich data source, including temporal records of communication patterns, digital traces in transaction logs, and positioning data that record individual movements [31]. These data provide behavioral proxies related to mobility, access, opportunity, and socioeconomic activity, and have supported applications ranging from epidemic modeling to disaster response and regional development analysis [32-39]. Open-access social media data from social media platforms such as Twitter, Facebook, and Google Trends have also been used to forecast economic activity and estimate poverty-related outcomes [40-44]. Although the utilization of a single data source can be informative, multisource approaches can improve predictive performance by combining complementary information. For example, high-resolution poverty maps of Bangladesh constructed from both mobile phone data and satellite imagery outperformed models based only on NTL or mobile phone data [45].

These developments, however, do not remove the methodological difficulty of poverty mapping. Poverty is typically inferred indirectly from imperfect proxies observed at different spatial and temporal scales. Data source differs not only in nominal resolution, but also in representativeness, validation requirements, transferability, and socioeconomic interpretability. The effective resolution of a poverty map is determined not simply by the input data, but by the scale at which predictions remain accurate, transferable, and meaningful in relation to established poverty concepts. A critical review of this field must consequently compare data sources and models not only by predictive performance, but also in terms of the mechanisms they represent, the biases they introduce, and the spatial heterogeneity they can resolve.

This review is motivated by the tension between expanding data availability and the unresolved problem of reliable socioeconomic inference. Since poverty is defined and measured through different conceptual frameworks, we first clarify the foundations of poverty concepts and indicators. We then examine how nontraditional data sources and computational models have been used to estimate and map poverty across different spatial and temporal scales. To organize this literature, we categorize representative studies according to data source, modeling strategy, and research purpose, complementing existing sector-specific reviews. The aim is to provide a structured account of what contemporary poverty mapping can infer, where it remains uncertain, and how it can complement, rather than replace conventional poverty statistics.

The remainder of the review is organized as follows. Section 2 introduces major definitions of poverty and the conceptual distinctions underlying poverty measurement. Section 3 summarizes unidimensional and multidimensional poverty indicators, including their assumptions and limitations. Sections 4–7 review poverty mapping methods based on remote sensing imagery, mobile phone data, social media data, and multisource data fusion. Each section discusses the relevant data sources, modeling approaches, representative applications, and methodological constraints. Section 8 synthesizes the main progress of the field and discusses open challenges, including data representativeness, uncertainty, transferability, effective spatial resolution, interpretability, and the evaluation of poverty estimates. The Appendix lists selected poverty indicators and publicly accessible datasets that can support empirical analysis (Tables S1 and S2).

## 2. Definition of poverty

Although poverty alleviation is widely recognized as a central objective of the Sustainable Development Goals (SDGs), the concept of poverty remains subject to debate in both academic scholarship and policy practice. To clarify how poverty can be defined and measured, this section reviews four major approaches to poverty conceptualization, i.e., the monetary, capability, social exclusion, and participatory approaches.

### *2.1. Conceptual approaches to poverty*

#### *2.1.1. Monetary approach*

The monetary approach defines poverty in terms of measurable material deprivation, identifying individuals or households as poor when their income or consumption expenditure falls below a specified poverty line [46, 47]. Although household income, consumption, and expenditure are widely used indicators of economic activity and can be standardized across countries, monetary measures alone cannot adequately capture other dimensions that are central to well-being, including social resources, health, and education.

#### *2.1.2. Capability approach*

The capability approach emphasizes individuals' freedom to live lives they have reason to value. From this perspective, poverty is not merely a condition of low income, but also reflects the deprivation of basic capabilities required to achieve minimum levels of well-being [48]. Within this framework, poverty is understood as "capability deprivation", individuals are unable to participate in essential aspects of human life because they lack the necessary capabilities, making poverty an inherently multidimensional problem [49].

*2.1.3. Social exclusion approach*

The social exclusion approach understands poverty as a relational and dynamic process of systemic marginalization. Unlike approaches that focus primarily on individuals, it emphasizes structural barriers and group-based mechanisms of deprivation [50]. From this perspective, institutional discrimination, spatial segregation, and unequal access to social networks can reinforce socioeconomic disadvantage across generations. Social exclusion also unfolds across multiple intersecting dimensions, including economic, political, and cultural domains, which together weaken both individual agency and social cohesion.

*2.1.4. Participatory approach*

The participatory approach stresses the importance of involving people affected by poverty in defining the problem and shaping responses to it. It recognizes that the experiences and perspectives of those living in poverty are essential for understanding poverty and developing effective strategies to address it. Participatory poverty assessments are therefore intended to empower individuals living in poverty, deepen their understanding of the conditions they face, and increase their influence over decisions that affect their lives. In the long term, this may support the design and implementation of more effective poverty alleviation policies [51].

Each approach to poverty rests on a distinct view of what constitutes a good life and a just society. As a result, who is identified as poor may vary depending on the approach adopted [52]. In practice, the monetary definition remains the most widely recognized. It typically involves establishing a poverty line or threshold, below which individuals or households are classified as poor on the basis of income or consumption expenditure. This poverty line is generally defined according to the resources required to meet basic needs and may be specified in either absolute or relative terms.

*2.2. Absolute and relative measures of poverty*

*2.2.1. Absolute poverty*

Absolute poverty is defined as a condition in which household income is below a certain level. It denotes a severe form of deprivation, in which individuals or households lack the basic necessities required to maintain a minimal standard of living, including food, shelter, clothing, and health care [53]. The income threshold used to define absolute poverty differs across countries and is periodically adjusted to reflect inflation and changes in living standards. International organizations such as the World Bank commonly adopt a global poverty line based on purchasing power parity exchange rates (PPPs). In 1990, the initial poverty line was proposed as "US$1 a day" using the 1985 PPPs. Subsequent updates to PPP estimates led to revisions of the threshold, US$1.08 per day based on the 1993 PPPs, US$1.25 following the 2005 PPP update, and US$1.90 using the 2011 PPPs. As of 2022, the World Bank raised the standard to US$2.15, which is the current benchmark [54].

*2.2.2. Relative poverty*

Relative poverty is defined in relation to the economic standards and social context of a particular society or community. Under this concept, individuals or families may be able to meet basic needs yet still experience poverty if their standard of living falls substantially below that of the broader population. Such disadvantage may result in social exclusion, restricted access to opportunities, and a lower quality of life relative to others in the same society. This approach recognizes that the meaning of basic needs depends partly on prevailing social standards and expectations. A commonly used relative poverty line classifies households as poor when their income is below half of the national median income. The European Union adopts a broader standard, setting the poverty line at 60% of median equivalized household income to better capture dimensions of social inclusion [55].

Absolute poverty focuses on universally meeting basic needs. It is relatively straightforward and facilitates comparisons across countries and regions, but it is less able to account for differences in quality of living. By contrast, relative poverty is more sensitive to social inequality within a given society, although it can complicate comparisons across diverse social contexts and may be affected by changes in income inequality over time. Taken together, absolute and relative measures offer a more comprehensive basis for understanding poverty dynamics across different contexts.

## 3. Measurement of poverty

Traditional approaches to poverty measurement generally rely on unidimensional indicators based primarily on income and consumption, and are therefore commonly described as “monetary measures” of poverty. Because they are simple and convenient to apply, these measures are widely available and recognized across countries (Fig. 1a). Nevertheless, poverty classifications can vary substantially depending on the measurement approach adopted, making appropriate poverty measurement essential for effective policy design and program targeting [12].

*3.1. Unidimensional poverty measurement*

*3.1.1. Headcount index*

The Headcount Index (Fig. 1b), often denoted by $P_0$, is widely used by governments, international organizations, and researchers to monitor changes in poverty across time and space. This index measures poverty as the proportion of all households whose income or consumption falls below a specified poverty line [56, 57]. It is defined as follows:

$$P_0 = \frac{1}{N}\sum_{i=1}^{N} I(y_i < z), \tag{1}$$

where $N$ is the total population, $I(\bullet)$ is an indicator function that takes on the value of 1 if the income $y_i$ is below the poverty line $z$ and 0 otherwise.

*3.1.2. Poverty gap index*

The headcount index cannot determine the severity of poverty because it does not change as those who live below the poverty line become even poorer. To fill this gap, the poverty gap Index (Fig. 1c), often denoted by $P_1$, was proposed [56]. It takes into account the depth of poverty experienced by individuals below the poverty line, calculating the average difference between the poverty line and the income of individuals or households living below it. Formally, it is expressed as follows:

$$\begin{cases} P_1 = \frac{1}{N}\sum_{i=1}^{N}\frac{G_i}{z} \\ G_i = (z - y_i) \times I(y_i < z) \end{cases}, \tag{2}$$

where $G_i$ is the poverty gap representing the extent of the actual income $y_i$ that lies below the poverty line $z$ for poor individuals.

*3.1.3. Poverty severity index*

The poverty severity index (Fig. 1d), often denoted by $P_2$, extends poverty-gap measurement by incorporating inequality among the poor through weighted poverty gaps [56]. More specifically, it is calculated as the average of the squared poverty-gap ratio, thereby assigning greater implicit weight to individuals or households located farther below the poverty line.

$$\begin{cases} P_2 = \frac{1}{N}\sum_{i=1}^{N}\left(\frac{G_i}{z}\right)^2 \\ G_i = (z - y_i) \times I(y_i < z) \end{cases} \tag{3}$$

*3.1.4. Gini coefficient*

Although the Gini coefficient is not itself a direct poverty measure, it is often used together with poverty indicators to characterize distributional inequality [58]. Ranging from 0 to 1, the coefficient measures income inequality within a given population. A value of 0 represents perfect equality, meaning that all individuals have the same income, whereas a value of 1 indicates complete inequality, in which all income is held by a single individual. The Gini coefficient is calculated from the Lorenz curve, which graphically represents the cumulative distribution of income or wealth. Specifically, it is defined as the ratio of the area between the Lorenz curve and the line of equality to the total area beneath the line of equality (Fig. 1e). Assuming the Lorenz curve is represented by the function $Y = L(X)$, the general formula for calculating the Gini coefficient can be expressed as

$$G = 1 - 2\int_0^1 L(X)dX. \tag{4}$$

*3.1.5. Sen index*

The Sen index [59] is a composite poverty measure designed to combine the number of the poor, the depth of poverty, and inequality among the poor in a single indicator. It can be written as

$$P_s = P_0 G^p + P_1(1 - G^p), \tag{5}$$

where $G^P$ is the Gini coefficient of inequality among the poor.

This formulation is grounded in Sen's axiomatic approach to poverty measurement, which holds that a poverty index should capture not only the incidence and average depth of poverty, but also inequality among the poor. The Sen index can therefore be understood as a Gini-based social evaluation function with a multiplicative decomposition structure, making it sensitive to the number of poor individuals, their average level of deprivation, and the distribution of poverty within the poor population. Despite its conceptual strengths, however, the Sen index is used less often in practice because it is less intuitive than simpler poverty measures and cannot be easily decomposed by subgroup [60].

*3.1.6. Sen-Shorrocks-Thon index*

The Sen-Shorrocks-Thon (SST) index extends the Sen index by offering a formulation better suited to decomposition and distributional analysis [61]. Whereas the Sen index evaluates inequality only among the poor, the SST index measures inequality using the distribution of poverty-gap ratios across the entire population. To do so, individuals above the poverty line are assigned a poverty-gap ratio of zero, allowing poor and non-poor individuals to be represented within a single distribution. Meanwhile, the average poverty gap is calculated only for the poor population, so that the measure is not diluted by zero values from non-poor individuals. This structure enables the SST index to capture both average deprivation among the poor and the broader distribution of poverty across society. It can be expressed as

$$P_{SST} = P_0 P_1^P (1 + \hat{G}_P), \tag{6}$$

where $P_1^P$ denotes the poverty gap index computed over the poor population only, and $\hat{G}^P$ is the Gini coefficient of the poverty gap ratios for the entire population.

*3.2. Multidimensional poverty measurement*

Although monetary poverty measures are useful for identifying the most underprivileged households or individuals, they are limited in their ability to capture the full spectrum of deprivation. Poverty is, in practice, a multidimensional problem that extends beyond financial scarcity [62]. It includes conditions such as hunger, inadequate shelter, and limited access to education and health care. For this reason, a range of multidimensional indicators has been developed to provide a more comprehensive account of poverty.

*3.2.1. Townsend deprivation index*

The Townsend Deprivation Index measures material deprivation using census data [63]. It evaluates poverty by combining four deprivation-related variables, i.e., lack of car ownership, absence of home ownership, unemployment, and overcrowding. Widely used in social policy and academic research, the index supports the identification of highly deprived areas and can inform the targeted allocation of resources. Its calculation involves several steps, including log transformation and standardization of the selected variables. Standardization is carried out using Z-scores relative to the reference population under study, typically across all spatial units within a country or study area rather than across countries. The detailed procedure for computing the index is presented in Table 1 [64].

**Table 1. Townsend Deprivation Index calculation method**

| ***Step 1: Calculate the percentages of four variables.*** |
|---|
| • Percentage non-car ownership: (Households with no car/Total households) × 100 |
| • Percentage of non-home ownership: (Households that are not owner-occupied [rented]/Total households) × 100 |
| • Percentage unemployment: (People who are unemployed/Total people economically active) × 100 |
| • Percentage overcrowding: (Households that are overcrowded/Total households) × 100 |
| ***Step 2: Apply a log transformation.*** |
| • Logged $p$(unemployed) = ln(unemployment + 1) |
| • Logged $p$(overcrowding) = ln(overcrowding + 1) |
| ***Step 3: Standardize the data.*** |
| • $Z$ score no car = (Percentage no car − Mean percentage no car)/SD percentage no car |
| • $Z$ score non-homeowner = (Percentage non-homeowner − Mean percentage non-homeowner)/SD percentage no homeowner |
| • $Z$ score unemployed = (Logged $p$[unemployed] − Mean logged $p$[unemployed])/SD logged p(unemployed) |
| • $Z$ score overcrowding = (Logged $p$(overcrowding) − Mean logged $p$[overcrowding])/SD logged $p$(overcrowding) |
| ***Step 4: Summarize the results.*** |
| Townsend Deprivation Index = $Z$ score no car + $Z$ score non homeowner + $Z$ score unemployed + $Z$ score overcrowded |

### *3.2.2. Human poverty index*

The United Nations Development Program (UNDP) developed the Human Poverty Index (HPI) to measure human deprivation and poverty at the national level [65]. This composite index focuses on the most deprived individuals and on basic human capabilities within a given country, using deprivation as a proxy for poverty. The HPI includes two main versions, i.e., HPI-1, developed for developing countries, and HPI-2, designed for Organisation for Economic Co-operation and Development (OECD) economies. This distinction was intended to better reflect socioeconomic differences and the distinct forms of deprivation observed in these two contexts.

HPI-1 measures human deprivation across three core domains, i.e., longevity, knowledge, and daily comforts. HPI-2 retains these dimensions and further incorporates social exclusion as an additional factor (Table 2). Both versions seek to broaden the assessment of poverty by integrating social and health indicators alongside economic indicators. However, the usefulness of the HPI is constrained by the fact that it combines only average deprivation levels for each dimension, making it impossible to associate the results with specific groups of people.

**Table 2. Comparison of HPI-1 and HPI-2**

| *Component* | *HPI-1* | *HPI-2* |
|---|---|---|
| Longevity $(x_1)$ | Probability of not surviving to the age of 40 years | Probability of not reaching the age of 60 years |
| Knowledge $(x_2)$ | Adult illiteracy rate | Percentage of adults (aged 16–65 years) that lack essential practical literacy skills |
| Daily comforts $(x_3)$ | Percentage of the population without access to clean water and health services, and underweight youngsters under the age of 5 years | Percentage of the population living below the income poverty line |
| Social exclusion $(x_4)$ | - | The rate of long-term unemployment (12 months or more) |
| ***Calculation*** | ***HPI-1*** | ***HPI-2*** |
| Equation | $\text{HPI-1} = \left( \frac{1}{3}\left( x_1^3 + x_2^3 + x_3^3 \right) \right)^{1/3}$ | $\text{HPI-2} = \left( \frac{1}{4}\left( x_1^3 + x_2^3 + x_3^3 + x_4^3 \right) \right)^{1/3}$ |

### *3.2.3. Multidimensional poverty index*

The Multidimensional Poverty Index (MPI) goes beyond assessing poverty based solely on income and encompasses other dimensions such as health, education, and living standards [66]. Specifically, the MPI employs ten indicators across three dimensions, as shown in Table 3 and Fig. 1(f). By using the MPI, individuals experiencing deprivation in a third or more of these ten (weighted) indicators are classified as “MPI poor,” with the intensity of their poverty measured by the percentage of deprivations they endure. This metric identifies who is poor, where they live, and which dimensions contribute most strongly to their poverty. This information provides a more nuanced understanding of poverty, helping to identify specific areas of deprivation and to design more targeted interventions. The MPI is used globally to assess and monitor poverty, offering a more holistic perspective on well-being [67].

**Table 3. Dimensions and indicators of MPI**

| **Indicator** | **Deprived if living in a household where…** | **Weight** | **SDG** |
|---|---|---|---|
| ***Health*** | | | |
| Nutrition | any person younger than 70 years of age is undernourished according to their nutritional information. | 1/6 | SDG 2: Zero Hunger |
| Child Mortality | a child younger than 18 years has died within the 5-year period preceding the survey. | 1/6 | SDG 3: Health and Well-being |
| ***Education*** | | | |
| Years of Schooling | no eligible household member has completed 6 years of schooling. | 1/6 | SDG 4: Quality Education |
| School Attendanc e | any school-aged child is not attending school up to the age at which they would complete class 8. | 1/6 | SDG 4: Quality Education |
| ***Standard of living*** | | | |
| Cooking Fuel | solid fuel such as dung, agricultural crop, shrubs, wood, charcoal, or coal is used for cooking. | 1/18 | SDG 7: Affordable and Clean Energy |
| Sanitation | sanitation facility is unimproved or not available, or is improved but shared with other households. | 1/18 | SDG 6: Clean Water and Sanitation |
| Drinking Water | the source of drinking water is not safe or safe drinking water is a 30-minute or longer walk from home, round trip. | 1/18 | SDG 6: Clean Water and Sanitation |
| Electricity | there is no electricity. | 1/18 | SDG 7: Affordable and Clean Energy |
| Floor | housing materials are inadequate in any of the three components: floor, roof, and walls. | 1/18 | SDG 11: Sustainable Cities and Communities |
| Assets | no car or truck is owned, and no more than one of the following assets is owned: radio, TV, telephone, computer, animal cart, bicycle, motorbike, or refrigerator. | 1/18 | SDG 1: No Poverty |

*3.2.4. Progress-out-of-poverty index*

The Progress-out-of-Poverty Index (PPI) was first proposed in Bosnia-Herzegovina through microfinance initiatives and was later introduced more broadly by the Grameen Foundation [68]. The PPI is a scorecard-based tool that assesses poverty levels using a set of simple and verifiable indicators, such as household assets, educational level, and access to basic services. It provides a practical and quickly implementable method for measuring the distribution of country-specific poverty among participants and monitoring changes over time.

As of 2021, more than 50 developing nations have participated in the development of the PPI, which is specifically designed to quantify poverty at the household level in a particular country. As an example, the Rwandan PPI [69] is a scorecard consisting of ten questions about living conditions, household composition, education, household conditions, and ownership of durable goods. Each option in these ten questions carries a predetermined point based on how well it helps distinguish poor from non-poor households. The cumulative points from all questions contribute to the computation of the PPI score, which is subsequently translated into the likelihood of a household living below the poverty line using dedicated look-up tables.

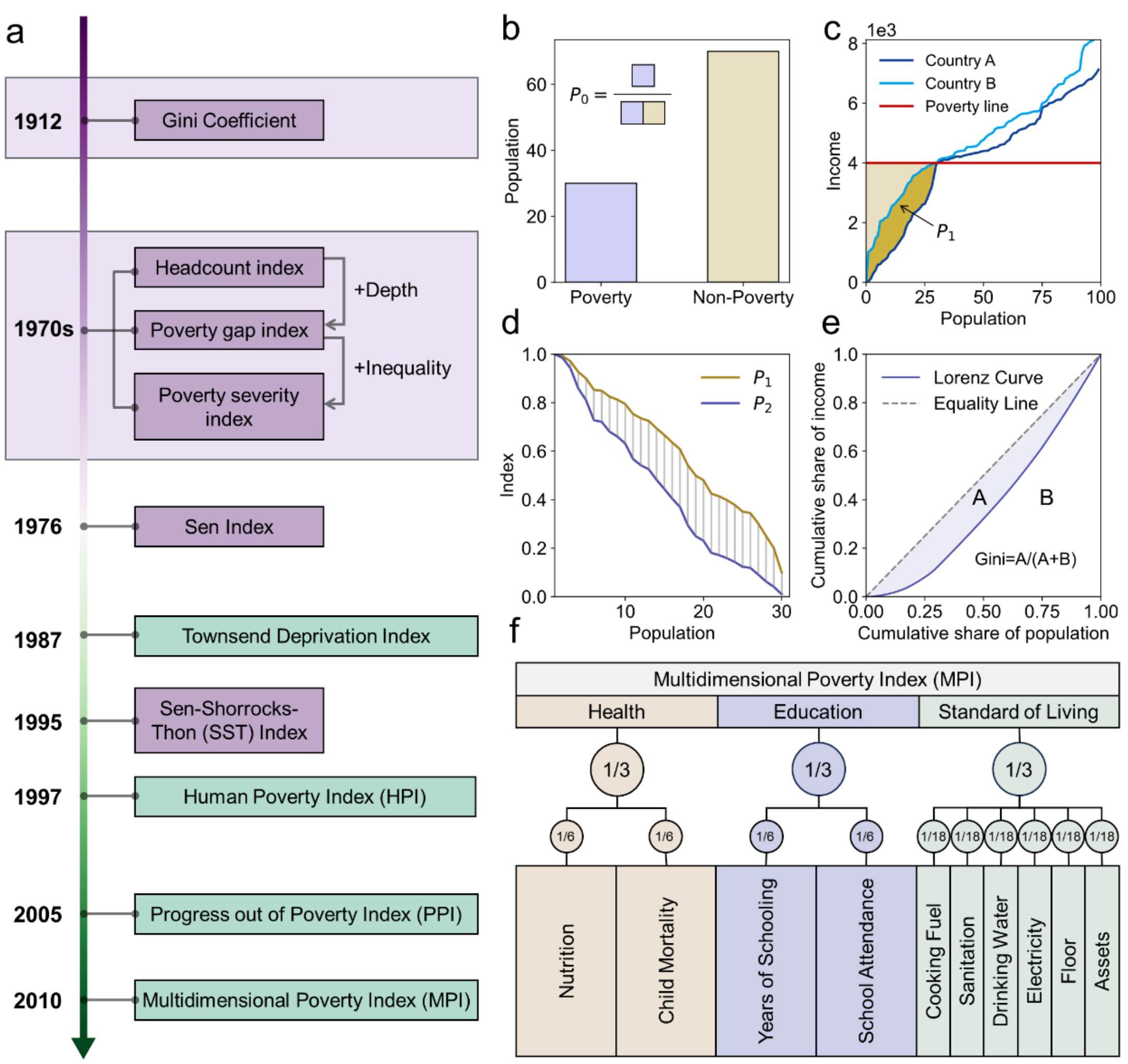

**Fig. 1. Overview of monetary-based poverty indicators and multidimensional poverty measurements.** (a) The timeline illustrates the proposed monetary-based poverty indicators (colored in purple) and multidimensional poverty measurements (colored in green). It should be noted that different indices are based on distinct conceptual frameworks and data inputs, and therefore may not yield directly comparable rankings or consistent correlations. (b) The Headcount Index ($P_0$) is calculated as the ratio of individuals below the poverty line to the total population. (c) Although countries A and B share the same $P_0$ in the chart, Country A exhibits a notably deeper level of poverty. The light tan shadow corresponds to the poverty gap index ($P_1$) of Country B, while the additional darker tan shade reflects a larger $P_1$, emphasizing a more extensive poverty depth in Country A. (d) For individuals experiencing the same level of poverty, the poverty severity index ($P_2$) has a larger value than $P_1$. $P_2$ considers inequality among the poor by weighing poverty gaps and assigns greater significance (highlighted in dash shadow) to observations further below the poverty line. (e) The Gini coefficient is calculated as the area between the Lorenz curve and the line of perfect equality divided by the total area under the line of perfect equality. (f) The MPI evaluates health, education, and standard-of-living deprivations that people may experience simultaneously in different aspects of their lives. Within each dimension, specific indicators are chosen to assess in which aspects individuals or households are deprived.

### *3.3. Data-driven poverty mapping methods*

Poverty measurement methods and indices, such as the MPI, define deprivation according to monetary, multidimensional, or capability-related criteria. Poverty mapping, by contrast, translates these measures into spatially explicit estimates. It should therefore not be understood as a distinct concept of poverty, but rather as an inferential task that connects established poverty indicators to geographical units. In practice, poverty mapping typically integrates ground-truth information from household surveys or censuses with auxiliary variables derived from census records, geospatial data, or nontraditional data sources, in order to estimate the spatial distribution of poverty across administrative or geographic scales.

The development of data-driven poverty mapping is motivated by the limited spatial and temporal coverage of conventional poverty statistics. Survey- and census-based methods, including small-area estimation, remain essential because they provide direct and interpretable measures of deprivation. Nevertheless, these approaches are costly, updated infrequently, and often spatially sparse, particularly in low- and middle-income countries and in contexts affected by conflict, displacement, rapid urbanization, or large-scale shocks.

Nontraditional data sources have broadened the empirical foundation for poverty mapping [70-72]. Remote sensing imagery, mobile phone records, social media traces, and multisource datasets can offer observations that are more frequent and spatially detailed than those provided by conventional surveys alone, although they capture poverty only indirectly. These data are commonly analyzed using regression models, CNNs, transfer learning, gradient boosting, and other supervised learning methods. Such algorithms are general-purpose machine-learning tools rather than physics-based methods in themselves. In many studies, especially those involving high-dimensional and heterogeneous data, their primary function is predictive. Accordingly, the physics- or network-related contribution of these studies is not determined solely by the final predictive algorithm, but by the features, relational structures, modeling assumptions, and interpretations through which poverty-related socioeconomic patterns are represented. Spatial interaction, mobility and communication networks, scaling behavior, spatial heterogeneity, and multiscale structure all provide lenses for describing the collective patterns through which deprivation becomes observable. In this sense, standard machine-learning models often function as estimators, while the input representations and interpretations may be informed by network-based or statistical-physics reasoning. The following sections review these data-driven approaches according to the main data sources from which poverty-related information is extracted.

## 4. Poverty mapping with remote sensing imagery

### *4.1. Data source and description*

Remote sensing (RS) refers to the acquisition of information about terrestrial objects without direct physical contact, typically through specialized sensors mounted on satellites, aircraft, or drones. As the demand for targeted poverty alleviation has grown, RS technologies have become increasingly important in poverty mapping. Their value lies not in the direct measurement of poverty, but in their capacity to capture physical, infrastructural, and activity-related proxies that are associated with different dimensions of deprivation. This role is particularly evident in the context of geographically disaggregated poverty data, which are essential for representing the spatial distribution of poverty at a finer scale, revealing the spatial heterogeneity and areal differentiation of deprivation across regions, and ensuring that development interventions are directed to the areas where they are most needed [73-76]. Typically, RS images are in the form of digital images, and the most commonly utilized types of RS data are optical satellite imagery, which includes both daytime and nighttime data obtained from space-borne platforms such as geostationary and polar orbiting satellites, and high- and very high-resolution imagery sourced from manned aircraft- or drone-based sensors. In addition, OSM provides a valuable complementary source of detailed geographic information that can be integrated with remotely sensed observations. Examples of RS imagery and OSM data are presented in Fig. 2.

#### *4.1.1. Nighttime light*

NTL data record emissions from artificial lighting at night, including those from cities, towns, villages, and gas flares, and therefore provide information on the location and population density of human settlements [77-80]. Since the 1970s, the Defense Meteorological Satellite Program (DMSP) has operated a series of satellites equipped with highly sensitive Operational Linescan System (OLS) sensors for detecting light emissions from the Earth's surface. The OLS data are acquired in two spatial-resolution modes, i.e., "fine" and "smoothed." The fine mode has a nominal resolution of 0.56 km, whereas the smoothed mode, produced by averaging five-by-five blocks of fine-resolution data, has a resolution of approximately 2.7 km. From 1992 to 2013, DMSP-OLS consistently produced a time series of annual global NTL data, and such multitemporal data have been widely used to monitor socioeconomic dynamics [81, 82]. Since 2014, the National Oceanic and Atmospheric Administration's National Geophysical Data Center (NOAA-NGDC) has advanced NTL observation by providing a separate dataset derived from the Visible Infrared Imaging Radiometer Suite (VIIRS) and Day-Night Band (DNB). Compared with DMSP-OLS, VIIRS-DNB represents a substantial improvement, offering lower detection limits, a wider dynamic range, and finer quantization [83-85]. VIIRS-DNB data are produced on a 15-arc-second geographic grid, with a spatial resolution of approximately 750 m at nadir, about twice that of the DMSP product, and provide monthly observations that continue to the present.

#### *4.1.2. Daytime satellite imagery*

Although NTL is a useful proxy for economic growth in many developing countries, it is less effective in identifying the poorest areas, where nighttime illumination is often weak or absent. Its temporal coverage is also limited, with most available datasets beginning in 1992. Daytime satellite imagery offers a promising alternative by providing a broader and more detailed geospatial perspective. Its use can be traced to early platforms such as Landsat, launched in the 1970s, which provided multispectral imagery but was constrained by relatively coarse spatial resolution and long revisit intervals. Subsequent advances in satellite technology have enabled access to higher-resolution imagery from commercial providers such as DigitalGlobe and Airbus, with such datasets available through platforms including Google Static Maps. A comparative study across five African countries found that daytime imagery outperformed NTL in poverty mapping. Specifically, daytime imagery was 81.3% more accurate in predicting poverty in areas below the absolute poverty line, and this advantage increased to 99.5% in areas where incomes were below three times the poverty line [86].

#### *4.1.3. Very high-resolution imagery*

Daytime and nighttime satellite imagery is available at relatively high spatial resolution across the global land surface. However, some applications require still finer spatial detail. This is especially the case for complex tasks such as monitoring land ecosystems for biodiversity conservation or identifying slums, which are often densely built and heterogeneously distributed within local areas. Since the launch of Ikonos, the first commercial satellite sensor with a resolution of up to 1 m, in 1999, a range of advanced sensors has become available, including the EROS series, WorldView series, and SPOT family of satellites. Equipped with state-of-the-art star trackers and onboard global positioning systems (GPS), these sensors not only provide high-spatial-resolution imagery but also allow frequent revisits to the same target area. Together with rapid data processing and broad aerial coverage, this high revisit capacity makes very high-resolution (VH-R) imagery both information-rich and globally scalable, supporting applications that require substantial spatial detail and accuracy, including precise poverty mapping.

#### *4.1.4. OpenStreetMap*

OpenStreetMap (OSM) is a collaborative, crowdsourced mapping platform designed to produce a comprehensive, freely accessible, and editable global map with detailed geographic information. Its data are generated by volunteers who draw on GPS devices, aerial imagery, and other sources to continuously update and expand the map. Because of its crowdsourced structure, OSM can be revised rapidly and thus provide relatively timely information. A key advantage of OSM is its integration of local knowledge with global geospatial data, offering insight into infrastructure, land use, and other features that are important for understanding socioeconomic contexts. Such locally informed information is particularly valuable in regions where official datasets or conventional data sources are limited or outdated.

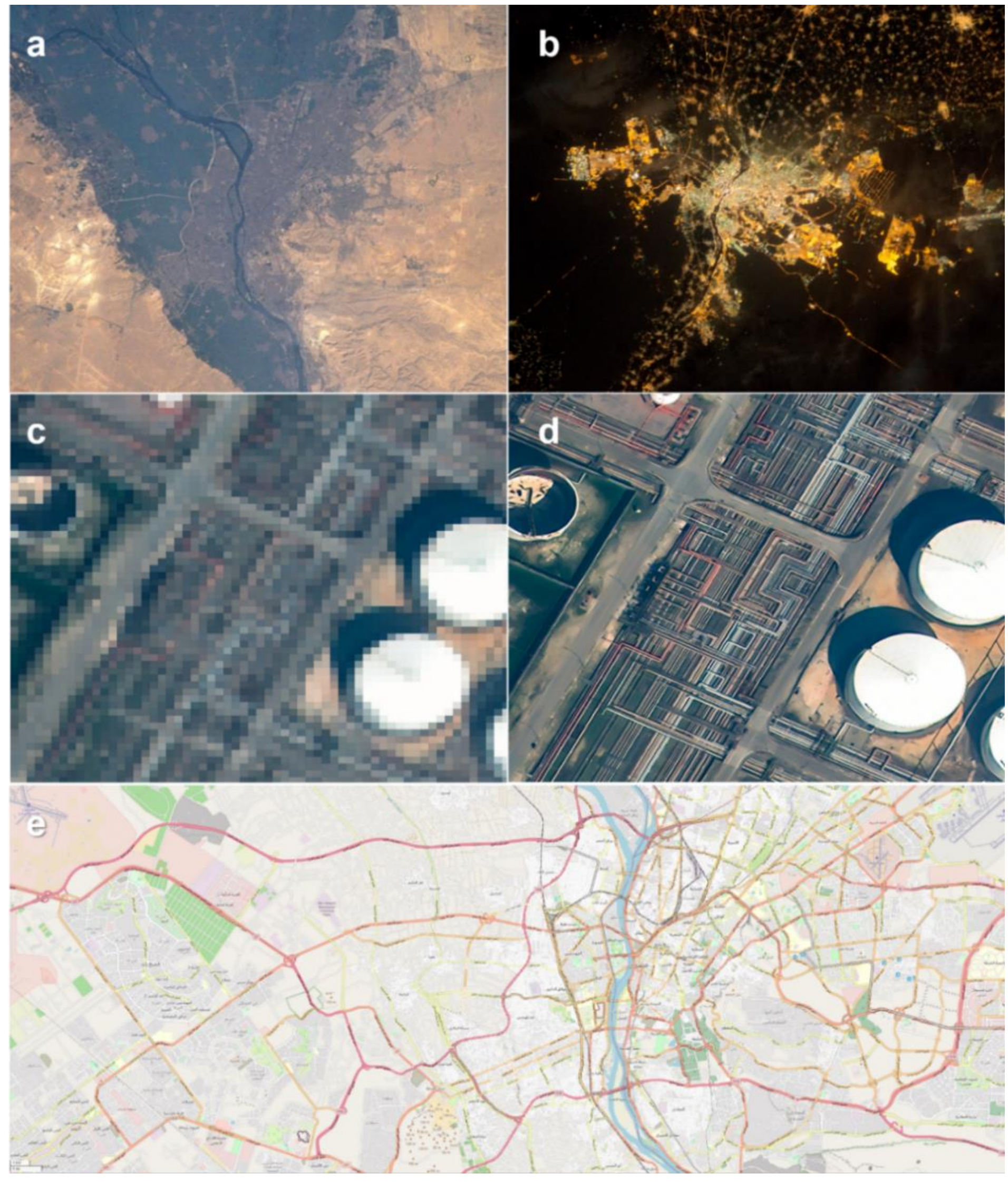

**Fig. 2. Sample images of remote sensing imagery**. Daytime satellite imagery (a) and NTL imagery (b) in the region of Cairo, the capital city of Egypt, which is the largest city in the Middle East and the second largest city in Africa after Lagos, Nigeria. Cairo has one of the highest population densities in the world (source: NASA earth observatory, *https://earthobservatory.nasa.gov/*). Oil Refinery Taranto of 1-m high resolution (c) compared to that of 30-cm very high-resolution satellite imagery (d), obtained from the commercial provider European Space Imaging (source: *https://www.euspaceimaging.com/true-30-cm-imagery/*). (e) Map of the central areas of Cairo (source: OSM, *https://www.openstreetmap.org/#map=13/30.0419/31.2010*).

### *4.2. Modeling approaches*

Remote sensing imagery provides spatially distributed observations of socioeconomic systems across multiple scales. Fine-grained pixel-level features, including luminosity, texture, settlement morphology, and infrastructure density, can be aggregated to identify broader spatial regularities associated with economic development and deprivation. In this sense, many remote sensing approaches implicitly draw on coarse-graining and scaling principles, whereby macroscopic poverty patterns arise from the spatial organization of numerous local environmental features. These regularities can then be quantified using statistical models, machine learning frameworks, and deep learning approaches for poverty estimation.

### *4.2.1. Feature-engineered regression and machine-learning models*

With prior knowledge of the potential links between satellite-derived features and economic outcomes, regression models such as linear regression and random forest (RF) [87] can be particularly effective for directly predicting pixel-wise estimates of poverty-related socioeconomic indicators, including the Wealth Index (WI) [88]. For instance, NTL is strongly correlated with GDP, population density, and energy consumption, making it a useful predictor of economic activity in linear regression models. When treated as a "structural" relationship, this association can support reliable poverty estimation through statistical and machine-learning methods. The typical procedure first involves calculating regional socioeconomic scores and extracting statistical measures of NTL values. A regression relationship is then established between these variables to evaluate their significance, with the coefficient of determination, $R^2$, commonly used to assess model fit. RF regression, although relatively simple, is often effective for identifying geospatial features and generating predictions based on correlates of poverty.

### *4.2.2. Image-based deep learning models*

Simple regression models are effective when there is substantial prior knowledge of the relationships between features and outcomes. Within certain domains of computer vision, particularly in tasks that pertain to image analysis, geospatial context can significantly enhance prediction accuracy. Various deep learning architectures have been proposed for satellite image recognition tasks, i.e., to identify objects (roof type, crop type, built-up areas, etc.), texture and spectral characteristics (NDVI [89], PanTex [90], intensity of lights, etc.). Convolutional neural networks (CNNs) , as an advanced class of deep learning models, are well suited to capturing geospatial relationships among pixels in satellite imagery [91, 92]. By automatically extracting spatial patterns and structures from imagery, these models can learn more complex features and support a more flexible, data-driven mapping from inputs to outputs. In addition, sequential images of the same location acquired over time can improve image-based inference and provide important information on temporal changes on the ground. Models such as long short-term memory networks (LSTMs) and convolutional LSTMs are particularly effective for processing image sequences, in which multiple images are fed into the model in temporal order before a prediction is made.

### *4.2.3. Transfer-learning and weakly supervised models*

High-resolution daytime satellite imagery is increasingly available at the global scale, yet reliable and geographically disaggregated poverty data remain limited. In the poorest areas, for example, nationally representative poverty surveys or DHS are often infrequent, incomplete, or unavailable, which restricts the number of labeled samples available for poverty prediction. This imbalance between abundant imagery and scarce poverty labels makes "end-to-end" training of deep neural networks challenging. Two main approaches have been used to address this problem. The first is transfer learning, which enables knowledge learned in one domain to be applied to a different but related task. In satellite-based poverty mapping, transfer learning can guide CNN models in learning image features while reducing the time and computational resources required to train a new model, even when labeled data are limited [93]. This approach was first introduced by Jean et al. [86], who used NTL intensity as a noisy but accessible proxy for economic activity. Their approach follows a three-stage process: (1) a CNN pretrained on ImageNet is fine-tuned to predict NTL values from high-resolution daytime satellite imagery; (2) the resulting intermediate CNN is then used as a feature extractor to encode visual cues associated with economic activity; and (3) these extracted features are finally used to train a regression model for predicting asset- or consumption-based poverty indicators. This strategy demonstrates the potential of transfer learning to bridge data gaps arising from inadequate survey coverage in several African countries (Fig. 3). In addition, some studies have adopted unsupervised or semi-supervised learning to exploit large volumes of unlabeled satellite imagery. These methods aim to improve model performance by making more effective use of limited labeled imagery [94, 95].

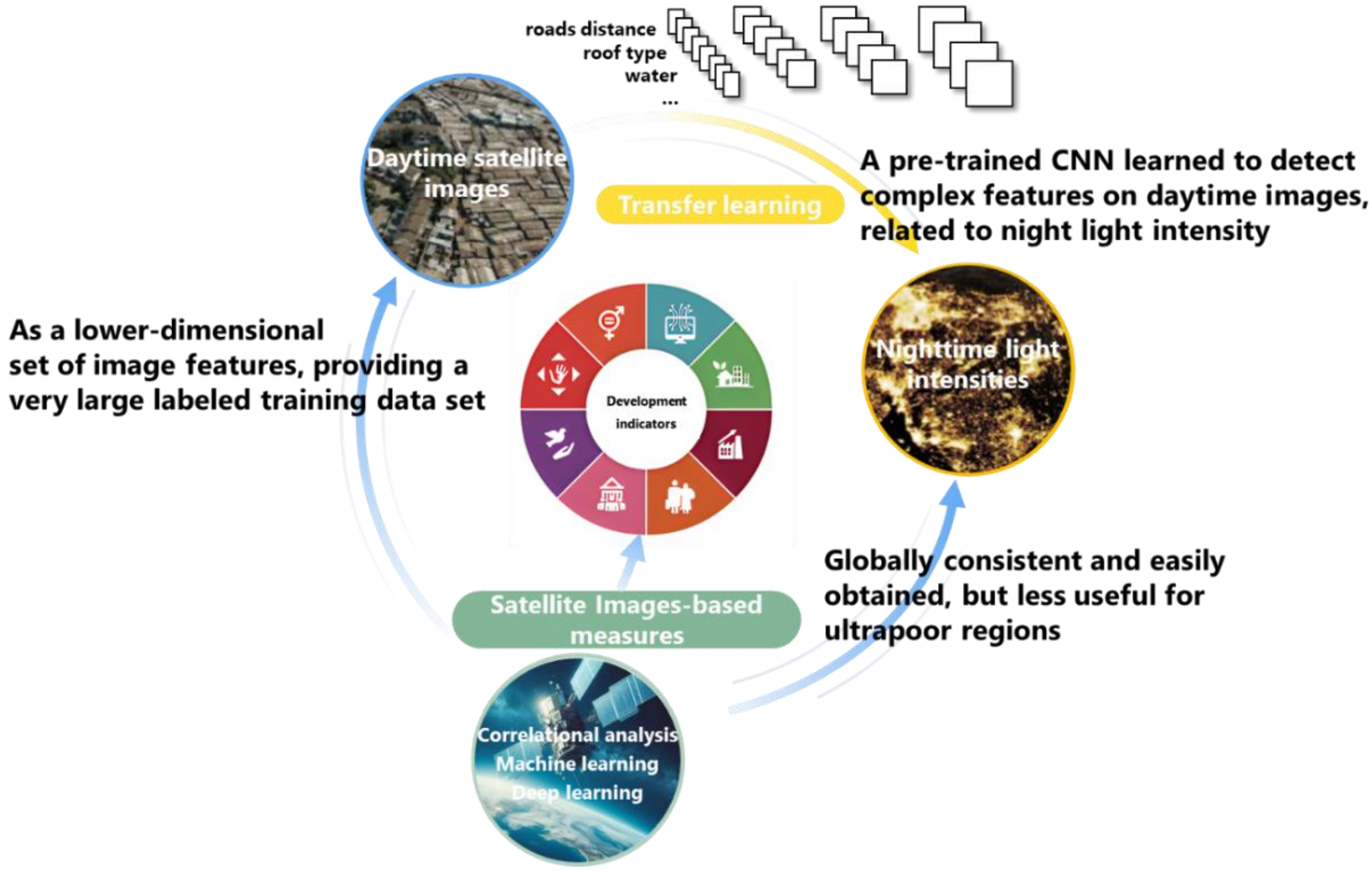

**Fig. 3. Methodology for predicting poverty using satellite imagery.** Transfer learning bridges the data gap by condensing high-dimensional daytime satellite images into lower-dimensional feature vectors, which are then used in a regularized regression model to predict wealth.

### *4.3. Related research*

#### *4.3.1. Nighttime light for economic activity and poverty estimation*

Satellites equipped with advanced cameras continuously capture detailed images of the Earth, prompting researchers to explore the potential of mapping poverty by analyzing the images. Early efforts in this field relied primarily on images of electric lights at night.

In 1997, Elvidge et al. [96] made a pioneering contribution by using NTL data as a proxy for socioeconomic development. At the national level, they identified strong correlations between lit (km$^2$) and population, per capita GDP, and electric power consumption across 21 countries. Similar associations have since been observed for other socioeconomic indicators, including gross regional product [97] and total carbon dioxide ($CO_2$) emissions [98], at both national and subnational scales (Fig. 4a–c). The COVID-19 pandemic further demonstrated the utility of NTL data for examining economic disruption, as changes in nighttime illumination were used to track economic activities affected by the pandemic [99-102].

These findings established that economic activity could be estimated using NTL data in combination with household-level expenditure data, census-derived population data, or survey-based information. Ebener et al. [103] proposed regression methods that used NTL to model the distribution of income per capita as a proxy for wealth across 171 countries at the national level. They argued that such spatial income models could inform health-related inference, given the well-established relationship between income and health outcomes. They also demonstrated the feasibility of applying this approach to subnational units in 26 countries, while emphasizing the need for country-specific models because of contextual heterogeneity. Sutton et al. [104] extended and compared the approach proposed by Ebener by applying both light-summing methods and a spatial analytic approach based on urban population estimation to predict GDP at subnational levels in China, India, Turkey, and the United States. Building on earlier work by Sutton and Costanza [105], which showed a strong correlation between aggregated NTL intensity and national GDP, they produced spatially explicit maps of marketed economic activity by combining satellite-derived light data with nationally aggregated GDP statistics. Noor et al. [106] examined the relationship between the asset-based WI and three NTL measures (mean brightness, mean distance to NTL, and proportion of area covered by NTL) at the Administrative 1 level across 37 African countries. Their results showed that mean NTL brightness had the strongest Spearman's rank correlations.

Elvidge et al. [107] later developed an NTL-based poverty index, calculated using population counts from LandScan 2004, to estimate poverty at national and subnational levels. They estimated that 2.2 billion people were living below the poverty line, approximately 15% lower than the World Bank estimate of 2.6 billion. On the basis of this satellite-derived poverty index, they also produced the first global poverty map covering 233 countries at 30-arc-second resolution (Fig. 4d). At the same time, the authors acknowledged that the accuracy of the index could be affected by regional and technical factors, including sensor saturation, population concentration, environmental lighting regulations, and interference from non-socioeconomic industrial light. Such factors may cause NTL-based proxies to diverge from actual socioeconomic conditions and generate regional anomalies in poverty estimates. Nevertheless, the study remains methodologically important because it offered an early globally consistent framework for approximating poverty patterns independently of national statistical reporting systems. In this respect, it represented an important step toward spatially disaggregated poverty mapping and more targeted poverty-alleviation interventions.

To improve the estimation of real GDP growth over time from NTL, Henderson et al. [108] developed a statistical framework that used NTL to measure the income growth for 36 countries in Africa with low-quality national income data, which is an empirical contribution to the debate about how NTL contributes to GDP growth in the countries or subnational regions. The correlation between NTL and the growth of true income is articulated as follows:

$$x_j = \beta y_j + \varepsilon_{x,j}, \tag{7}$$

where *x is* the growth of observed NTL, and *y* is the growth rate in real GDP for country *j*. $\beta$ is the elasticity of lights with respect to income. The error term in equation (7) is the noise in the way measured light growth reflects GDP growth, and the variance of $\varepsilon_x$ is denoted as $\sigma_x^2$.

Since Henderson et al. [108] brought the capacity of NTL to track changes in economic activity to wider attention, NTL has become a widely accepted proxy for economic activity. Mellander et al. [109] combined correlation analysis with Geographically Weighted Regression (GWR) to examine whether NTL could serve as a proxy for economic activity at fine geographic scales (250 by 250 meters in urban areas and 1,000 by 1,000 meters in rural areas). The GWR model is expressed as follows:

$$y_i = \beta_0(i) + \beta_1(i)x_{1i} + \beta_2(i)x_{2i} + \ldots + \beta_n(i)x_{ni} + \varepsilon_i, \tag{8}$$

where *i* denotes location, *y* is the dependent NTL variable, *x* is a population- or industry-based socioeconomic variable, and $\varepsilon$ is the error term. The model used $\beta$ as a local estimator

$$\beta'(i) = \left(\mathrm{X}^T W(i)\mathrm{X}\right)^{-1} \mathrm{X}^T W(i) Y, \tag{9}$$

where W(*i*) is a weight-specific matrix to location *i*. Results from industrial economic data in Sweden show that the correlation between NTL and economic activity is strong enough to make it a relatively good proxy for population and establishment density, but the correlation is weaker in relation to wages, which, in theory, are closely related to productivity.

In rural areas with little or no artificial lighting, McCallum et al. [110] combined NTL data with the 2015 World Settlement Footprint (WSF) to identify and quantify unlit settlements globally. Using Naïve Bayes classification and 10-fold cross-validation, they predicted wealth classes—poorer, average, and richer—based on the percentage of unlit settlements within household clusters derived from approximately 2.4 million geolocated DHS households across 49 countries in Africa, Asia, and the Americas. Their model achieved an overall accuracy of 86.6%, with higher precision in Africa (93%) than in Asia and the Americas, both of which reached 86%, as shown in Fig. 4e. To address the limitation that NTL values cannot distinguish levels of asset wealth across different unlit areas, Wang and Li [111] developed an RF prediction model that integrates multidimensional features derived from NTL and WSF data to estimate asset wealth at 500 m resolution in Africa. By extracting information such as the shortest distances to nearby light sources at different brightness thresholds and combining these measures with brightness values, the model captures variations in luminous context even within unlit areas, that is, areas with zero NTL values. The model explained 71% of the variation in asset wealth across settlements. For unlit clusters, reduced mean absolute error (MAE) and root mean square error (RMSE) by 31.35% and 31.43%, respectively, compared with models using only NTL brightness at settlement locations. Tang et al. [112] addressed the same issue from the perspective of NTL data reconstruction. They developed a global NTL dataset covering 1992–2022, with improved representation of weak light signals in sparsely lit regions. By enhancing temporal continuity and sensitivity in low-light areas, this dataset provides a more reliable basis for future poverty-related applications in rural and other dimly lit regions.

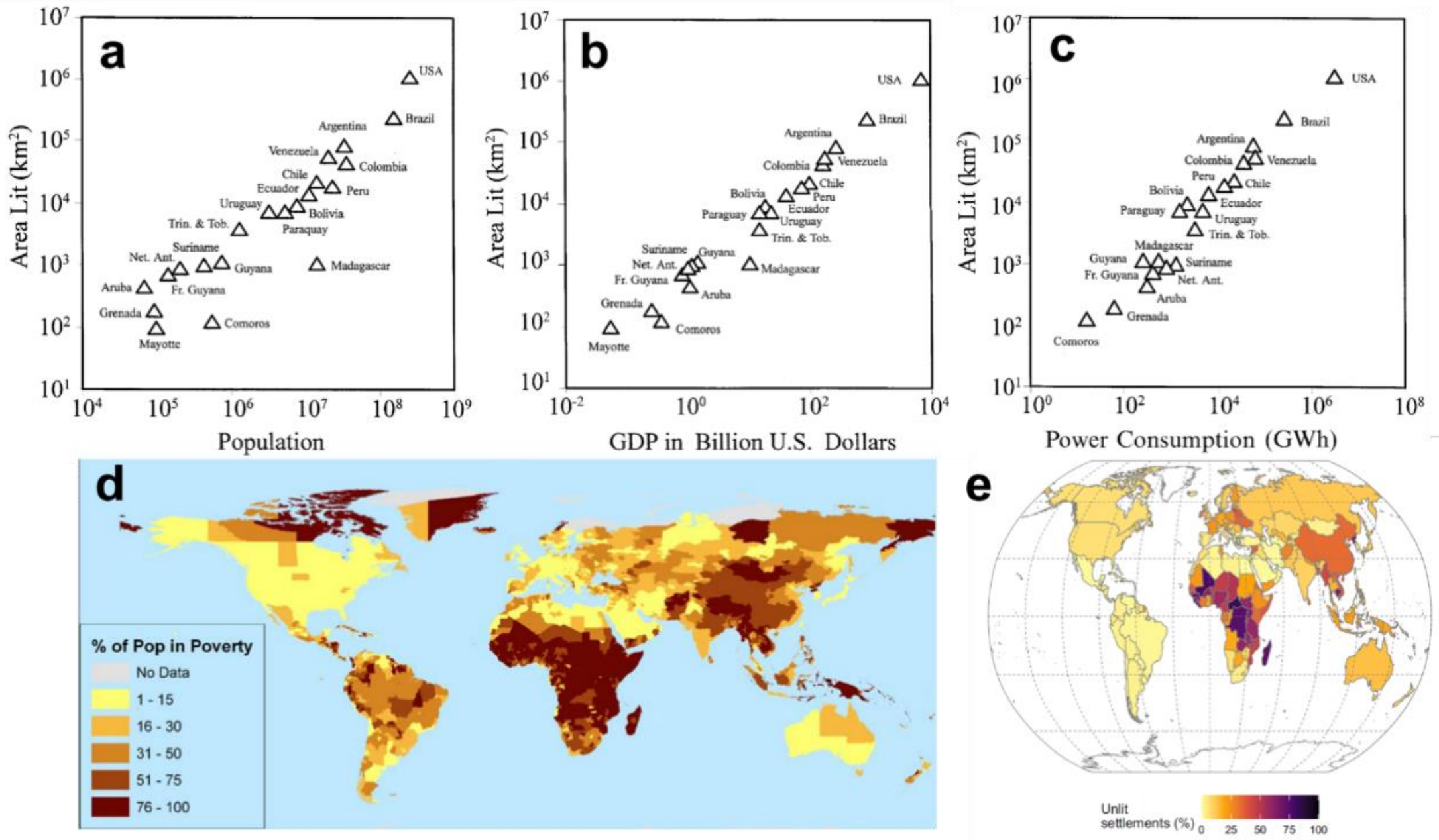

**Fig. 4. Nighttime lights reflect economic activity and are useful for observing poverty.** (a)–(c) Log-log plots showing strong relationships between lit area and population, electric power consumption, and GDP (adapted from Elvidge et al. [96]). (d) A global poverty map at the subnational level for 2543 administrative units, estimated using a satellite-derived poverty index (adapted from Elvidge et al. [107]). (e) A map showing countries categorized by the percentage of settlements, represented by building footprints, without corresponding satellite-derived nighttime radiance data across both urban and rural areas (adapted from McCallum et al. [110]).

#### *4.3.2. Daytime satellite imagery for transfer-learning-based wealth prediction*

While NTL is widely recognized as a reliable indicator of economic prosperity worldwide, areas without detectable nighttime illumination often remain insufficiently examined. The absence of nighttime radiance in these regions poses substantial challenges for generating accurate economic estimates, primarily because reliable training data are scarce.

Jean et al. [86] proposed a multistep transfer learning approach to address the problem of limited training data by training a deep learning model with an accessible but imperfect indicator of poverty. Using NTL as a noisy yet globally consistent proxy, they trained a CNN to extract features useful for poverty prediction. Through this procedure, the model was able to identify livelihood-related attributes in daytime satellite images, such as roads and urban areas, which are highly informative for poverty mapping. Across five African countries, the CNN successfully learned image features from daytime imagery that explained up to 75% of the variation in local economic conditions. Because this approach relies only on publicly available data, it has considerable potential for tracking and targeting poverty and can be scaled across countries at very low cost, including in settings where recent survey data are unavailable. Xie et al. [113] further extended this framework through deep feature transfer learning, substantially reducing the dependence on labeled data. Their method improved the accuracy of economic-condition predictions from remote sensing imagery and enhanced overall model performance in poverty mapping.

Head et al. [114] built on this line of work and showed that the method is both cost-effective and scalable, achieving strong performance in predicting asset-based wealth indices ($R^2 = 0.73$). However, using a limited sample of 11 development indicators from four countries, they found that no single country consistently produced the highest predictive accuracy across all indicators. The authors therefore emphasized that the framework is not readily transferable across countries or development outcomes, given its sensitivity to hyperparameter settings and context-specific variation. The technical details of the framework are shown in Fig. 5.

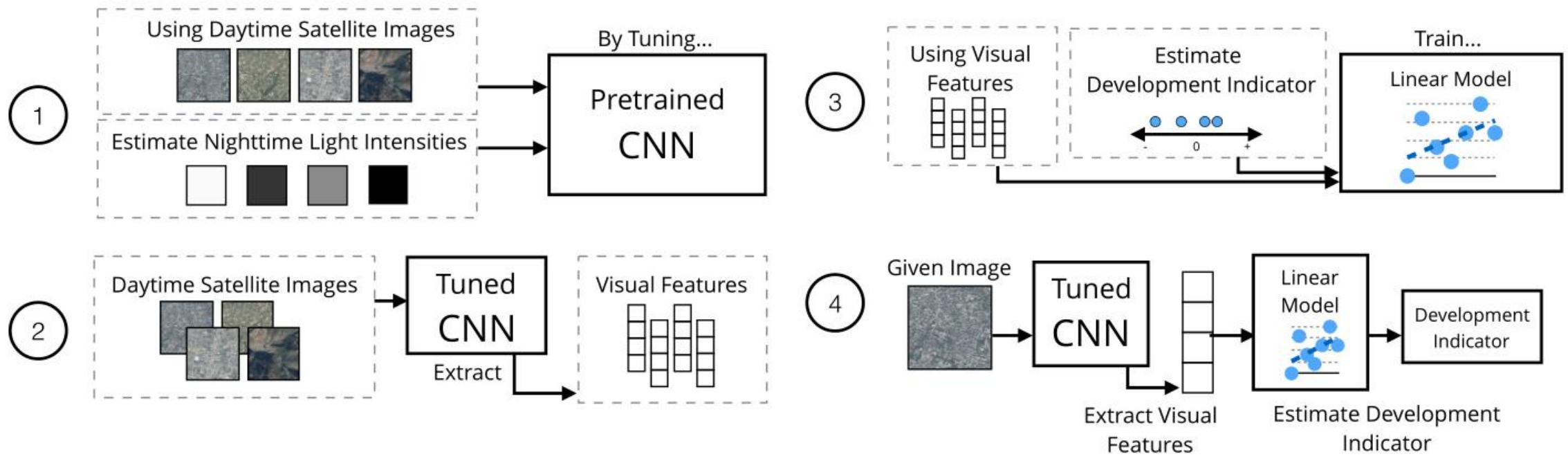

**Fig. 5. In the transfer learning process, a deep neural network is retrained on NTL data and then uses those features to predict wealth.** In Step 1, a pretrained CNN is fine-tuned to predict NTL intensity from daytime satellite images. In Step 2, satellite-derived features from the top layers of the tuned CNN are calculated. In Step 3, a linear model is trained to estimate a development indicator using ridge regression and the extracted visual features. In Step 4, given arbitrary daytime satellite images, the extracted visual features are fed into the trained linear model to predict the development indicator. The figure is adapted from Head et al. [114].

Yeh et al. [27] introduced a satellite-based deep learning approach and demonstrated its scalability by generating a wealth map for 19,669 villages across 23 African countries. Unlike earlier methods that used NTL as intermediate labels in a transfer learning framework, their approach incorporated both daytime multispectral imagery and NTL as direct inputs. Each input was processed through a dedicated CNN (ResNet-18), and the resulting feature vectors were concatenated and passed through a final ridge regression layer for wealth prediction. Their dual-input model explained up to 70% of the variation in village-level asset wealth and up to 50% of the variation in district-level changes in asset wealth over time. This approach demonstrates strong potential for tracking localized economic well-being across space and time in data-sparse regions. Ratledge et al. [115] developed a two-stage methodology to estimate the causal impact of rural electrification on livelihoods in Uganda. They first trained a CNN on daytime satellite imagery, excluding NTL, to predict village-level asset wealth. They then applied machine-learning-based causal inference techniques, including matrix completion and synthetic control with elastic net, to estimate treatment effects. Their results showed that communities that received grid access in 2011–2012 experienced a statistically significant increase of 0.15 standard deviations in asset wealth by 2016. These findings highlight the potential of daytime satellite imagery for scalable policy evaluation in data-sparse settings.

#### 4.3.3. *VH-R imagery for fine-scale deprivation mapping*

VH-R satellite imagery provides a fine-grained view of the Earth's surface, making it possible to observe pockets of extreme poverty that are often embedded within wealthier neighborhoods. One major application of such data is the construction of composite indicators that offer a more comprehensive representation of poverty. As a potential complement to census or household survey data, VH-R imagery helps detect small-scale features associated with poverty, including housing quality, infrastructure conditions, and land-use patterns.

By combining household survey data with environmental variables, Rogers et al. [116] used 1-km AVHRR satellite data and discriminant analysis to map poverty levels in Uganda. Their method, based on environmental predictors, produced estimates comparable to those generated by traditional small-area methods and included statistical measures of model precision. However, the approach performed more accurately at coarser spatial resolutions. Engstrom et al. [117] used high-spatial-resolution satellite imagery from Sri Lanka to extract features such as building density, shadow area (as a proxy for height), car counts, road infrastructure, farmland type, and roof material. Using a Lasso-regularized linear model, they found that these features were highly predictive of both poverty rates and average per capita consumption at the village level. Among these variables, building density, roof type, and especially shadow area were strongly correlated with economic well-being, with the model explaining more than 60% of the variation in poverty and consumption. Watmough et al. [118] examined the relationship between environmental characteristics and household poverty in rural Kenya using high-resolution satellite imagery. They developed a multilevel modeling framework that incorporated remote sensing features at four spatial levels, i.e., the homestead, agricultural land, village cluster, and wider periphery. Using a classification tree, their model identified the poorest 40% of households with 62% accuracy, representing a 10% improvement over traditional single-level approaches.

Another important application for VH-R satellite imagery is the observation and mapping of informal settlements, or slum areas, together with their populations and changes at a global scale. This concept was first proposed at an expert meeting in 2008 [119].

Supported by methodological and technological advances, various algorithms have been developed to rapidly inventory the locations and physical characteristics of slums with limited operator intervention. Based on analytical approach, these algorithms can be broadly grouped into multiscale analysis, which distinguishes features according to properties emerging across different spatial scales, including fractal and lacunarity measures, and object-structure analysis, which extracts information about the spatial arrangement and structure of objects in an image through structural, statistical, model-based, or transform methods [120-122]. Kuffer et al. [123] conducted a systematic review of slum mapping studies across four dimensions, i.e., contextual factors, physical slum characteristics, data and requirements, and slum extraction methods. They reported that machine learning approaches were among the most promising methods for developing a global slum inventory because of their flexibility and capacity to handle large sets of indicators. Given the diversity of slum forms, the authors emphasized the need to establish more systematic linkages between higher-level image elements and slum characteristics in order to improve algorithmic robustness and transferability.

Kit et al. [124] applied the concept of lacunarity to identify slums in Hyderabad, India, using VH-R satellite imagery. They compared two preprocessing methods, i.e., principal component analysis (PCA) and line detection, to generate binary images suitable for lacunarity analysis. Their results showed that the line detection algorithm produced more reliable binary datasets and improved slum identification, particularly within the lacunarity value range of 1.10–1.15, where classification accuracy reached 83.3%. Kit and Lüdeke [125] then improved their lacunarity-based slum detection algorithm using Canny- and LSD-based imagery preprocessing routines. Their approach proved useful for capturing spatiotemporal patterns of slum development in the urban area of Hyderabad between 2003 and 2010, indicating that higher lacunarity values were associated with informal settlements.

Kohli et al. [126] developed a Generic Slum Ontology (GSO) based on expert input from 16 countries, organizing slum characteristics into three spatial levels. This ontology supports object-oriented image analysis and helps clarify slum definitions for systematic classification. Kuffer et al. [127] evaluated the use of gray-level co-occurrence matrix (GLCM) variance to identify slums in three cities, i.e., Mumbai, Ahmedabad, and Kigali, based on VH-R imagery. They achieved classification accuracies of 84–88% with a rule-based approach. In Mumbai, the accuracy increased to 90% when GLCM was combined with spectral data and NDVI in an RF classifier. These findings demonstrate the potential of the method for transferable slum detection, particularly when applied at the HUP level. More recently, Stark et al. [128] highlighted the complexity of slum classification and proposed an efficient approach for classifying intricate urban patterns in noisy datasets using transfer learning and Monte Carlo dropout. Unlike traditional classification methods that produce a single prediction, Monte Carlo dropout enables the modeling of a predictive distribution.

$$\bar{y} = \frac{1}{T}\sum_{t=1}^{T} y^{(t)}, \tag{10}$$

where $\bar{y}$ is the averaged prediction divided by the number of Monte Carlo runs, $T$ is the total number of Monte Carlo runs, and $y^{(t)}$ is the prediction of the *t*-th forward pass. They also introduced a custom CNN, the Slum Transfer Network (STnet). STnet achieved a high F1 score of 86.24%, producing results comparable to well-known models such as ResNet50 and Xception while substantially reducing processing time.

As volunteered geographic information (VGI) platforms such as OSM have become increasingly popular, Mahabir et al. [129] conducted a comparative case study in Nairobi, Kenya, to assess the spatial coverage and utility of VGI platforms, including OSM and Google Map Maker, relative to authoritative datasets from RCMRD. They found that RCMRD provided the most comprehensive but outdated road data, whereas OSM offered denser coverage in slum areas because of targeted mapping initiatives such as Map Kibera. Google Map Maker, by contrast, provided better coverage in some rural and green areas. Their study highlights the complementary value of VGI for addressing data gaps in dynamic urban environments, especially where authoritative sources are incomplete or outdated. Mahabir et al. [130] further suggested that other emerging geospatial data sources, such as volunteered geographic information, should be considered alongside technological advances and growing data trends when developing a more comprehensive framework for slum detection and mapping.

Table 4 provides a summary of the research reviewed on poverty mapping using remote sensing imagery.

**Table 4. Summary of research studies on mapping poverty based on remote sensing imagery.**

| Authors (Year) | Data Period | Target Variable | Study Area | Modeling Approach | Performance |
|---|---|---|---|---|---|
| ***Nighttime lights for economic activity and poverty estimation*** | | | | | |
| Elvidge et al. (1997) | 1994–1995 | GDP and electric power consumption | 21 countries | Linear regression | $R^2$ = 0.97(GDP), 0.96(power) |
| Ebener et al. (2005) | October 1994–March 1995 | GDP per capita | 171 countries | Multivariate regression | $R^2$ = 0.826 |
| Sutton et al. (2007) | 1992–1993, 2000 | Sub-national GDP | India, China, Turkey, and U.S. | Linear regression | $R^2$ = 0.72–0.96 |
| Sutton and Costanza (2002) | 1995-1996 | Gross State Product | U.S. states | Linear regression | $R^2$ = 0.86 |
| Noor et al. (2008) | 2000 | WI | 37 countries in Africa | Correlation analysis | $r$ = 0.64 |
| Elvidge et al. (2009) | 2003 | Poverty index | 233 countries | Linear regression | $R^2$ = 0.722; RMSE = 4.22 |
| Henderson et al. (2012) | 1992–2003 | GDP growth | 187 countries | Panel fixed-effects regression | $R^2$ = 0.661 |
| Mellander et al. (2015) | 2006 | Wage income, establishments, and population | Sweden | Correlation analysis and GWR | $R^2$ = 0.700(wage income), 0.757(establishment), 0.763 (people) |
| McCallum et al. (2022) | 2015 | Community wealth class | 49 countries | Naive Bayes classification model | Accuracy = 86.6% |
| Wang and Li (2024) | 2015 | WI | 33 countries in Africa | RF | $R^2$ = 0.71 |
| ***Daytime satellite imagery for transfer-learning-based wealth prediction*** | | | | | |
| Jean et al. (2016) | - | Consumption expenditure and asset wealth | Nigeria, Tanzania, Uganda, Malawi, and Rwanda | CNN and ridge regression | $R^2$ = 0.37-0.55 (consumption), 0.55-0.75 (assets) |
| Xie et al. (2016) | 2015 | Poverty probability | Uganda | CNN and logistic regression | AUC = 0.761, Accuracy = 0.716 |
| Head et al. (2017) | 2010–2013 | Asset wealth index | Rwanda, Nigeria, Haiti and Nepal | CNN and ridge regression | $R^2$ = 0.51-0.74 |
| Yeh et al. (2020) | 2009–2017 | Asset wealth index | 23 countries in Sub-Saharan Africa | CNN | $R^2$ = 0.70 (village), 0.83 (district) |
| Ratledge et al. (2022) | 2006–2016 | WI | Sub-Saharan Africa and Uganda | CNN | $R^2$ = 0.64 (SSA); $R^2$ = 0.63 (Uganda) |
| ***VH-R imagery for fine-scale deprivation mapping*** | | | | | |
| Rogers et al. (2006) | 1992–1996 | Household expenditure | Uganda | Discriminant analysis | $R^2$ = 0.973 |
| Engstrom et al. (2021) | - | Consumption per capita | Sri Lanka | CNN and Lasso regression | $R^2$ = 0.608 |
| Watmough et al. (2019) | September 2004 | Household-level wealth | Sauri, Kenya | Classification tree | Accuracy = 0.62 for poorest households |
| Kit et al. (2012) | 2003 | Slums | Hyderabad, India | PCA and line detection algorithms | Accuracy = 83.33% |
| Kit and Lüdeke (2013) | 2003-2010 | Slums | Hyderabad, India | Canny edge detection and line-segment detection | Accuracy = 87% |
| Kohli et al. (2012) | - | Slums | Kisumu, Kenya | Generic Slum Ontology | - |
| Kuffer et al. (2016) | 2004, 2009 | Slum area, formal area, vegetation, and bare soil | Mumbai, Ahmedabad and Kigali | GLCM variance and RF | Accuracy = 0.84-0.88 |
| Stark et al. (2024) | 2021 | Slums | 8 cities of the Global South | STnet CNN and Monte Carlo dropout | $F_1$ = 0.862, precision = 0.872, recall = 0.864 |

## 5. Poverty mapping with mobile phone data

### *5.1. Data source and description*

While remote sensing (RS) offers efficient and scalable observations of physical environments and natural resources, it provides relatively limited information on population behavioral dynamics. This limitation constrains its ability to capture the fine-grained behavioral processes that shape socioeconomic conditions. Large-scale mobile phone data, by contrast, provide behaviorally grounded information on human activities at unprecedented spatial and temporal resolution [131]. They therefore constitute a valuable complement for poverty mapping, particularly in regions where conventional socioeconomic data are sparse, outdated, or unavailable [36, 132].

Mobile phone data include several modalities, such as call detail records (CDRs), cellular signaling data (CSD), base station logs, device usage information, and location traces derived from positioning technologies including GPS and location-based services (LBS). Beyond their use as raw records, these data can be transformed into communication and mobility networks, allowing interaction patterns and movement behaviors to be examined in a structured way. Such representations provide insights into socioeconomic disparities and can inform targeted interventions in economically disadvantaged regions [36, 37].

#### *5.1.1. Call detail records*

CDRs are routinely collected by mobile network operators for billing and network management. They record communication events generated when users make or receive calls, or when they use messaging services such as SMS or MMS. A typical record contains a timestamp, anonymized identifiers for the caller and recipient, call duration, and the identifiers of the cellular towers that handle the interaction (Fig. 6a) [133, 134].

#### *5.1.2. Cellular signaling data*

CSD are produced through continuous interactions between mobile devices and network infrastructure. Unlike CDRs, they are not restricted to active communication events; they also capture passive signaling processes, including device registration, handovers between base stations, and periodic network updates. This gives CSD substantially higher temporal resolution for observing user presence and mobility. Location information is recorded whenever a device connects to a base station, including during idle states. Depending on operator protocols, periodic updates may occur after a specified period of inactivity, such as 15–30 minutes, enabling more continuous observation of user trajectories than event-driven CDR data.

#### *5.1.3. Mobile communications network*

In poverty estimation studies, mobile communications networks are commonly constructed from CDR data to represent patterns of social interaction. In these networks, nodes correspond to individual users, and edges represent communication events such as calls or messages between them. At an aggregated level, communication can also be modeled between spatial units by treating cellular towers as nodes and defining edges according to communication volume between towers. The spatial coverage of each tower is commonly approximated using Voronoi tessellation based on tower locations (Fig. 6b). These network representations allow structural properties such as connectivity, centrality, and community organization to be analyzed.

#### *5.1.4. Mobility network*

Mobility networks are constructed from location-based data sources and represent human movement across space. They can be analyzed at multiple scales, from individual trajectories to aggregated population flows [135, 136]. Individual mobility paths may be reconstructed from data such as CDRs, GPS traces, or smartcard records (Fig. 6c), while aggregated trajectories reveal collective mobility patterns and flows between locations (Fig. 6d). In network representations, nodes correspond to individuals or spatial units, and edges denote movements between them. These networks capture important dimensions of human behavior, including commuting patterns, travel routines, migration processes, and social interaction structures [137].

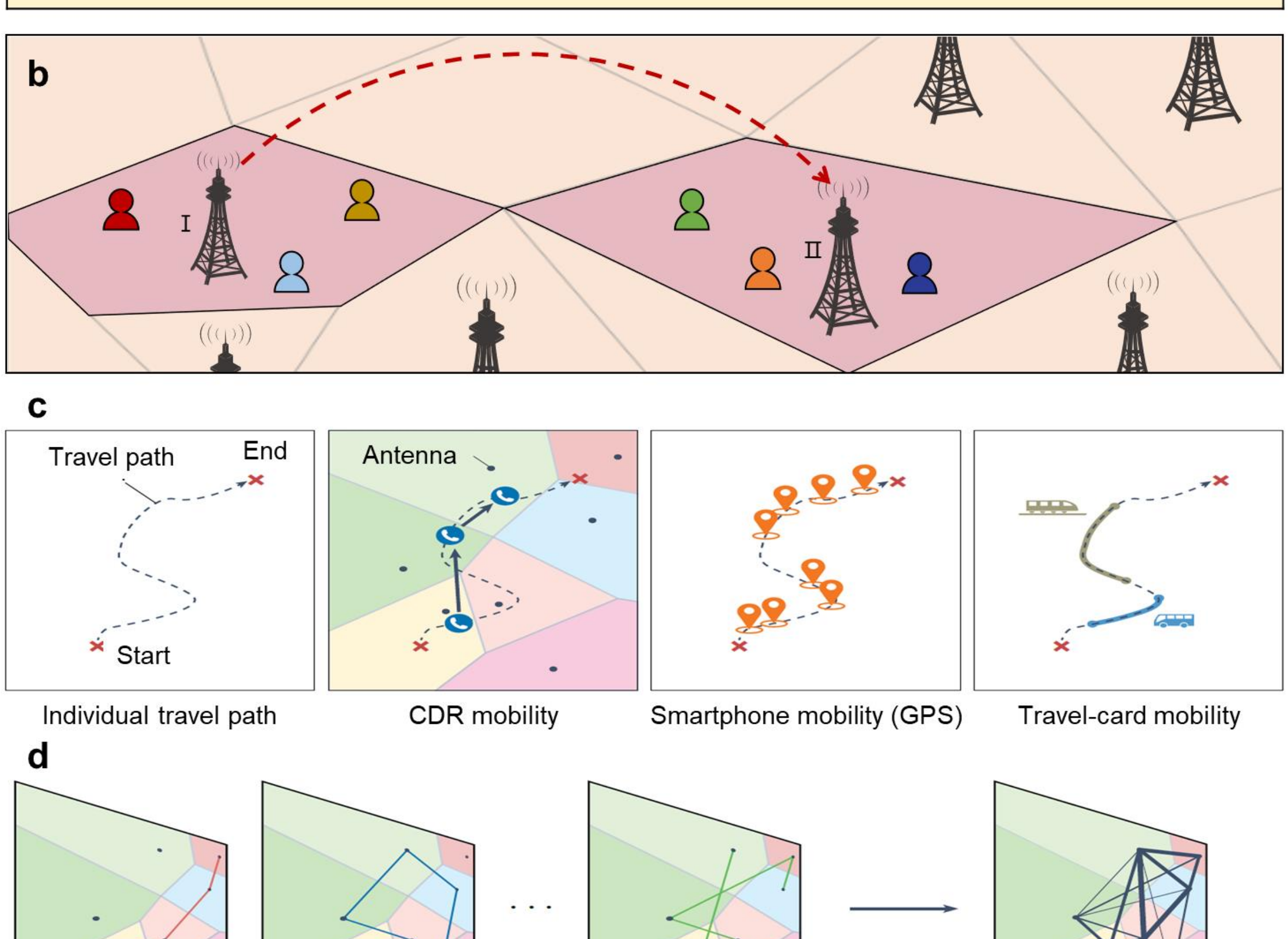

**Fig. 6. CDRs, mobile communications network, and mobility network.** (a) Example of CDRs. (b) Example of a mobile communication network distributed across cellular coverage areas ("cells"), whose spatial boundaries are determined by fixed base stations and commonly approximated using Voronoi tessellation. (c) Individual mobility trajectories reconstructed from location-based data. (d) Collective mobility patterns obtained by aggregating individual trajectories. Panels (c) and (d) are adapted from Pappalardo et al. [138].

## *5.2. Modeling approaches*

Mobile phone data record large-scale human interactions and mobility processes at high spatial and temporal resolution. Individual communication events and movement trajectories collectively form dynamic social and mobility networks, through which population-level regularities can emerge from large numbers of decentralized behaviors. For example, differences in mobility diversity, communication structure, activity regularity, and spatial accessibility often become increasingly organized at aggregated scales, revealing persistent socioeconomic inequalities across regions and populations. To quantify these relationships, existing studies typically extract statistical features from communication behaviors, mobility patterns, and network structures, and then combine them with correlation analysis and machine learning to estimate socioeconomic indicators.

### *5.2.1. Correlation- and regression-based approaches*

Mobile phone data, including CDRs and airtime credit purchases, provide useful behavioral signals associated with socioeconomic conditions [36, 139]. Based on domain knowledge and statistical analysis, various features can be constructed to characterize individual mobile phone use. These include call-related patterns, such as call frequency, duration, and temporal distribution; social network characteristics, such as network size, communication frequency, and contact diversity; mobility indicators, such as the number of activity locations, activity entropy, and radius of gyration; and proxies for economic activity, such as airtime expenditure or top-up amounts. These features can be incorporated into linear regression or related statistical models to construct composite indicators that function as proxies for SES. Correlation analysis is then used to evaluate how these indicators relate to conventional socioeconomic metrics. A major strength of this approach is its interpretability, as it allows individual features to be examined directly in terms of their direction, magnitude, and statistical significance in relation to socioeconomic outcomes.

### *5.2.2. Feature-based machine-learning approaches*

Machine learning models, including logistic regression, RF, gradient boosting decision tree [140], and LightGBM [141], are widely applied to predict socioeconomic indicators from features derived from mobile phone data. A typical workflow begins with mobile phone data obtained from telecommunications operators and ground-truth socioeconomic indicators collected from government statistics, household surveys, or other authoritative datasets. Behavioral features, including mobility patterns and social interaction measures, are then extracted at the individual level. Home locations are subsequently inferred, commonly from the most frequently used cell tower during nighttime hours [142, 143] or via residential location algorithms [144]. When regional-level prediction is required, these individual-level features are aggregated to predefined spatial units.

Prediction models are then trained using the extracted features, with the target variable specified according to the application. Depending on the type of outcome, the task may be formulated as classification, such as distinguishing poor from non-poor populations, or as regression, such as predicting continuous indicators including DHS-derived wealth indices, per capita income, or consumption. After training and validation, the models can be applied to estimate socioeconomic conditions in regions or populations where direct measurements are unavailable (Fig. 7). Compared with correlation-based approaches, machine-learning methods are generally better able to capture nonlinear relationships and complex interactions among features, which often improves predictive performance. This gain in accuracy, however, may be accompanied by lower interpretability, particularly when more complex model architectures are used.

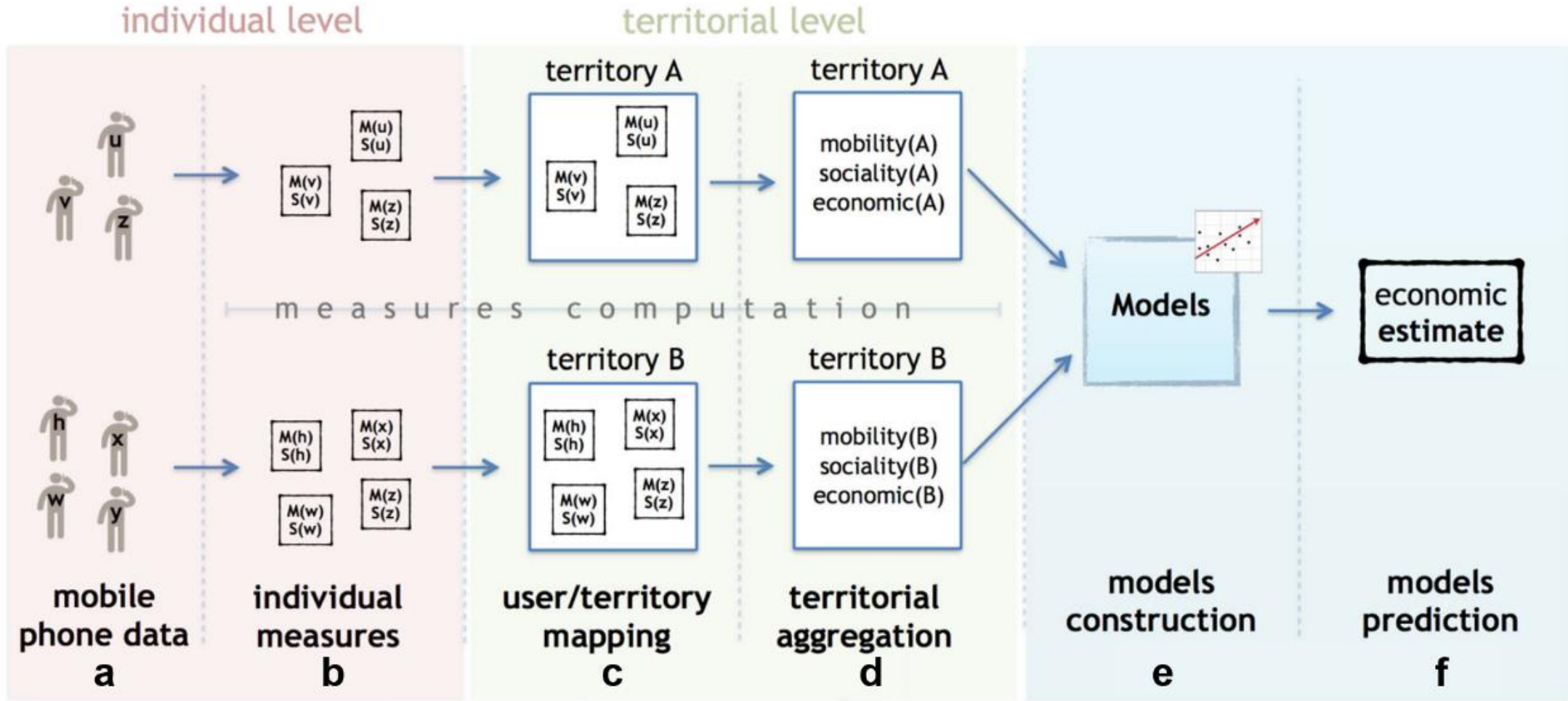

**Fig. 7. Framework for estimating poverty using mobile phone data.** The framework begins with the collection of mobile phone data, particularly CDRs, which capture large-scale communication activities (a). These data are then processed through feature engineering, through which individual-level statistical indicators, social network metrics, and mobility measures are extracted (b). Based on these data, individuals' locations are inferred, typically to identify residential or home areas (c). The derived individual-level features are subsequently aggregated to regional units to enable population-level analysis (d). These aggregated features are then used to construct predictive models (e), which are ultimately applied to estimate socioeconomic indicators or poverty levels in areas lacking ground-truth data (f). The figure is adapted from Pappalardo et al. [134].

### *5.3. Related research*

#### *5.3.1. Mobile phone usage for socioeconomic and poverty inference*

Mobile phone data constitute a rich source of digital traces that capture multiple dimensions of human behavior. As mobile communication infrastructure has expanded, a substantial share of the global population has come within the coverage of mobile networks, making it possible to collect large-scale behavioral data such as CDRs and airtime transactions. Early studies in this area examined whether socioeconomic inequality is reflected in measurable patterns of everyday digital behavior. By analyzing communication intensity, mobility range, and social-network structure in mobile phone records, these studies found that individuals and regions with different socioeconomic conditions often display distinct activity patterns and spatial behaviors. This body of work provided an important empirical basis for subsequent research on poverty mapping, socioeconomic inference, and mobility-based assessments of social inequality.

Blumenstock and Eagle [145] offered one of the earliest large-scale empirical analyses linking mobile phone usage patterns to socioeconomic inequality in Rwanda. By integrating survey data with operator-level CDRs, they showed that phone ownership and usage were strongly stratified. Phone users were substantially wealthier and better educated than the general population, with predicted annual expenditures more than twice the national average ($1,725 versus $753). Among phone owners, wealthier individuals also showed more intensive phone activity, including larger communication networks, longer calls, and higher airtime consumption. These findings demonstrated that behavioral traces contained in mobile phone records systematically reflect underlying socioeconomic inequality.

Extending this line of work from interview-based analysis to population-scale census integration, Frias-Martinez and Virseda [146] investigated the relationship between mobile phone usage and socioeconomic conditions using nationwide census data from a Latin American country. They combined five months of CDR and SMS records from more than 10 million subscribers with census indicators across 12 cities, constructing consumption, social-network, and mobility features at the base transceiver station level. Their results showed that higher socioeconomic areas were associated with more intensive phone use, stronger communication reciprocity, geographically wider social ties, and larger mobility ranges. Mobility-related indicators, including the number of visited base transceiver stations and radius of gyration, showed particularly strong positive correlations with socioeconomic level. The study further showed that mobile phone features could closely approximate census variables, with $R^2$ values of 0.83 for socioeconomic level, 0.82 for primary education, and 0.86 for household goods ownership. These findings underscored the potential of combining large-scale mobile phone metadata with census information for scalable socioeconomic mapping without relying on costly household surveys.

Rubio et al. [147] examined cross-country variation in mobility behavior using large-scale CDR data from developing and advanced economies. Based on four months of mobile phone records from 220,000 users in an advanced economy and 190,000 users in a developing economy, they compared average travel distance, area of influence, and the geographic sparsity of social networks. Users in the developing economy generally travelled shorter distances, moved within smaller activity areas, and maintained geographically closer social ties than those in the advanced economy. At the same time, a small group of long-distance travelers in the developing economy displayed mobility ranges even larger than those observed in the advanced economy. These results indicate that mobility patterns and the geography of social networks differ across economic contexts.

Xu et al. [148] analyzed the relationship between human mobility and socioeconomic status from a comparative urban perspective. Using large-scale mobile phone data from Singapore and Boston, they derived six mobility indicators, i.e., radius of gyration, number of activity locations, activity entropy, travel diversity, k-radius of gyration, and unicity. These indicators were then linked to neighborhood-level socioeconomic proxies. The study found that the relationship between mobility and SES was not consistent across cities. Wealthier users tended to travel shorter distances in Singapore but longer distances in Boston, a difference likely related to the central location of affluent neighborhoods in Singapore and their more peripheral location in Boston. By contrast, indicators capturing the diversity and regularity of daily activities, such as the number of activity locations, activity entropy, and travel diversity, showed limited variation across socioeconomic groups in both cities. These findings suggest that mobility–SES relationships are shaped not by wealth alone, but also by urban spatial structure, housing distribution, and access to employment and activity opportunities.

*5.3.2. Communication networks for poverty estimation*

Beyond individual mobility and phone usage intensity, an important line of research has examined the structural properties of communication networks embedded in mobile phone records. Mobile phone interactions naturally generate large-scale social graphs, in which calls and messages capture not only communication frequency but also the organization of social relationships, including network size, tie strength, reciprocity, diversity, and community structure. Network diversity, interregional connectivity, and communication segregation therefore offer useful perspectives for analyzing how socioeconomic inequality is reflected in large-scale communication systems.

The seminal study by Eagle et al. [149] identified a strong association between social network diversity and community-level economic development in the United Kingdom. Using a large-scale national mobile communication network matched with the Index of Multiple Deprivation (IMD), the study quantified both social diversity and spatial diversity through entropy-based measures of communication distributions across social ties and geographic regions (Fig. 8a and b). Their analyses showed that communities with more diverse communication networks consistently exhibited lower deprivation and higher socioeconomic status, with social diversity ($r = 0.73$) showing a stronger association with IMD than communication volume or number of contacts alone. The relationship remained robust using Burt's structural hole metric ($r = 0.72$), and a composite diversity indicator constructed via PCA further improved the correlation ($r = 0.78$). The study provided one of the earliest large-scale demonstrations that the structural organization of communication networks contains measurable signatures of regional socioeconomic inequality.

Smith-Clarke et al. [150] further examined the potential of aggregated communication networks for fine-grained poverty estimation in developing countries. Using large-scale CDR data from Côte d'Ivoire and another developing region, they proposed several interpretable network-based indicators, including communication activity, gravity residuals, network entropy, degree centrality, and introversion. Their analyses showed that these features were strongly associated with regional poverty levels. Areas with lower communication activity, lower network entropy and degree centrality, and weaker interregional interactions tended to have higher poverty rates. This study highlighted the value of interpretable network metrics derived from aggregated call flows, showing that mobile communication networks can support low-cost, privacy-preserving, and spatially detailed poverty mapping in data-scarce regions.

Mao et al. [151] investigated how mobile communication networks reflect regional socioeconomic disparities in Côte d'Ivoire using 2.5 billion call and SMS records from the Orange D4D dataset. They introduced "CallRank," a PageRank-based measure of regional centrality in communication networks, and found that economically developed regions consistently had higher CallRank values. The study also showed that regions initiating a larger proportion of interregional calls tended to have higher income and lower poverty rates, with correlations reaching 0.80 and -0.83, respectively. Rich-club analysis further revealed pronounced communication segregation, whereby wealthier regions communicated disproportionately more with one another than with poorer regions. These findings show that large-scale communication network topology can capture regional economic hierarchy and social segregation in developing countries.

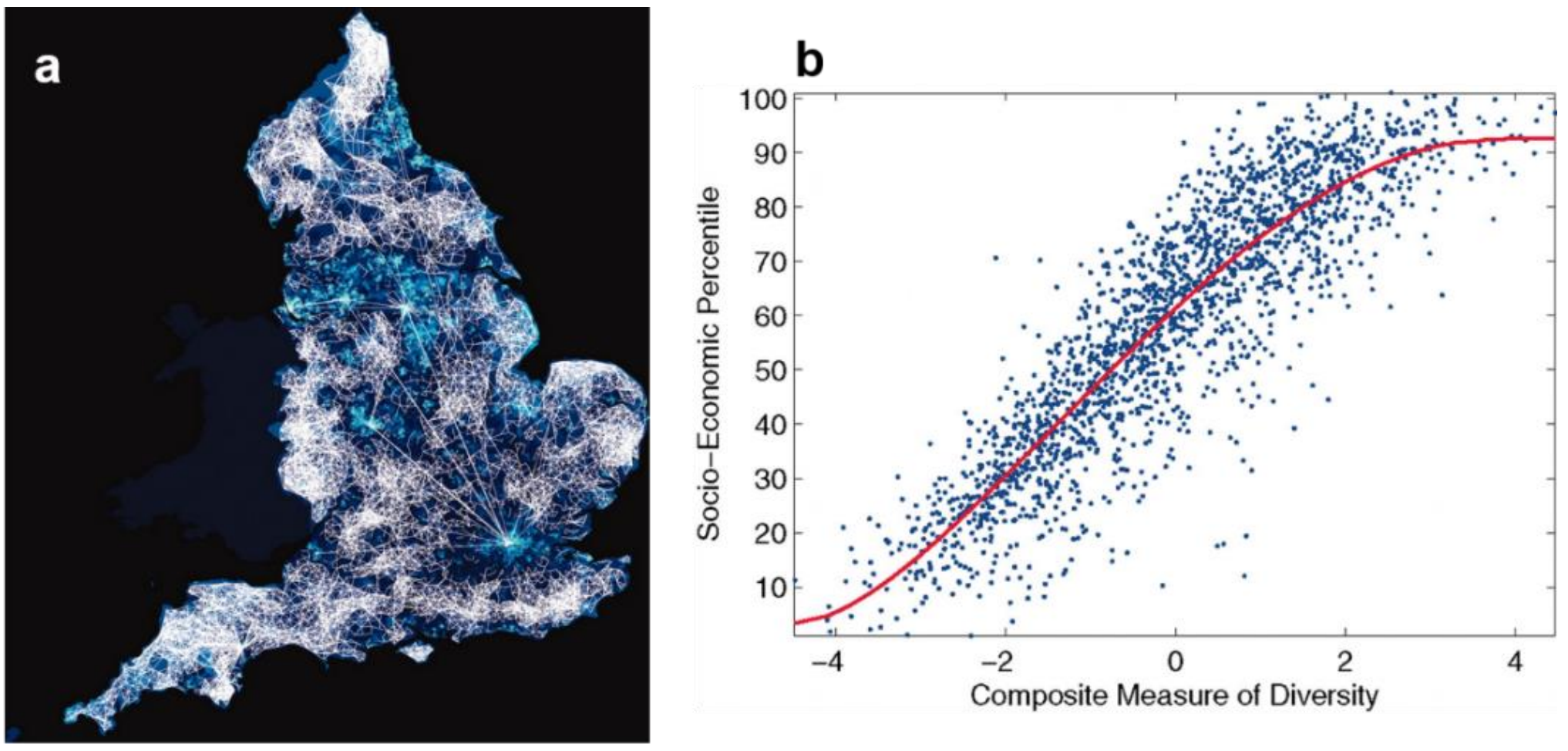

**Fig. 8. Communication network diversity and socioeconomic inequality in the United Kingdom.** (a) Regional communication network illustrating the relationship between communication diversity and socioeconomic ranking across communities in the United Kingdom. Communities are ranked according to the diversity of their communication patterns, ranging from insular to highly diverse, and are represented from light to dark blue. (b) Relationship between social network diversity and socioeconomic rank. Panels (a) and (b) are adapted from Eagle et al. [149].

*5.3.3. Human mobility measures for socioeconomic inference*

Beyond the structure of communication networks, another important body of research has examined how large-scale human mobility patterns derived from mobile phone data are associated with socioeconomic development. Mobile phone records provide continuous observations of population movement, allowing researchers to quantify mobility range, travel regularity, activity space, commuting behavior, and spatial interaction patterns at both individual and regional scales. These mobility characteristics are closely linked to access to employment, transportation infrastructure, economic opportunities, and urban organization. As a result, differences in mobility diversity, activity patterns, and spatial interactions have become important behavioral signals for interpreting socioeconomic disparities and regional development.

Pappalardo et al. [134] examined the relationship between large-scale human mobility patterns and socioeconomic development using nationwide mobile phone data from France. By introducing mobility volume and entropy-based measures of mobility diversity derived from CDR trajectories (Fig. 9a and b), they showed that mobility diversity was substantially more strongly correlated with socioeconomic indicators than mobility volume alone. Areas with more diverse mobility patterns tended to have higher income levels and lower deprivation. The study further incorporated social-network measures and predictive modeling, showing that mobility diversity consistently ranked among the most informative predictors of regional well-being and socioeconomic status (Fig. 9c). These findings emphasized mobility diversity as a behavioral signature of socioeconomic development and pointed to the potential of mobile phone data for real-time socioeconomic monitoring and nowcasting.

Although many studies have emphasized the predictive value of frequent, regular, and habitual mobility patterns, Wang et al. [152] showed that infrequent, irregular, and exploratory activities may contain disproportionately strong socioeconomic signals. Using large-scale mobility data from Boston, Chicago, and Miami, they found that a small subset of activities, representing less than 2% of total visits, explained more than 50% of the variation in median household income and property values. Visits associated with aspirational behaviors and non-essential consumption, including French restaurants, golf courses, theaters, and sports venues, were particularly predictive of economic outcomes. By contrast, routine activities such as commuting and residential visits showed relatively weak associations. These results suggest that the strongest socioeconomic signals are not necessarily embedded in routine behaviors, but rather in selective and long-tail activity patterns that reflect lifestyle preferences, consumption capacity, and latent socioeconomic characteristics.

Gao et al. [153] investigated income estimation from human mobility patterns using large-scale public transit data from Shenzhen together with a series of machine-learning and deep-learning models. To better represent the spatiotemporal organization of mobility, the study proposed three complementary mobility representations, i.e., mobility indicators, activity footprints, and travel graphs. Their predictive performance was compared using XGBoost, CNN, and graph convolutional recurrent networks (GCRN). The results showed that graph-based mobility representations combined with deep-learning models achieved the highest predictive accuracy, with the GCRN model reaching $R^2$ values of 0.910 and 0.976 for two income-related indicators. Feature-importance analyses further indicated that lower-income groups tended to live farther from city centers, travel longer distances with lower efficiency, start trips earlier, return later, and participate in fewer non-mandatory activities.

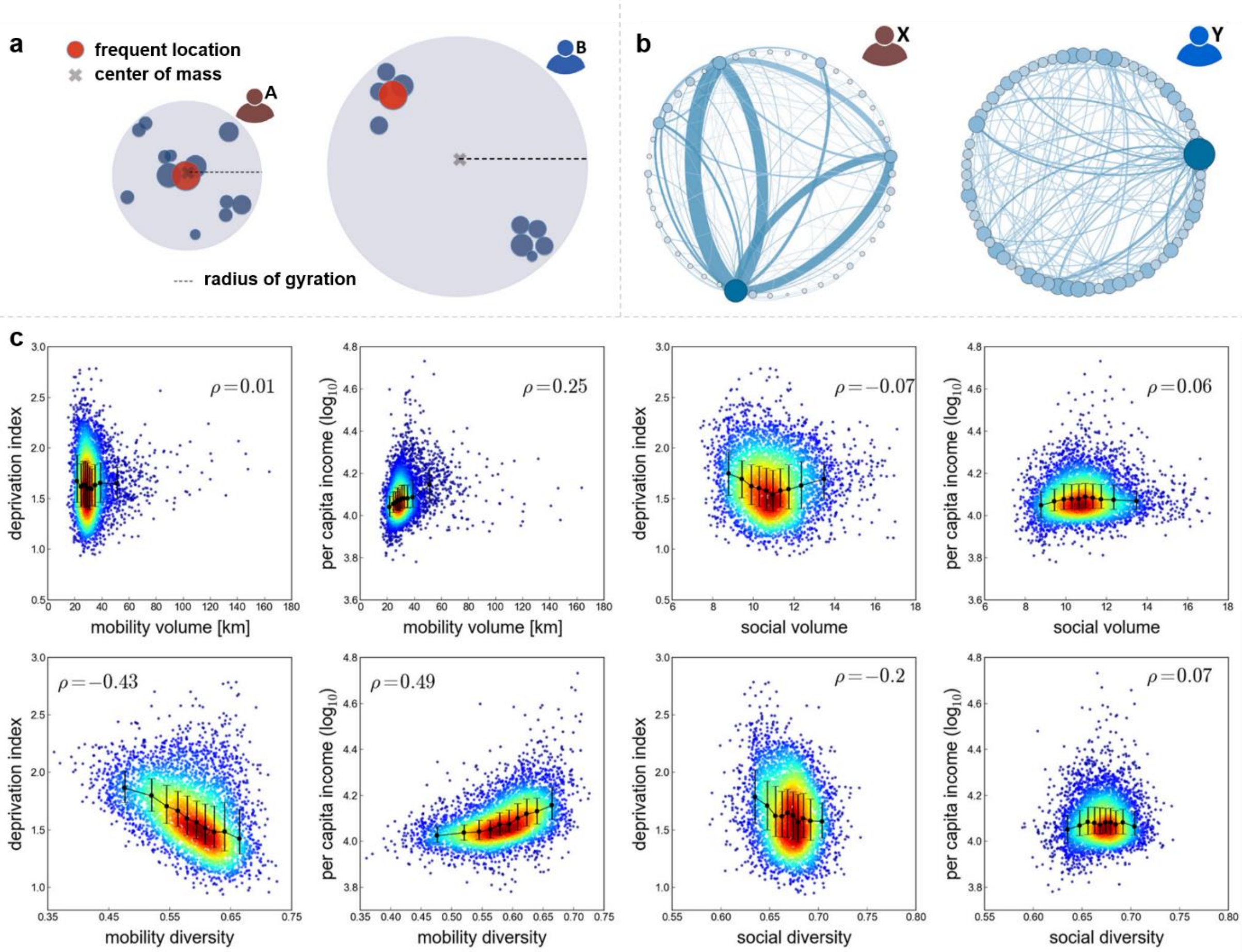

**Fig. 9. Interpretation of human mobility measures and their correlations with socioeconomic indicators.** (a) Radius of gyration for two users. Circle size is proportional to visitation frequency. The red circle denotes the most frequent location, while the cross marks the center of mass. The black dashed line

represents the radius of gyration. User A has a small radius of gyration because the visited locations are spatially concentrated. (b) Mobility entropy of two users. Nodes represent phone towers, and edges denote trips between towers. Node size reflects the volume of calls facilitated by each tower for the user, while edge thickness indicates the frequency of trips performed by the user on the edge. User X exhibits lower mobility entropy because their movements are concentrated along a limited set of preferred routes. (c) Correlations between human mobility measures and socioeconomic indicators. Figures are adapted from Pappalardo et al. [133, 134].

*5.3.4. Mobile phone records for wealth and socioeconomic prediction*

Mobile phone records contain rich behavioral information that extends beyond mobility trajectories and communication network structure. Everyday phone use generates diverse digital traces, including call frequency, airtime purchases, recharge behavior, communication timing, intensity of phone activity, and patterns of service use. These traces reflect several dimensions of daily life, such as consumption capacity, social engagement, routine regularity, and access to digital resources. This has led to increasing interest in whether mobile phone usage patterns can serve as behavioral proxies for socioeconomic status at both individual and regional scales.

Blumenstock et al. [154] proposed one of the earliest individual-level frameworks for estimating wealth and poverty using mobile phone metadata. Using billions of mobile phone transactions from Rwanda's largest telecom operator combined with phone survey data, the study showed that behavioral features derived from call records, including communication intensity, mobility patterns, social network structure, and airtime expenditure, could be used to predict individual socioeconomic status with substantial accuracy ($r = 0.68$). By aggregating these individual-level predictions, the authors further produced high-resolution poverty maps whose spatial patterns closely aligned with national DHS statistics (Fig. 10). This demonstrated the potential of mobile phone metadata for fine-grained and low-cost poverty estimation in data-scarce regions.

Blumenstock [155] later examined the generalizability of this approach across national contexts using datasets from Rwanda and Afghanistan. Although models trained and tested within the same country achieved moderate predictive performance ($R^2 \approx 0.33$), cross-country transferability was highly limited. When models trained in one country were applied to another, predictive performance declined to $R^2$ values of only 0.05–0.07. The study showed that behavioral patterns embedded in mobile phone usage are strongly context-dependent and shaped by local social and technological practices, indicating that socioeconomic prediction models based on digital traces require substantial contextual adaptation across geographic settings.

Soto et al. [156] investigated whether aggregated mobile phone usage patterns could predict socioeconomic levels in urban areas. Using six months of CDR data from nearly 500,000 users in a large Latin American city, the study developed socioeconomic prediction models at the base transceiver station level using aggregated behavioral, mobility, and social-interaction features. Mobility-related variables, including the number of distinct towers visited, radius of gyration, and travel distance, were among the strongest predictors of socioeconomic status, while reciprocity and SMS communication patterns also showed strong associations with socioeconomic level. Using SVM and RF models, the study achieved classification accuracies above 80% in distinguishing high-, middle-, and low-socioeconomic areas. These findings indicate that aggregated mobile phone usage patterns can capture fine-scale spatial socioeconomic differentiation in urban environments.

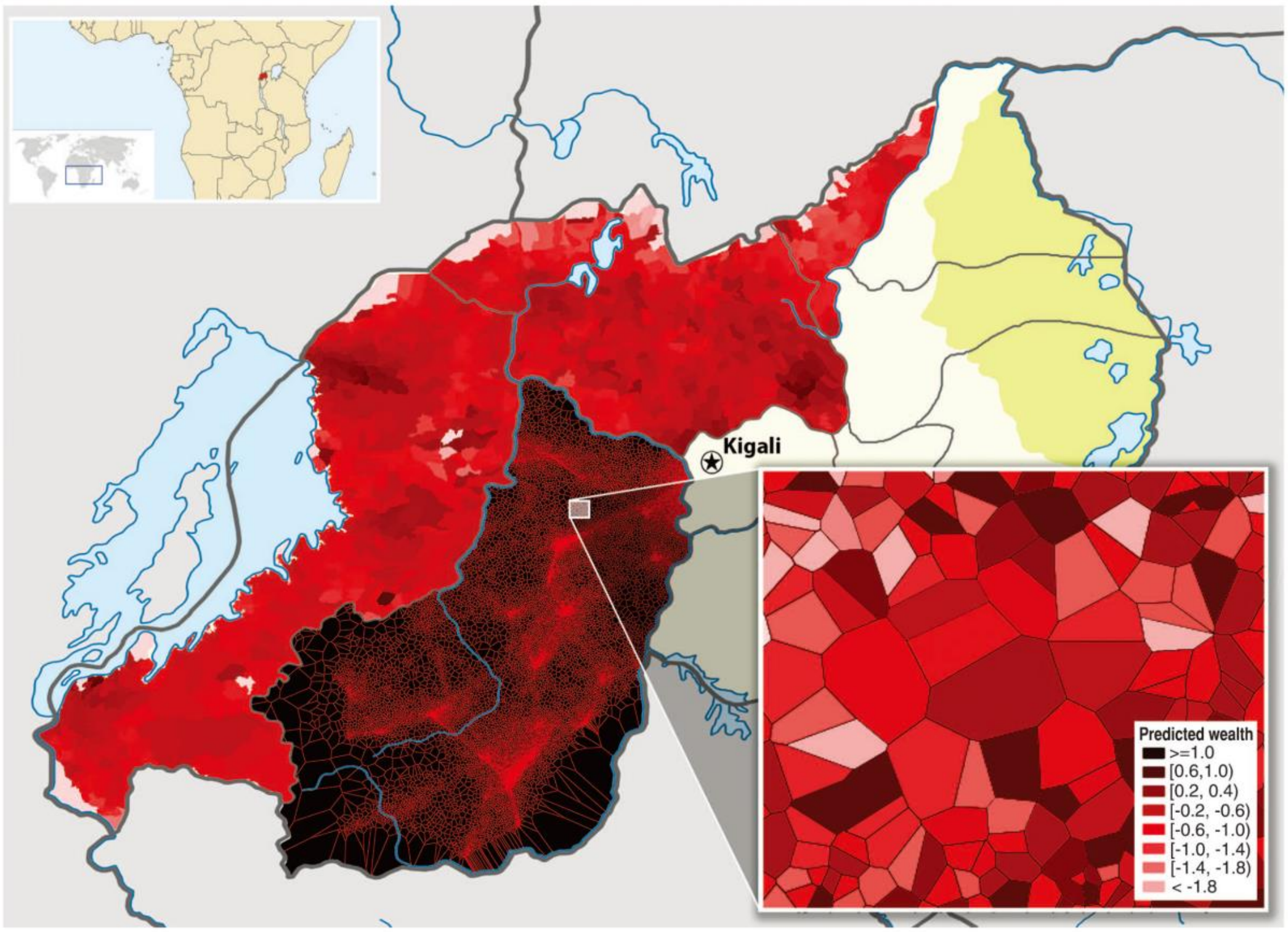

**Fig. 10. High-resolution poverty mapping from mobile phone metadata in Rwanda.**
Spatial distribution of predicted wealth inferred from mobile phone records of 1.5 million subscribers in Rwanda. Wealth estimates are aggregated at multiple spatial scales, including administrative cells and microregions composed of small groups of subscribers. Darker colors indicate lower predicted wealth levels. Adapted from Blumenstock et al. [154].

*5.3.5. Mobile phone data for poverty targeting and humanitarian assistance*

Mobile phone data are increasingly used not only to characterize socioeconomic conditions, but also to inform practical decision-making in poverty alleviation and humanitarian response. Because mobile phone records provide large-scale and continuously updated information on population activity and mobility, they create opportunities to identify vulnerable populations, monitor disruptions caused by crises or disasters, and improve the allocation of limited humanitarian resources.

A representative large-scale application was demonstrated by Aiken et al. [157] during the COVID-19 pandemic in Togo, where mobile phone data and machine learning were incorporated into the Novissi emergency cash transfer program to support humanitarian aid targeting. The study combined satellite-derived regional poverty maps with mobile phone metadata from national telecommunications operators to identify economically vulnerable populations at both geographic and individual levels. More specifically, machine-learning models were trained to infer subscriber-level poverty from behavioral indicators embedded in call detail records, including communication activity, recharge behavior, and mobility-related patterns. The resulting framework was deployed in a real-world policy setting to guide the distribution of emergency financial assistance to vulnerable populations affected by pandemic-related lockdowns. Evaluation results showed that the phone-based targeting approach substantially outperformed conventional geographic targeting strategies. In the rural Novissi program, the machine-learning approach achieved an AUC of 0.70, compared with 0.64 for prefecture-level targeting and 0.59 for canton-level targeting, while exclusion errors were reduced by 4–21% relative to geographic approaches. The study further showed that the proposed framework could improve aggregate social welfare under fixed-budget conditions and did not generate substantial demographic disparities across gender or ethnic groups. These findings highlight the potential of mobile phone data to support rapid, scalable, and operationally deployable humanitarian targeting in crisis settings where traditional socioeconomic data are incomplete or outdated.

Aiken et al. [158] further assessed the feasibility of using mobile phone data for anti-poverty program targeting in Afghanistan. Using household survey data from the government's "Targeting the Ultra-Poor" (TUP) program together with mobile phone metadata from a major telecommunications operator, the study evaluated whether machine-learning models based on CDR features could accurately identify households eligible for social assistance. The behavioral indicators extracted from mobile phone usage records included communication intensity, recharge behavior, social network characteristics, and spatial mobility metrics. Results showed that the CDR-based approach achieved targeting performance comparable to traditional survey-based poverty measures. For identifying ultra-poor households, the machine-learning model reached an AUC of 0.68, close to asset-based targeting (0.73) and consumption-based targeting (0.71). Combining mobile phone data with survey-based indicators further improved performance, achieving an AUC of 0.78. The study also noted important practical constraints, especially the exclusion of households without mobile phone access, emphasizing that digital targeting systems may unintentionally miss the most marginalized populations when phone ownership is limited.

Table 5 provides a summary of the studies reviewed on socioeconomic assessment, poverty targeting, and humanitarian applications using mobile phone data.

**Table 5. Summary of research studies that estimated poverty based on mobile phone data.**

| Authors (Year) | Data Period | Target Variable | Study Area | Modeling Approach | Performance |
|---|---|---|---|---|---|
| ***Mobile phone usage for socioeconomic and poverty inference*** | | | | | |
| Blumenstock and Eagle (2012) | 2005–2008 | Asset ownership and predicted expenditures | Rwanda | Polynomial regression | $R^2 \approx 0.577$ |
| Frias-Martinez and Virseda (2012) | - | Socioeconomic level, education level, household goods' ownership | 12 cities in a Latin American country | Multivariate linear regression | $R^2 \approx 0.82$-0.86 |
| Rubio et al. (2010) | - | Economic development level | One advanced economy and | Statistical analysis | - |
| | | (advanced vs. developing economy) | one developing economy | | |
| Xu et al. (2018) | 50 days in 2011 (Singapore); 4 months in 2009 (Boston) | House price (Singapore); per capita income (Boston) | Singapore and Boston | Correlation analysis | - |
| ***Communication networks for poverty estimation*** | | | | | |
| Eagle et al. (2010) | August 2005 | Index of multiple deprivation | United Kingdom | Fractional polynomial regression | $r$ = 0.78 |
| Smith-Clarke et al. (2014) | December 1, 2011–April 28, 2012 (Côte d'Ivoire); 6 weeks in early 2012 (Region B) | Poverty rate (Côte d'Ivoire); and assets-based poverty index (Region B) | Côte d'Ivoire and an anonymous developing region (Region B) | Correlation analysis, OLS regression | Pearson $r$ = 0.83 |
| Mao et al. (2015) | December 1, 2011–April 28, 2012 | Annual income, poverty rate, and Gini index | Côte d'Ivoire | Correlation analyses | Spearman $\rho = -0.83$ |
| ***Human mobility measures for socioeconomic inference*** | | | | | |
| Pappalardo et al. (2016) | 45 days in 2007 | Deprivation index and per capita income | France | Multiple linear regression, generalized nonlinear models, and RF classifier | $R^2$ = 0.43, accuracy = 0.61 |
| Wang et al. (2024) | September 2016 – April 2017 | Median household income and property values | Boston, Chicago, and Miami, USA | Elastic net regressions, gradient boosting regression | $R^2 \approx$ 0.7-0.8 |
| Gao et al. (2024) | 6 days in November 2016 | Low-income ratio (LR) and dissimilarity index (D-index) | Shenzhen, China | CNN, XGBoost, and GCRN | $R^2$ = 0.910 (LR), $R^2$ = 0.976 (D-index) |
| ***Mobile phone records for wealth and socioeconomic prediction*** | | | | | |
| Blumenstock et al. (2015) | May 2008–May 2009 | Composite WI, poverty status, and | Rwanda | Elastic net regularization, tree-based ensemble models | $r$ = 0.68 for (WI), $r$ = 0.92 (DHS) |

| | | asset ownership | | | |
|---|---|---|---|---|---|
| Blumenstock (2018) | 2009 (Rwanda); 2015–2016 (Afghanistan) | WI | Rwanda, Afghanistan | Gradient boosting algorithm | $R^2$ = 0.33 |
| Soto et al. (2011) | February–July 2010 | Socioeconomic level (A/B/C) | A main city in a Latin-American country | SVM, RF, and SVR | Accuracy = 0.824 |
| ***Mobile phone data for poverty targeting and humanitarian assistance*** | | | | | |
| Aiken et al. (2022) | March–September 2020 | Consumption expenditure, WI, and PPI | Togo | LightGBM | AUC = 0.70–0.73 |
| Aiken et al. (2023) | November 2015 –April 2018 | Ultra-poor household status | Balkh Province, Afghanistan | Gradient boosting model, logistic regression | AUC = 0.78 |

## 6. Poverty mapping with social media data

### *6.1. Data source and description*

Because of their real-time and large-scale characteristics, social media data have become an increasingly important resource for monitoring and predicting real-world phenomena [159]. The behavioral heterogeneity observed among social media users is shaped in part by underlying socioeconomic conditions, making it a useful source of signals for inferring economic status at both individual and regional levels [160]. Such information can help inform policies and interventions aimed at reducing socioeconomic disparities and improving social well-being.

As of October 2025, there were 5.66 billion active social media users worldwide, representing 68.7% of the global population [161]. This widespread adoption has produced vast quantities of heterogeneous data across platforms. In broad terms, social media data can be grouped into three main categories, i.e., user profiles, user-generated content, and network interaction data.

#### *6.1.1. User profile*

User profiles contain structured metadata that describe individual attributes, including demographic characteristics such as age, gender, marital status, and education, as well as geolocation and device-related information. Device-related attributes include mobile operating systems, such as iOS and Android, device types and brands, network access modes, including 2G, 3G, 4G, 5G, and Wi-Fi, and language preferences. These attributes are usually accessible through platform-specific data interfaces. One representative example is the Facebook Marketing API, which enables advertisers to define target audiences according to demographic, behavioral, and interest-based criteria. Through this interface, aggregated and anonymized audience estimates can be queried programmatically for specific geographic regions, offering a flexible data source for socioeconomic inference [162].

*6.1.2. User-generated content*

User-generated content (UGC) refers to text, images, videos, reviews, and other media created and shared by users on social media platforms [163]. These data capture diverse forms of online activity, including posts, comments, and discussions, and are available across platforms such as X, formerly Twitter, Facebook, and blogs. They contain detailed information about individuals, including interests, emotions, writing styles, location, occupation, and other relevant attributes. Although UGC is largely unstructured, it provides a rich source of information from which behavioral and contextual signals can be extracted. Linguistic patterns, expressed sentiments, topical interests, and activity contexts, for example, may reflect aspects of users' socioeconomic environments.

*6.1.3. Online social network*

Online social networks (OSNs) describe the structure of interactions and relationships among users on digital platforms (Fig. 11). These networks encode both the existence and intensity of social ties, commonly represented through interaction frequency, communication patterns, and network topology. Examining their structural properties, including connectivity, tie strength, and community organization, provides an additional perspective for understanding socioeconomic stratification.

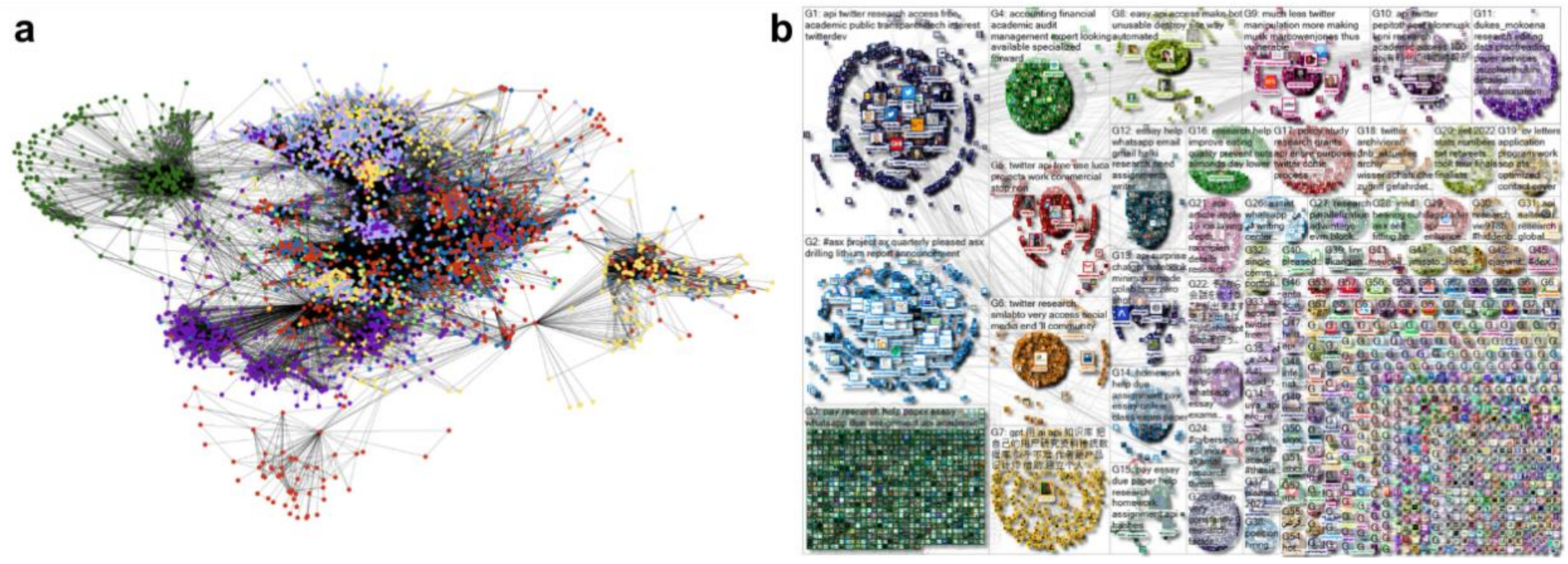

**Fig. 11. Example online social networks derived from social media platforms.** (a) Facebook friendship network (undirected, unweighted) comprising 4,039 nodes (users) and 88,234 edges (friendship ties), constructed from the friends list of ten seed users [164]. (b) Example of X network analysis using NodeXL. NodeXL provides an interface for exploring and visualizing interaction networks on the X platform (source: https://www.smrfoundation.org/networks/x-twitter-analytics/).

*6.2. Modeling approaches*

Social media platforms can also be understood as large-scale interacting systems in which information, attention, and social influence circulate through complex networks. Collective online behaviors, including information diffusion, community formation, and activity concentration, provide observable traces of broader socioeconomic structures. On this basis, large-scale social media data offer cost-effective opportunities for poverty estimation.

The analytical workflow typically begins with data acquisition, most often through official platform APIs (Table 6), although web- or application-based crawlers may also be used where appropriate ethical and legal conditions are met. After data collection, features are extracted according to the type of data involved (Table 7). Structured profiles, textual content, and network interactions require different analytical methods, producing features through statistical analysis, natural language processing, and network analysis. These features are then used as inputs to machine-learning models, such as regression, RF, and neural networks, to predict SES at the individual or regional level (Fig. 12). Once trained, the models can be applied to new data to produce updated estimates of socioeconomic conditions.

**Table 6. API references for selected social media**

| Social Media Platform | URL of the API reference |
|---|---|
| X (Twitter) | https://developer.x.com/ |
| Facebook | https://developers.facebook.com/docs/graph-api |
| Instagram | https://developers.facebook.com/docs/instagram-api |
| LinkedIn | https://developer.linkedin.com/ |
| Reddit | https://www.reddit.com/dev/api/ |
| YouTube | https://developers.google.com/youtube/v3/docs/ |
| Pinterest | https://developers.pinterest.com/docs/api/v5/introduction |
| Sina Weibo | https://open.weibo.com/ |

#### *6.2.1. Statistical and regression-based approaches*

Structured profile data, such as age, gender, and education, allow features to be extracted through descriptive statistics, including means, medians, and distributions. At the aggregate level, these features describe population structure, such as age composition, gender balance, and spatial distribution. At the individual level, they represent observable attributes associated with user profiles. Because such demographic features are often correlated with regional economic indicators and individual SES, they are frequently incorporated as inputs in poverty mapping models.

#### *6.2.2. Text-mining and natural language processing approaches*

Textual UGC is processed using natural language processing (NLP) techniques such as Word2vec [165], TF-IDF (Term Frequency-Inverse Document Frequency) [166], GloVe (Global Vectors for Word Representation) [167], and BERT (Bidirectional Encoder Representations from Transformers) [168]. These methods support the extraction of semantic and contextual features for applications such as sentiment analysis, topic modeling, and text embedding. Linguistic features derived from UGC, including vocabulary use, syntactic patterns, and expression complexity, may reflect behavioral and contextual differences across populations. In some cases, sentiment signals are associated with indicators of financial stress or well-being, although these relationships are generally indirect and context-dependent. Geotagged content further allows spatial variation in language use and sentiment to be examined, thereby supporting socioeconomic inference at the regional level.

### 6.2.3. Network-based approaches

OSNs are typically represented as graphs, with nodes denoting users and edges representing interactions or relationships. Network analysis methods can then be used to characterize structural properties and patterns of interaction. Node-level metrics, such as degree centrality, betweenness centrality, and closeness centrality, capture dimensions of user connectivity and positional importance. Network-level measures, including density, average path length, and clustering coefficient, describe the broader organization of the network. Community detection methods, such as the Louvain algorithm [169], Girvan-Newman algorithm [170], and Infomap algorithm [171], identify groups of nodes that are more densely connected internally than externally. Such structures can reveal patterns of social grouping and interaction that may be associated with differences in access to information, resources, and opportunities.

**Table 7. Social media user level features (adapted from Preoţiuc-Pietro et al. [172])**

| ***User profile features*** | |
|---|---|
| $u_1$ | number of followers |
| $u_2$ | number of friends |
| $u_3$ | number of times listed |
| $u_4$ | Follower-to-friend ratio |
| $u_5$ | number of favorites the account made |
| $u_6$ | average number of tweets/day |
| $u_7$ | total number of tweets |
| $u_8$ | proportion of tweets in English |
| ***User psycho-demographic features*** | |
| $d_1$ | gender (male or female) |
| $d_2$ | age (18–70 years) |
| $d_3$ | political (independent, conservative, liberal, or unaffiliated) |
| $d_4$ | intelligence (>average, average, ≤average, >>average, or <<average) |
| $d_5$ | relationship (married, in a relationship, single, or other) |
| $d_6$ | ethnicity (Asian, African American, Indian, Hispanic, other, or Caucasian) |
| $d_7$ | education (bachelor, graduate, or high school) |
| $d_8$ | religion (Christian, Jewish, Muslim, Hindu, unaffiliated, or other) |
| $d_9$ | children (yes or no) |
| $d_{10}$ | income (below average, above average, or very high) |
| $d_{11}$ | life satisfaction (satisfied, dissatisfied, very satisfied, very dissatisfied, or neither) |
| $d_{12}$ | optimism (optimist, pessimist, extreme optimist, extreme pessimist, or neither) |
| $d_{13}$ | narcissism (agree strongly, agree, disagree, disagree strongly, or neither) |
| $d_{14}$ | excited (agree strongly, agree, disagree, disagree strongly, or neither) |
| $d_{15}$ | anxious (agree strongly, agree, disagree, disagree strongly, or neither) |
| ***User emotion features*** | |
| $e_1$ | proportion of tweets with positive sentiment |
| $e_2$ | proportion of tweets with neutral sentiment |
| $e_3$ | proportion of tweets with negative sentiment |
| $e_4$ | proportion of joy tweets |
| $e_5$ | proportion of sadness tweets |
| $e_6$ | proportion of disgust tweets |
| $e_7$ | proportion of anger tweets |

| $e_8$ | proportion of surprise tweets |
|---|---|
| $e_9$ | proportion of fear tweets |
| ***Shallow textual features*** | |
| $s_1$ | proportion of nonduplicate tweets |
| $s_2$ | proportion of retweeted tweets |
| $s_3$ | average number of retweets/tweet |
| $s_4$ | proportion of retweets done |
| $s_5$ | proportion of hashtags |
| $s_6$ | proportion of tweets with hashtags |
| $s_7$ | proportion of tweets with @-mentions |
| $s_8$ | proportion of @-replies |
| $s_9$ | number of unique @-mentions in tweets |
| $s_{10}$ | proportion of tweets with links |

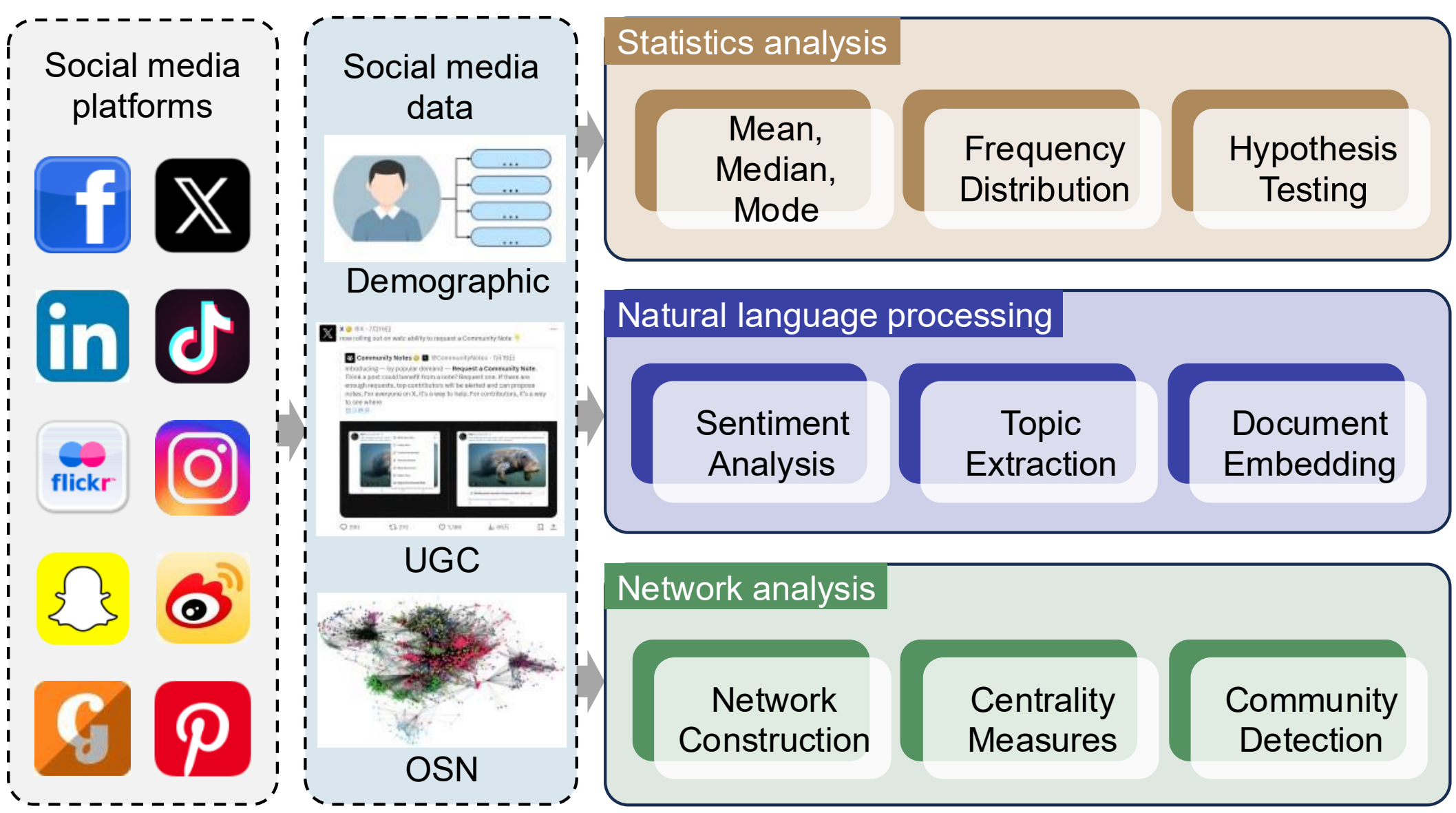

**Fig. 12. Techniques of assessing poverty using social media data.** The process begins with collecting data from social media platforms, including user demographic information, UGC, and OSN. This is followed by the application of statistical analysis, natural language processing, and network analysis to the collected data.

### *6.3. Related research*

#### *6.3.1. Platform-derived user attributes for poverty and welfare inference*

Social media advertising data, especially Facebook audience estimates, have increasingly been used as proxy indicators for assessing SES in settings where conventional data are limited. Fatehkia et al. [173] illustrated this approach by using audience estimates from the Facebook Marketing API to predict socioeconomic conditions in the Philippines and India. Drawing on attributes such as device type, including iPhone versus Android, and network access, such as 4G versus 2G connectivity, they estimated regional variation in the DHS-derived WI. Using Facebook features alone, their gradient boosting tree models achieved cross-validated $R^2$ values of 0.608 in the Philippines (Fig. 13a) and 0.563 in India (Fig. 13b). Model performance increased further, to 0.630 and 0.728, when these features were combined with population density and regional indicators. The study demonstrates that patterns of digital connectivity encode meaningful socioeconomic information. Wealthier areas were more likely to have larger shares of iOS users, WiFi access, 4G connectivity, and high-end smartphone adoption, whereas lower-SES areas were more closely associated with older devices and lower-quality connections, including 2G/3G access. These patterns indicate that material inequalities are reflected in uneven access to digital infrastructure and device quality.

In a subsequent study, Fatehkia et al. [174] applied a similar strategy to examine the SES of Syrian refugees in Lebanon. Their results revealed consistent patterns that variables related to network access and operating system use were associated with poverty and income levels at the subnational scale (Fig. 13c–f). Notably, even a single indicator, such as the share of iOS users, explained a substantial proportion of regional variation in poverty. At the same time, the study also exposed marked demographic and educational biases in platform representation. Refugee women were substantially underrepresented relative to survey-based population distributions, while Facebook users within the refugee population appeared to be more educated than the broader refugee population, pointing to potential sampling distortions.

Building on this foundation, Piaggesi et al. [175] conducted a comparative study across four cities, i.e., Atlanta, Bogotá, Santiago, and Casablanca. Using more than 35 Facebook-derived features, including demographic, cultural, and behavioral attributes, they showed that Facebook audience estimates could accurately distinguish high- and low-SES urban districts, defined according to percentile-based thresholds of socioeconomic indicators, in both high- and lower-income settings. However, predictive performance declined markedly when models were transferred across cities, underscoring the limited generalizability of learned relationships. Feature-importance analysis further showed that the meaning of profile attributes depends strongly on context. Technology-related variables were among the most informative predictors across cities, but their relative importance varied by setting. For instance, a higher share of iOS users was associated with higher SES in Atlanta, Santiago, and Bogotá, whereas Android adoption and older devices were more common in lower-SES areas. In Casablanca, by contrast, 3G/4G connectivity and Android use were more closely linked to higher SES.

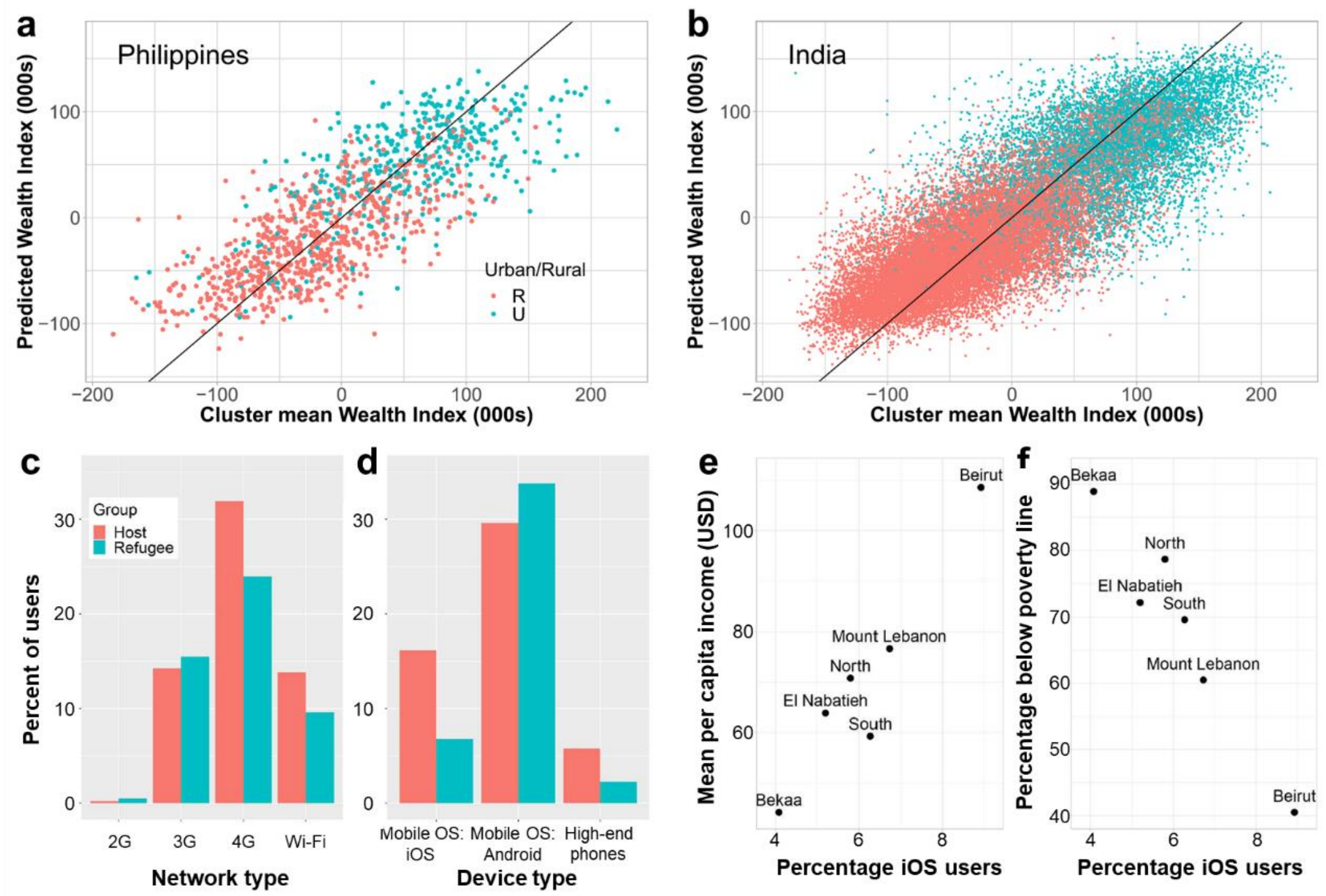

**Fig. 13. Socioeconomic assessment using Facebook advertising data in the Philippines, India, and Lebanon.** (a)–(b) Scatter plots displaying the correlation between model-predicted and actual Wealth Index values at the cluster level, based on DHS data in the Philippines (a) with 1205 clusters and India (b) with 28,043 clusters. (c)-(d), Bar charts showing the percentage of Facebook users aged 13+ years from host and refugee communities in Lebanon according to network connection type (c) and device type (d). (e)-(f), Scatter plots illustrating the relationship between the percentage of Facebook-using refugees with iOS devices and mean per capita income (e) and the percentage living below the poverty line (f). Panels (a) and (b) are adapted from Fatehkia et al. [173]. Panels (c)-(f) are adapted from Fatehkia et al. [174].

*6.3.2. Textual content and sentiment-based assessment of well-being and deprivation*

Textual content on social media offers indirect evidence of emotional and psychological states, providing a scalable lens through which regional well-being can be assessed. Quercia et al. [176] examined this relationship at the community level using tweets from residents of 51 London communities. They applied two sentiment analysis methods, a word-count approach and a maximum entropy classifier, to derive community-level sentiment measures. When aggregated into a metric termed "Gross Community Happiness," these scores showed a statistically significant negative correlation with the UK Index of Multiple Deprivation ($r \approx -0.35$). As illustrated in Fig. 14a, users in more deprived areas tended, on average, to express more negative sentiments than those in less deprived communities. Similarly, Preoţiuc-Pietro et al. [172] used Twitter data to develop an income prediction model based on Gaussian Processes, achieving a Pearson correlation of up to 0.633. Their analysis incorporated features related to emotions, sentiment, perceived psycho-demographics, and language use (Table 7). The results revealed systematic associations between these feature categories and income. Higher-income users tended to show greater social influence, reflected for example in larger numbers of followers and retweets, and were more likely to discuss topics related to politics and corporate activity. Lower-income users, by contrast, more often used informal or personal language and expressed more subjective content. Affective differences were also observed, with higher-income users displaying relatively higher levels of fear and anger.

Building on occupation-based proxies, Preoţiuc-Pietro et al. [172] also introduced a dataset linking Twitter users to job titles, which enabled the estimation of occupation-level income. Using this dataset, Flekova et al. [177] examined the relationship between linguistic style and income, considering features related to surface form, readability, syntax, and stylistic expression. Their results showed that stylistic features contained meaningful predictive signals of income, with a Pearson correlation of approximately 0.38. High-income users tended to write in a more complex and formal style, whereas low-income users more frequently used interjections, pronouns, and informal expressions (Fig. 14b). In a related contribution, Hasanuzzaman et al. [178] proposed a two-stage framework based on temporal patterns in language use. They first developed a classifier to categorize tweets as past-, present-, or future-oriented expressions, which were then aggregated into user-level temporal orientation profiles. These profiles were subsequently used for income prediction, revealing a positive association between future-oriented language and income (Pearson r $\approx$ 0.51).

Whereas the studies above largely frame income inference as a regression task using occupation-linked labels, Lampos et al. [179] treated SES prediction as a multi-class classification problem. Using Twitter data from UK users labeled through occupation-derived SES, they developed a Gaussian Process classifier that combined five groups of features, i.e., profile keywords, tweet n-grams, latent topics, user interactions, including retweets and mentions, and platform influence, such as follower count. Their model achieved an accuracy of 75% for three-class SES classification, distinguishing upper, middle, and lower groups, and 82% for binary classification between upper and non-upper classes. Beyond text-based content, Hristova et al. [180] examined social media images related to cultural events. Analyzing approximately 10 million geo-tagged Flickr images from London and New York between 2007 and 2015, they constructed a hierarchical taxonomy of cultural activities and quantified “cultural capital” using the share of culturally tagged images in each neighborhood. Their regression analysis showed that both cultural and economic capital were associated with urban development indicators, including housing prices and social vulnerability, while cultural diversity was linked to neighborhood improvement. This study suggests that digital traces of cultural expression can complement textual signals in explaining socioeconomic dynamics.

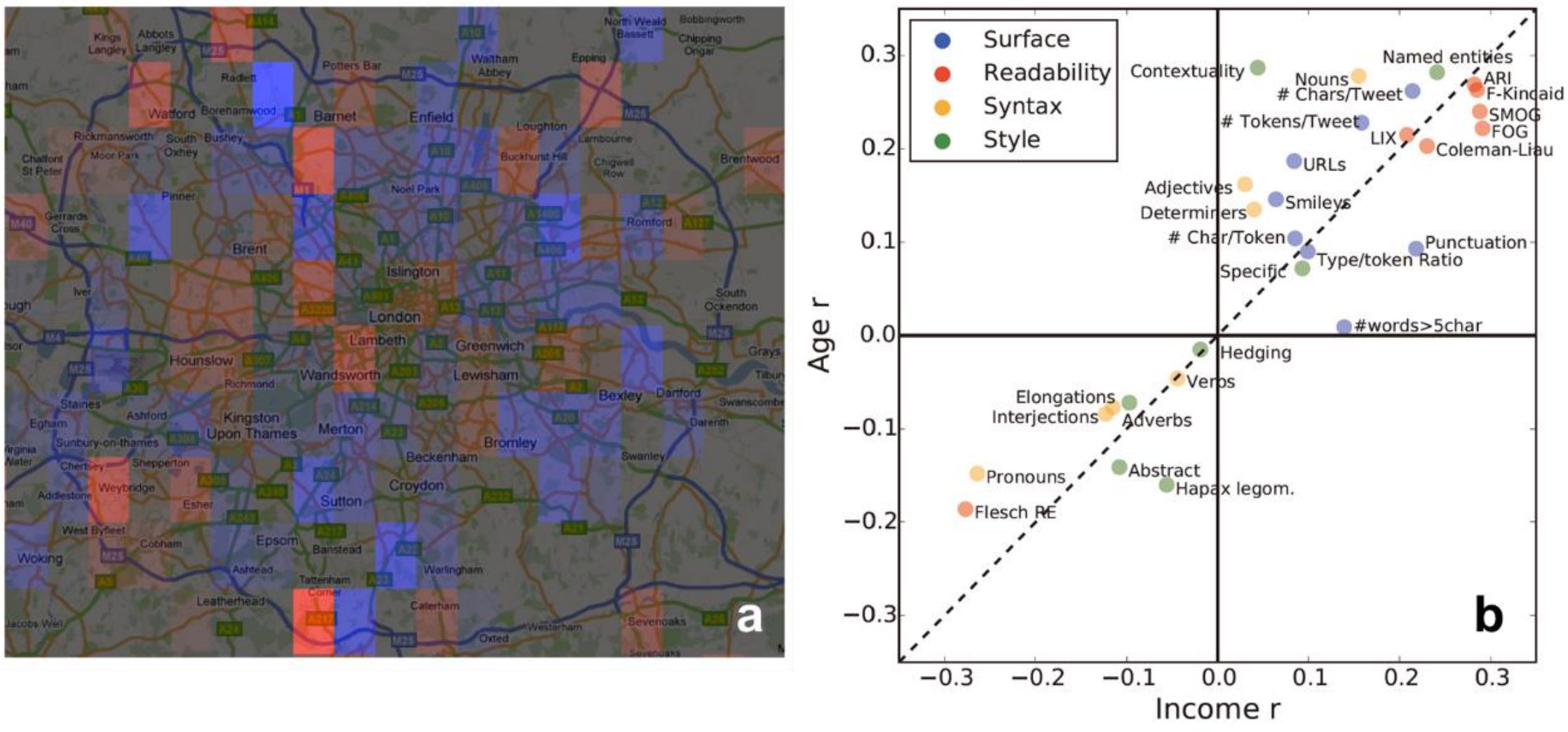

**Fig. 14. Social well-being inferred from language usage in social media platforms.** (a) Mapping "gross community happiness" in Greater London, with red indicating happy sentiments and blue representing unhappy sentiments. (b) Predictive performance of linguistic features for income and age. Panel (a) is adapted from Quercia et al. [176]. Panel (b) is adapted from Flekova et al. [177].

*6.3.3. Online social networks for socioeconomic estimation*

Interactions and relational structures within online social networks provide another avenue for examining wealth and poverty, offering complementary insight into socioeconomic organization [181]. Vaca-Ruiz et al. [182] utilized the sociological concept of "glocality" [183] to estimate city-level GDP from social media data. Using 1.4 million repost cascades from Yahoo Meme, a Brazilian microblogging platform, they constructed a set of attention-based metrics, including global and local attention, brokerage centrality, and cascade-based diffusion measures. Several of these indicators were positively and significantly correlated with GDP per capita, with local attention ($r$=0.56) and cascade depth ($r$=0.41) showing relatively strong associations. Incorporating these metrics into a regression model, alongside census variables such as population and Internet penetration, yielded a highly predictive performance (adjusted $R^2 = 0.94$) across 45 Brazilian cities.

Similarly, Norbutas and Corten [184] conducted a large-scale empirical analysis of social capital using complete friendship network data from the Dutch social platform Hyves, encompassing over 10 million users across 438 municipalities. They examined multiple structural dimensions of social capital, i.e., network density, diversity and distance of bridging ties, and modularity, and their relationship with municipal-level income. Their findings showed that higher network diversity and broader bridging ties are positively associated with disposable income, while greater network fragmentation is negatively associated.

Tóth et al. [185] investigated the joint relationship between urban geography, social network structure, and income inequality in 474 Hungarian towns, using large-scale data from the iWiW social network, which included approximately 2 million users and 300 million ties. Their analysis showed that stronger network fragmentation was associated with greater income inequality (Fig. 15). Through a two-stage instrumental variable approach, they further assessed the role of geographic factors, including amenity concentration, peripheral housing, and physical barriers such as rivers and railways. The results suggested that these spatial features may contribute to inequality indirectly by increasing social fragmentation, thereby indicating a possible spatial-social mechanism underlying persistent socioeconomic disparities.

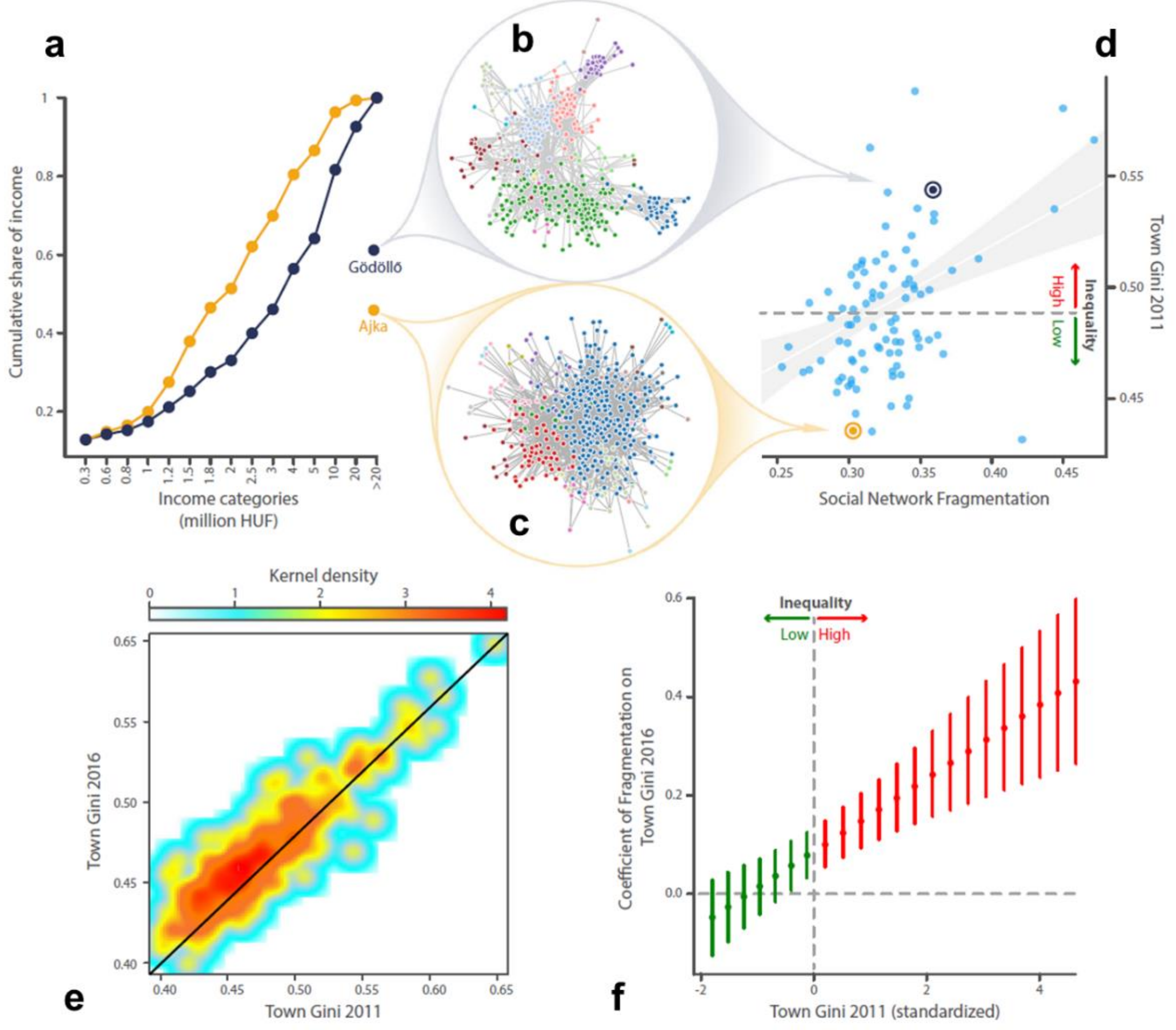

**Fig. 15**. **Relationship between income inequality and social network fragmentation in Hungary.** (a) Cumulative income distribution in Ajka (lower inequality) and Gödöllő (higher inequality). (b)-(c) Sampled social networks in Gödöllő (b) and Ajka (c), showing higher and lower levels of fragmentation, respectively. (d) Correlation between income inequality and social network fragmentation in 2011. (e) Correlation between the town Gini scores in 2011 and 2016. (f) Increased inequality in 2016, with higher initial inequality and network fragmentation in 2011. Figure adapted from Tóth et al. [185].

### *6.3.4. Integrated social-media features for socioeconomic prediction*

Combining multiple types of social media data allows socioeconomic indicators to be examined through behavioral, linguistic, and relational dimensions simultaneously. Lerman et al. [186] integrated geo-tagged Twitter data with U.S. census information to analyze emotional expression, social interactions, and socioeconomic conditions at the census-tract level. Using tweets from Los Angeles, they inferred social ties from "@" mentions and measured average tie strength as

$$S_i = \frac{\sum_{j=1}^{k_i} w_j}{k_i}, \tag{12}$$

where $w_j$ is the weight of the $j^{\text{th}}$ edge (i.e., the number of times user A mentioned user B), and $k_i$ is the total number of distinct users mentioned in tract *i*.

They further measured spatial diversity using a normalized entropy index,

$$D_i = \frac{-\sum_{j=1}^{n_i} p_{ij} log(p_{ij})}{log n_i},$$
$$p_{ij} = \frac{T_{ij}}{\sum_{j=1}^{n_i} T_{ij}}, \quad (13)$$

where $n_i$ denotes the number of tracts from which users who tweeted from tract *i* also posted tweets, and $T_{ij}$ is the number of tweets posted in tract *j* by users who tweeted from both tracts *i* and *j*.

Their findings showed that areas characterized by stronger and more localized ties tended to display less positive sentiment and lower socioeconomic status. By contrast, areas with weaker but more spatially diverse connections were associated with more positive emotions and higher levels of income and education.

Beyond these descriptive measures, a number of studies have examined representation learning approaches that combine network structure with behavioral signals. Aletras and Chamberlain [187] proposed a neural graph embedding method that represents the extended ego-networks of Twitter users as low-dimensional vectors. When evaluated on datasets annotated with occupational class and income, their approach outperformed models relying only on textual features, including those used by Hasanuzzaman et al. [178] and Preoţiuc-Pietro et al. [172].

Similarly, Vaganov et al. [188] combined data from the Russian social media platform VK.com with monthly expenses derived from bank profiles to infer user-level SES. Their feature space included demographic attributes, network embeddings, such as VERSE [189] and ARCTE [190], text features, and subscription-based features. Models that integrated graph embeddings with other feature groups achieved the best performance, reaching a macro-F1 score of 0.687 for classifying low- versus high-expense users. In subsequent work, Vaganov et al. [191] conducted a feature-importance analysis and found that community-level topics and structural properties of friendship networks were consistently among the strongest predictors of both SES and consumption behavior.

From a different but related perspective, Liao et al. [192] conceptualized online clickstream behavior as mobility within a cyber-activity space, treating websites as locations and user clickstream transitions as flows. They measured three dimensions of cyber behavior, namely diversity, size, and preference, and related these measures to regional socioeconomic indicators using OLS and GWR models. Their models explained up to 50.3% of the variance in regional SES, indicating that online activity patterns can reflect underlying offline disparities.

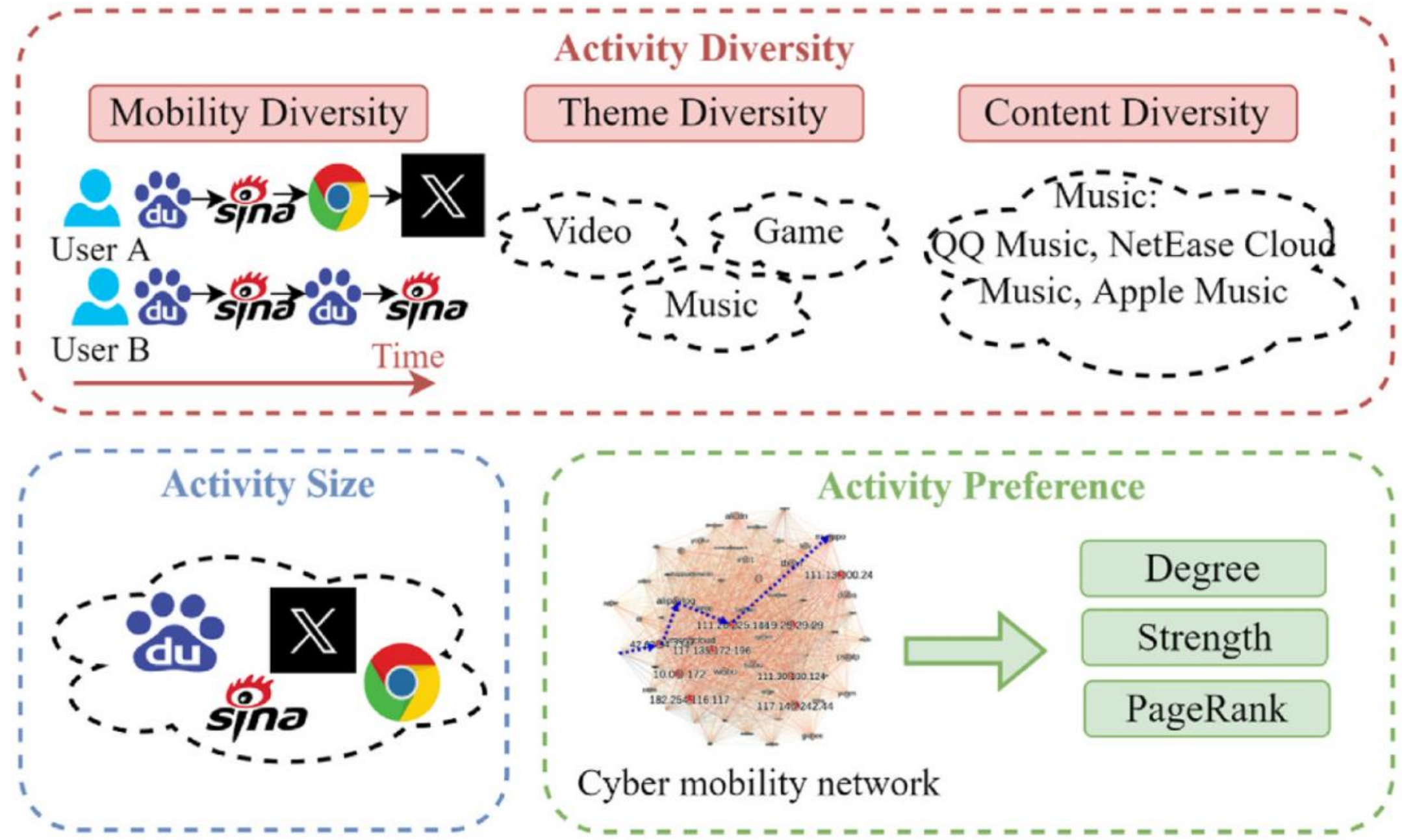

**Fig. 16. A hierarchical framework that characterizes cyber human activity.** Figure adapted from Liao et al. [192].

Table 8 provides a summary of studies on poverty estimation using social media data.

**Table 8. Summary of research studies that inferred poverty based on social media data.**

| Authors (Year) | Data Period | Target Variable | Study Area | Modeling Approach | Performance |
|---|---|---|---|---|---|
| ***Platform-derived user attributes for poverty and welfare inference*** | | | | | |
| Fatehkia et al. (2020) | March-April 2019 (Philippines); June-September 2019 (India) | WI | Philippines and India | LASSO regression and gradient boosting trees | $R^2$ = 0.728 (India); $R^2$ = 0.630 (Philippines); |
| Fatehkia et al. (2022) | November 2019 | Poverty rate and per capita income | Lebanon | Correlation analysis and single-variable regression | $R^2$ = 0.895 (poverty), 0.801 (income) |
| Piaggesi et al. (2022) | January-September 2020 | Income, socioeconomic strata, poverty rate | Atlanta, Bogotá, Santiago, and Casablanca | RF classification | ROC-AUC up to 0.925-0.993 |
| ***Textual content and sentiment-based assessment of well-being and deprivation*** | | | | | |
| Quercia et al. (2012) | September 27 – December 10, 2010 | Index of multiple deprivation | London | Sentiment-based analysis (LIWC + Maximum Entropy) and | Pearson $r$ = 0.350-0.365 |

| | | | | correlation analysis | |
|---|---|---|---|---|---|
| Preoţiuc-Pietro et al. (2015) | Around August 5, 2014 | Income | UK | GP regression | Pearson $r$ = 0.633 |
| Flekova et al. (2016) | - | Income | - | Linear regression and SVM | Pearson $r$ = 0.38 |
| Hasanuzzaman et al. (2017) | January 2015 | Income | - | Logistic regression and GP regression | Pearson $r \approx 0.51$ |
| Lampos et al. (2016) | February 1, 2014–March 21, 2015 | SES label | UK | GP regression | Accuracy = 75% (3-class), 82% (binary) |
| Hristova et al. (2018) | 2007–2015 | Multiple deprivation index and the social vulnerability index | London and New York | Linear regression | $R^2$ = 0.56-0.81 |
| ***Online social networks for socioeconomic estimation*** | | | | | |
| Vaca-Ruiz et al. (2014) | 2009–2012 | GDP | Brazil | Linear regression | Adj. $R^2$ = 0.94 |
| Norbutas and Corten (2018) | 2010 | Income per capita | Netherlands | OLS regression | Adj. $R^2$ = 0.635 |
| Tóth et al. (2021) | 2011–2016 | Gini index | Hungary | 2SLS/OLS regression | $R^2$ = 0.23-25 |
| ***Integrated social-media features for socioeconomic prediction*** | | | | | |
| Lerman et al. (2016) | July–October 2014 | Income | Los Angeles | Correlation analysis and linear regression | $r$ = −0.36-0.35 |
| Aletras and Chamberlain (2018) | - | Income | UK | Linear regression, support vector regression, and GP classifiers | Accuracy = 52%; $r$ = 0.64; MAE = £9048 |
| Vaganov et al. (2020) | - | Monthly expenses | Russia | Logistic regression | $F_1$ score = 0.687 |
| Liao et al. (2024) | 2016 | Economy per capita | Jilin province, China | OLS regression and GWR | Adj. $R^2$ = 0.503 |

## 7. Multisource data fusion and model improvements

Poverty, with its multifaceted dimensions, defies accurate quantification through a single data modality, both in theoretical constructs and in practical applications. Integrating multimodal data from multiple sources can bolster the accuracy of poverty mapping and enhance the understanding of the underlying factors by combining complementary information on the built environment, human mobility, social activity, accessibility, and administrative conditions. To this end, several recent studies have evaluated the predictive power of integrating the aforementioned and other data sources, illuminating considerable potential for improving poverty estimation and for characterizing the environmental, behavioral, and infrastructural correlates of deprivation.

Different data streams provide partially independent, noisy observations of living standards, and model structure can be used to combine them while accounting for spatial variation and uncertainty. RS and CDR data, generated at different spatial scales, capture different aspects of living conditions and mobility. Integrating RS, CDRs, and traditional survey-based data from Bangladesh, Steele et al. [45] employed hierarchical Bayesian geostatistical models to predict three prevalent poverty indicators (WI, PPI, and household income). Their results showed that models using multisource data generally outperformed models based on either RS-only or CDR-only data sources across varying spatial scales: urban areas ($r^2 = 0.78$), rural areas ($r^2 = 0.66$), and at the national level ($r^2 = 0.76$).

Pokhriyal and Jacques [193] used a related probabilistic approach, applying GP regression, a Bayesian machine-learning technique, to combine CDR-derived and environmental information for MPI prediction in Senegal. Initially, they employed GP regression to predict poverty from a single CDR or environmental data source, and each model produced a posterior Gaussian distribution. Subsequently, the combined poverty estimate was assumed to be a mixture distribution consisting of two Gaussians, with weights assigning greater importance to a single source. Their GP regression model, incorporating call volume, per capita mobile ownership, and NTL data, outperformed conventional linear regression partly because it allowed nonlinear relationships between predictors and MPI. The combination of CDR-derived features and environmental indicators produced accurate MPI estimates, yielding a correlation coefficient of r = 0.91. Njuguna and McSharry [194] addressed a more constrained setting, where only sparse or aggregated digital traces were available. Integrating a small set of aggregated CDR features with NTL and LandScan population data, they constructed sector-level MPI estimates in Rwanda. Their use of interpretable features and regularized regression produced a cross-validated correlation coefficient of $r = 0.88$, suggesting that limited CDR information can still be useful when combined with publicly available geospatial data.

Several studies have treated multisource variables as engineered feature sets for supervised learning, with RF regression and related ensemble methods commonly used to construct poverty-related proxies. Niu et al. [195] developed a multisource data poverty index using an RF algorithm to measure urban poverty at the community level in Guangzhou, China, on the basis of a combination of social media and RS data. The results showed strong consistency between the MDPI and the General Deprivation Index (GDI) derived from traditional census data. Moreover, the proposed MDPI improved the precision of the classification of community-level poverty with similar characteristics, which is useful for identifying poverty in fast-growing urban areas. Combining NTL and geolocated tweet data as inputs to a simple RF regression model, Kondmann et al. [42] predicted local wealth in 20,000 villages across sub-Saharan Africa. The results showed that the simple combination of these disparate data sources was already competitive, explaining up to 65% of the variance in local poverty, leading to an improvement of 10 percentage points over the Twitter-only result. This performance was competitive compared with the NTL-only result based on the CNN-based approaches by Yeh et al. [27], without requiring large-scale training. Zhao et al. [29] conducted a similar study but combined richer data sources as inputs to an RF regression model, including NTL, Google satellite images, land cover data, road map, and division headquarters location data. Compared with models using individual data sources, the multisource RF model achieved stronger predictive performance, with an $R^2$ of 0.70 between observed and estimated WI. Its application to Nepal further indicated both the potential and the limits of transferring feature-based geospatial models across national contexts.

At the county level, Zheng et al. [196] combined nighttime-light and multisource geospatial variables to estimate a county-level MPI in Fujian, China. Their RF model achieved the best predictive performance among several machine-learning models, with $R^2$=0.928, MAE=0.030, and RMSE=0.037, suggesting that nighttime-light and facility-related geospatial features can serve as useful proxies for approximating county-level multidimensional poverty where statistical updates are delayed.

Other multisource studies have combined image representations with structured geographic metadata. Tingzon et al. [28] successfully replicated the approach of Jean et al. [86] in the Philippines, using a VGG16-based transfer-learning model to extract image features and RF regression to incorporate OSM-derived features. Furthermore, they proposed an OSM-nightlights hybrid model that combines volunteered geographic information from OSM and NTL for estimating socioeconomic indicators. The best model achieved an $R^2$ = 0.63 for estimating asset-based wealth. Espín-Noboa et al. [197] proposed a pipeline of machine learning models to improve poverty map inference for subpopulations within Sierra Leone and Uganda and to capture local variation in wealth. These models used seven distinct, freely accessible feature sources derived from satellite imagery and metadata collected via online crowd-sourcing and social media. A CNN was trained on daylight satellite images, while the CatBoost model (CB) was trained with metadata features such as demographics and mobility. These models culminated in a combined model (CNN + CB), in which the third-to-last layer of the CNN (low-dimensional feature vector representation of the satellite images) was fed into the CB model. A distinctive contribution of this study is that it predicts not only the mean wealth of each place but also the standard deviation of local wealth, thereby moving poverty mapping from point estimation toward a distributional description of local socioeconomic heterogeneity. The results demonstrated that combined metadata features are the most reliable predictors of wealth in rural areas, outperforming image-based models, and highlighted the importance of integrating multiple models to provide a more comprehensive understanding of poverty distribution.

After these country-level and regional applications, several studies examined whether multisource models could be extended to broader geographic settings. Chi et al. [198] presented a complete and publicly accessible collection of microestimates of the distribution of relative wealth across 135 LMICs, rendered at a 2.4-km resolution, relying on several nontraditional data sources, including satellite imagery, mobile phone networks, geographic data, and Facebook connectivity data. Four types of features (satellite, connectivity, demographic, and geography) were used to train a popular and flexible supervised machine-learning model (i.e., the gradient-boosted regression tree) that omits preliminary feature selection prior to fitting the model (Fig. 17). The estimates showed strong predictive validity, explaining roughly 56%–70% of variation in survey-based wealth measures under different validation settings.

Marty and Duhaut [199] investigated the use of multiple data sources to estimate levels and changes in economic status, with particular attention to the contexts in which models trained on specific source combinations performed best. Their study provided a comprehensive analysis of global poverty estimation by developing a robust asset-based WI across 59 countries through the integration of multiple public and private sector data sources, including satellite imagery, Facebook marketing data, OSM information, weather, and climate indicators, among others. On average across countries, their models explained 55% of the variation in wealth levels at the survey-cluster level and 59% at the district level, whereas changes in wealth were much harder to predict. However, performance varied significantly, with the models performing best in lower-income countries and in countries with higher variance in wealth but with lower accuracy in some regions due to data limitations and other contextual factors.

Other work shifts attention from the scale of prediction to the poverty measure being predicted. Tennant et al. [200] shifted the prediction target from relative asset wealth to structural poverty measures in southern and eastern Africa. Their two-stage framework first estimates structural consumption from durable household assets and then uses Earth-observation and geospatial covariates, such as NTL, building footprints, remoteness, climate, and vegetation indicators, to predict cluster-level FGT poverty measures. This approach preserves some of the stability of asset-based measures while translating the output into policy-relevant quantities such as poverty headcount, poverty gap, and poverty severity. In pooled multicountry models, the method explained approximately 72%–78% of cluster-level variation, and 40%–54% when predicting out of country, while also revealing remaining bias and transferability limits at the lower end of the distribution.

More recent studies use multisource poverty estimates in targeting and allocation exercises. Jung et al. [201] examined whether multimodal spatial data could improve household-level poverty targeting in Brazzaville, Congo, an urban setting without conventional georeferenced income or consumption surveys. They combined social registry information with features derived from satellite imagery, social media, points of interest, connectivity, and administrative data to predict household-level multidimensional poverty. The predicted poverty measures were then used to compare machine-learning-based targeting with proxy means testing and community-based targeting. In a related study on Zambia, Jung et al. [202] combined satellite imagery, social media, and crowd-sourced geographic information to generate a 1 km × 1 km poverty map, which was further used to simulate geographic transfer allocation for the Food Security Pack. Together, these studies show how multisource poverty estimates can be linked to FGT-based allocation objectives and used beyond prediction, particularly in targeting and resource-allocation settings.

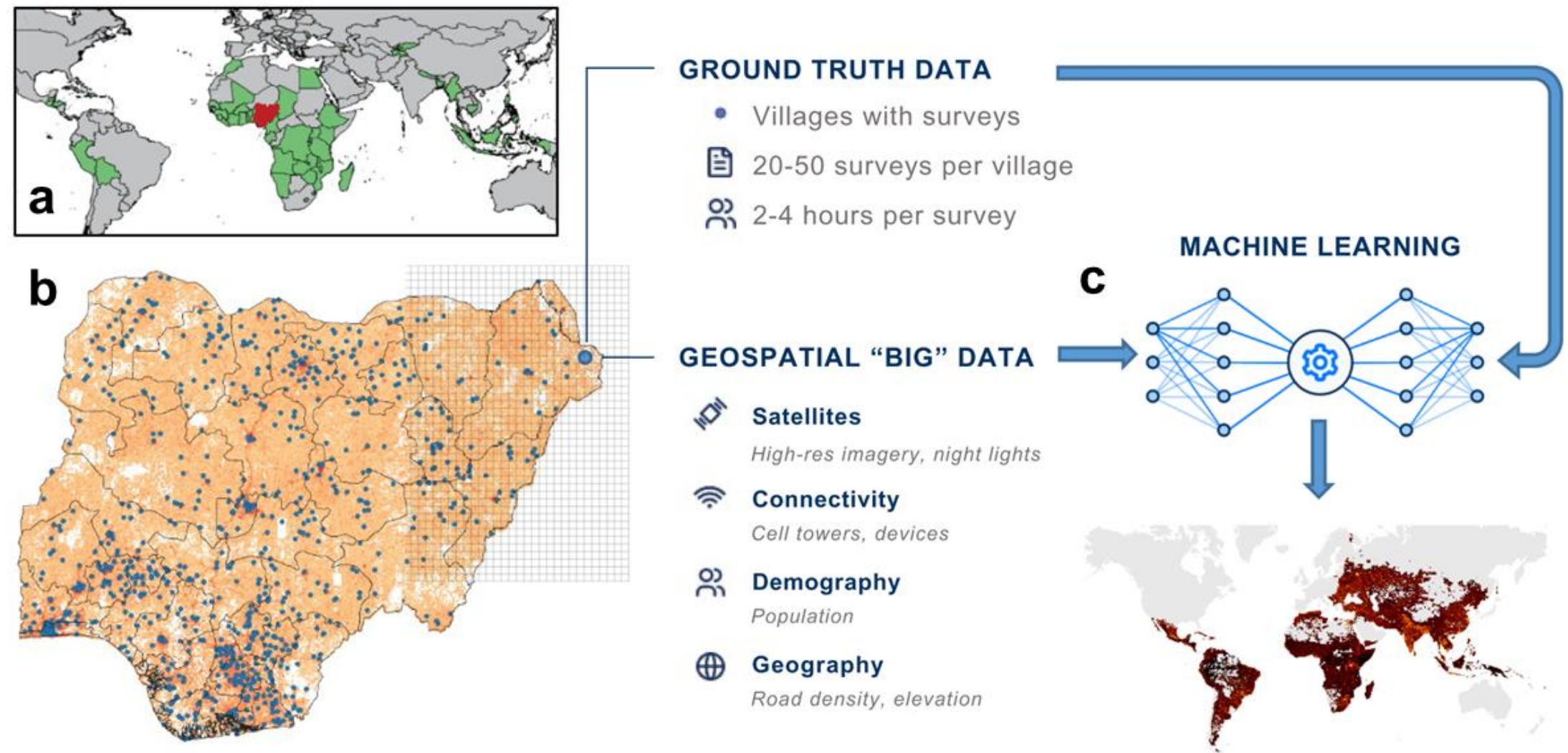

**Fig. 17. An illustration of the method for predicting microregional poverty using nontraditional data with machine learning algorithms.** (a) The geographic distribution of 1,457,315 unique households living in 66,819 villages from 56 LMICs that provide nationally representative household survey data. (b) The distribution of villages in Nigeria that provide ground truth data and geospatial "big" data. (c) The multisource data used to train a machine-learning algorithm that predicts microestimates of wealth for LMICs. The figure is adapted from Chi et al. [198].

Table 9 provides a summary of the research reviewed on poverty inference using multisource data.

**Table 9. Summary of research studies that inferred poverty based on multisource data.** "Samples" indicates the regions of training examples when available.

| Authors (Year) | Target Variable | Study Area | Nontraditional Data | Data Period | Modeling Approach | Performance |
|---|---|---|---|---|---|---|
| Steele et al. (2017) | WI, PPI, and household income | Bangladesh | RS and CDRs | 2011 (DHS WI); 2013–2014 (CDR/income survey); 2014 (PPI); | Hierarchical Bayesian geostatistical models | $R^2$ = 0.66–0.78, RMSE = 0.394 |
| Pokhriyal and Jacques (2017) | Multidimensional poverty index (MPI) | Senegal | CDRs and environmental data | 2013 (CDR/census/MPI); 1960–2014 (environmental data) | GP regression | $r$ = 0.91, RMSE = 0.08 |
| Njuguna and McSharry (2017) | Multidimensional poverty index (MPI) | Rwanda | CDRs, NTL and LandScan population density | 2014–2015 (CDR); 2014(NTL)；2012（MPI) | Manual feature engineering and selection (lasso | $r$ = 0.88 |
| | | | | | regression, ridge regression, and backwards stepwise regression) | |
| Niu et al. (2020) | MDPI, GDI | Guangzhou | RS and social media data | 2015 (RS/NTL) | RF regression | Spearman r = 0.954, Median relative error = 18.3% |
| Kondmann et al. (2020) | Local wealth | Sub-Saharan Africa | NTL and geolocated tweets data | 2009–2016 (DHS wealth); 2013 (NTL); 2018–2020 (Twitter) | RF regression | $R^2$ = 0.65 (Combined model) / $R^2$ = 0.55 (Twitter only) |
| Zhao et al. (2019) | WI | Bangladesh | NTL, Google satellite images, land cover data, road map, and division headquarter locations. | 2014 (DHS); 2015–2017 (Google satellite images); 2015 (NPP-VIIRS NTL) | RF regression | 10-fold CV $R^2$ = 0.70, Out-of-country $R^2$ = 0.61 |
| Zheng et al. (2024) | County-level MPI | Fujian, China | NTL and geospatial data | 2022 (statistics/NTL/geospatial data); 1999–2000 (DEM) | RF, SVM, AdaBoost, XGBoost, LightGBM, and LR | $R^2$=0.928, MAE=0.030, RMSE=0.037 |
| Tingzon et al. (2019) | WI, education, electricity access, and water access | Philippines | NTL, daytime satellite imagery, High Resolution Settlement Layer (HRSL), and OpenStreetMap | 2016 (NTL); 2017 (DHS/HRSL); 2018(daytime satellite imagery) | VGG16, RF regression | $R^2$ = 0.63 (Wealth), $R^2$ = 0.49 (Education), $R^2$= 0.36 (Electricity) |
| Espín-Noboa et al. (2023) | International WI | Sierra Leone and Uganda | Metadata collected via online crowd-sourcing, social media, and RS | 2016–2019 (DHS/MIS); 2022 (metadata/image features) | Combined model (CNN+ CatBoost regression model) | NRMSE (mean) = 0.45, NRMSE (std. dev.) = 0.79 |
| Chi et al. (2022) | Relative WI | 135 LMICs | RS, connectivity data from Facebook, data from mobile phone networks, and topographic maps | 2000–2020 (DHS) | Gradient-boosted regression tree | $R^2$ = 0.56–0.71 |
| Marty and Duhaut (2024) | WI | 59 countries | NTL, daytime satellite imagery, Facebook advertising data, OpenStreetMap, land | 2000–2020 (DHS); 2000–2022 | XGBoost, regularized linear regression | Cluster-level $R^2$ = 0.55, District-level $R^2$ = 0.59 |

| | | | cover characteristics, weather and climate data, pollution data, and synthetic aperture radar data | (RS/geospatial data) | (lasso, ridge, and elastic net), and SVM | |
|---|---|---|---|---|---|---|
| Tennant et al. (2025) | Structural FGT poverty measures | Ethiopia, Malawi, Tanzania, and Uganda | EO and geospatial data | 2008–2020 (LSMS) | OLS, RF, and EO-based ML models | $R^2$ = 0.72–0.78; out-of-country $R^2$ = 0.40–0.54 |
| Jung et al. (2026) | Household-level MPI | Brazzaville, Congo | Satellite imagery, Twitter, POI, connectivity, OSM, and administrative data | 2021 (social registry/MPI); 2023 (Landsat); 2021–2022 (NTL); 2019–early Sept. 2021 (Twitter); 2014 (OSM); | XGBoost, MLP, BART, SVR, and regularized linear models | $R^2$ = 0.709 (MLP), Targeting error = 7.5 |
| Jung et al. (2026) | WI and poverty allocation | Zambia | Satellite imagery, social media, OSM/POI, and connectivity data | 2018 (DHS WI); 2015 (LCMS); 2022–2023 (FSP); 2019–2023 (multimodal data) | Unimodal, multimodal, and stacked multimodal models | $R^2$=0.801, MAE=0.367 |

## 8. Discussion and perspectives

Poverty mapping has gradually moved from static statistical reporting toward the inference of deprivation within complex socioeconomic systems. Poverty-related conditions are shaped by coupled spatial, behavioral, infrastructural, and networked processes, including settlement morphology, access to services and opportunities, human mobility, communication networks, digital connectivity, and economic activity. Nontraditional observational data make some of these processes empirically visible at spatial and temporal scales that conventional surveys cannot easily capture, supporting more dynamic and fine-grained analyses of socioeconomic inequality and deprivation. Such data, however, do not measure poverty directly. They record observable traces of environments, movements, interactions, and activities from which poverty-related socioeconomic conditions can be indirectly inferred. Remote sensing, mobile phone data, social media traces, and multisource datasets therefore capture different components of the same socioeconomic system and should be interpreted together with, rather than as substitutes for, conventional survey- and census-based poverty statistics.

The policy relevance of such fine-scale socioeconomic inference is illustrated by China's experience with targeted poverty reduction. Over four decades of reform and poverty-reduction policy, China substantially reduced poverty and, by the end of 2020, announced the elimination of extreme poverty according to its national poverty threshold, with approximately 770 million people lifted out of poverty since 1978 [203-205]. This experience shows why poverty mapping should not be evaluated only as a predictive exercise, but also by whether estimates are spatially detailed, locally interpretable, and reliable enough to support targeted policy action.

For practical applications, the central issue is not prediction alone, but the conditions under which poverty estimates can be used with confidence. A key challenge is to determine when the observable traces used as poverty proxies are representative, transferable, interpretable, and reliable for local decision-making. These requirements depend on the coverage and quality of nontraditional data, the spatial scale at which poverty can be meaningfully inferred, and the way model performance is evaluated. The following subsections use these three dimensions to organize the discussion of data representativeness and uncertainty, spatial heterogeneity and effective resolution, and evaluation of poverty estimates.

## 8.1 Discussion

### *8.1.1 Data representativeness and uncertainty*

Although nontraditional data sources can improve the temporal and spatial coverage of poverty mapping, the reliability of such estimates remains constrained by two related sources of uncertainty. The first concerns data representativeness. CDR-based studies, for example, typically rely on data from one or a limited number of mobile operators and therefore capture only a subset of the national population. Specific demographic groups, including children, older adults, low-income populations, displaced populations, and individuals with limited access to digital infrastructure, may be systematically underrepresented in mobile-phone or platform-based datasets. Blumenstock and Eagle showed that mobile phone users in Rwanda were substantially wealthier and better educated than the general population, highlighting potential demographic bias in mobile-phone-based analyses [145]. Similar concerns have been reported in studies using Facebook advertising data, where online populations may underrepresent refugee women and lower-education groups [174]. Such biases can affect both the magnitude and the spatial distribution of inferred poverty estimates, particularly when model outputs are treated as population-representative evidence. Bias assessment and validation across different populations or regions are therefore necessary before these data are used for policy-oriented poverty mapping.

A second source of uncertainty concerns the reliability of observed proxies. Because nontraditional data capture signals through observation, they may be affected by sensor performance, data acquisition mechanisms or context-dependent biases. Social media data may contain noise, duplicated content, misinformation, or platform-specific behavioral biases. Mobile phone and remote sensing data also require choices about spatial aggregation, temporal windows, and linkage between observations and survey labels. Some studies have addressed these issues using methods informed by statistical physics. For example, coarse-graining and scale-dependent aggregation can reduce stochastic noise by summarizing satellite features, mobile-phone records, or base-station-level activities into grid units or population groups [86, 197]. However, aggregation alone cannot fully address uncertainty, which still needs to be quantified and tested. Entropy-based measures and spatial random-field models can further characterize behavioral variability and provide uncertainty estimates [45, 128, 154], while null models and cross-region validation can help assess whether inferred poverty-related patterns remain robust [86, 175]. Notably, the use of these methods remains uneven across data sources, spatial scales, and study contexts in poverty-mapping research. A systematic framework for handling and evaluating uncertainty in a unified manner has not yet been established. Given that these issues can fundamentally affect the reliability of poverty estimates, integrating uncertainty-aware methods into policy-relevant frameworks remains an important open challenge.

*8.1.2 Methodological transferability, effective resolution, and interpretability*

The effectiveness of data-driven poverty mapping depends not only on the type and quality of data sources, but also on how different modeling approaches balance spatial heterogeneity and transferability. Traditional statistical and machine learning models based on engineered features, such as linear regression and RF, are often easier to interpret and are commonly used for broad regional assessment. These models use predefined physical or behavioral indicators to identify broad socioeconomic trends. However, their ability to capture intra-urban heterogeneity is limited. Coarse NTL resolution, light saturation effects, or uneven cell-tower density may smooth sharp socioeconomic gradients and obscure localized pockets of poverty [107, 108, 110], limiting the effective resolution to the sub-kilometer level. In addition, model performance may decline across regions with different cultural, infrastructural, or urban conditions.

Compared to traditional feature-based models, representation learning models, including CNNs, LSTMs, and GCRNs, can capture more complex spatial and temporal patterns by automatically extracting visual, geometric, or sequential representations. These “end-to-end” approaches are more sensitive to fine-scale urban variations, making them particularly effective for identifying informal settlements and intricate slum-like morphologies, and in some applications, they can support effective analysis at the community or even individual-building scale [123, 124, 128]. Notably, their effective resolution is frequently constrained by the ground-truth label uncertainty, such as DHS coordinate displacement, or by privacy-related aggregation thresholds, rather than the algorithms themselves [27, 114, 174].

Multisource data fusion models can combine physical, behavioral, and digital dimensions of socioeconomic conditions within a unified analytical framework. By integrating data sources with different spatial supports, such frameworks can enable adaptive poverty mapping from 2.4 km global grids to 500 m regional units [152, 197]. With suitable input data and application-specific settings, they may support finer-grained spatial analysis [28]. This helps reduce the limitations of single data source proxies and improves the robustness of socioeconomic inference.

However, as deep learning and fusion models become increasingly complex,

interpretability becomes more critical for policy applications. Current research emphasizes predictive performance, yet the importance of different features and their mechanistic links to poverty determinants remain poorly understood. This opacity may conceal errors or biases, and without understanding how features drive predictions, the applicability of these findings is limited, often informing data science rather than policy [204, 205]. Decision-makers need to understand why a model identifies certain areas as poor and how reliable these results are. Otherwise, black-box predictions may weaken public trust. This is particularly important when model outputs are used to guide local resource allocation.

*8.1.3 Evaluation of poverty estimates*

Another methodological issue in poverty mapping concerns how model performance is evaluated. Existing studies have not adopted a uniform evaluation protocol, partly because they differ in the data sources they use, the outcomes they predict, the spatial units at which estimates are produced, the validation data available, and the policy settings in which the maps are intended to be used. Some studies produce continuous estimates of welfare, wealth, or deprivation, whereas others identify deprived settlements, distinguish poor from non-poor groups, or rank places for intervention. As a result, reported performance scores cannot always be read as directly comparable evidence of model quality across studies.

Measures based on explained variance or correlation, such as $R^2$, are widely reported in poverty-mapping studies because they provide a common way to assess whether a model captures the main spatial variation in poverty-related outcomes. These metrics are particularly useful when the purpose is to examine the relationship between external data sources and poverty-related conditions, or to evaluate continuous estimates of wealth, consumption, or deprivation. Yet these measures do not show the size of prediction errors in specific places. For continuous poverty estimation, mean absolute error (MAE) and root mean square error (RMSE) therefore provide a more direct indication of the magnitude of prediction errors, with RMSE placing greater weight on large deviations [193, 196]. When the task is to identify deprived areas, delineate slums, or distinguish poor from non-poor groups, evaluation should further consider classification-oriented measures. Accuracy, area under the curve (AUC), F1 score, sensitivity, and specificity are more suitable for assessing whether areas or populations in need are correctly identified [157, 188].

For targeted poverty alleviation, aggregate model performance is insufficient unless local estimates are accompanied by uncertainty information. A poverty map may show good overall fit, but this does not guarantee that estimates are equally reliable across all areas, especially in data-sparse regions, among populations with limited digital coverage, or in locations close to eligibility thresholds. In such settings, small estimation errors may affect whether an area or population is selected for intervention. Standard errors and confidence intervals can indicate the expected range of sampling or prediction error, while credible intervals, posterior predictive uncertainty, and spatial uncertainty maps can show where the model is less certain about its estimates [45, 193]. These measures help identify areas where poverty estimates should be interpreted with caution and where additional validation or data collection may be needed. For policy use, evaluation should therefore connect aggregate performance with local reliability, so that poverty maps are not treated as equally certain across space [74, 157].

## 8.2 Perspectives

Despite the aforementioned caveats, the breadth of novel data and algorithmic approaches is clearly promising in measuring poverty. We lay out some potential challenges and directions for future work.

First, a standardized data access and sharing ecosystem should be developed. A conducive environment for data collaboration will benefit big data analytics. The government can actively promote an environment through, for example, the open data challenge D4D [207]. On the basis of protecting data security and user privacy, a data access and sharing mechanism should be established, including data acquisition, preservation, and usage, and involving the government, private sector, charities, and academia.

Second, novel algorithms should be designed to accurately identify individuals in poverty. The essence of precisely alleviating poverty hinges upon assisting the truly impoverished, rather than upon adopting an interventionist approach that merely performs on average. However, classifying all household members as poor or non-poor on the basis of household-level status lacks precision, often resulting in overlooked intra-household nuances, such as the elevated poverty rates faced by women and children because household-level measures may obscure intra-household differences in access to resources [208]. Moreover, models trained on male-headed households achieve high predictive accuracy, while those based on female-headed households perform less well, a gap partly attributable to the smaller representation of female-headed households in the survey sample [209]. Granular and up-to-date data provide critical input for designing algorithms to achieve this goal with minimal operator intervention. For example, mobile phone datasets with demographic information have led to many discoveries about patterns of human behavior [210-213], which may contribute to modeling the features of the poor at the individual level.

Third, future work should move toward more dynamic and resilient poverty alleviation strategies. The frequency and intensity of natural hazards have increased globally in recent years. In particular, the almost three-year COVID-19 pandemic demonstrated the frightening susceptibility of all societies to global crises, with their cascading negative effects on global health, the environment, education, and livelihoods, as well as a severe global recession that jeopardizes the commitment of all nations to the SDGs [214-217]. It is crucial and challenging to create a dynamic and resilient poverty alleviation framework that can incorporate large datasets of historical events and environmental variables. This generalizable framework could be designed to model and predict the occurrence and impact of natural disasters, population movements, and shifting economic conditions, which would facilitate the constant adjustment and optimization of strategies based on dynamic conditions and needs.

Fourth, novel indices should be developed to evaluate aid approaches. Although there are some effective methods for mapping and predicting poverty, most evaluations have primarily focused on correlating these estimates with ground-based measures of specific outcomes. However, less attention has been devoted to assessing the long-term consequences of these methodologies, resulting in limited practical use of existing methods. A rigorous mechanism for real-time monitoring at all stages of long-term impacts on aid programs should be established for a thoughtful, detailed examination of variations. Policymakers should continuously monitor, evaluate, and adapt poverty alleviation strategies on the basis of changing environmental conditions and community needs.

We believe this review underscores the significant potential of data-driven modeling approaches to produce insights that can contribute to the fight against poverty. The increasing availability of nontraditional data sources, coupled with advances in statistical physics, network science, and machine learning, has been shown to improve decisions, outcomes, and the development of evidence-based policy [218]. However, such data and methodologies should not be viewed as a panacea for all problems with data collection and measurement [219]. Instead of exploring solutions from the standpoint of a data source and placing blind trust in data-driven algorithms, it would be wise to first consider them from a problem-driven perspective. When utilized in conjunction with traditional data sources, these data-driven approaches can serve as valuable complements to, rather than replacements for, traditional measures.

## CRediT authorship contribution statement

**Suoyi Tan:** Writing – review & editing, Writing – original draft, Visualization, Supervision, Software, Project administration, Methodology, Investigation, Funding acquisition, Data curation, Conceptualization. **Mengning Wang:** Writing – review & editing, Writing – original draft, Methodology, Investigation, Software, Formal analysis, Data curation. **Yixiu Kong:** Writing – review & editing, Writing – original draft, Methodology. **Huimin Bai:** Writing – review & editing, Writing – original draft. **Jianguo Liu:** Writing – review & editing, Writing – original draft, Conceptualization. **Dirk Brockmann:** Methodology, Conceptualization. **Yicheng Zhang:** Methodology, Conceptualization. **Xin Lu:** Writing – review & editing, Conceptualization, Funding acquisition.

## Declaration of competing interest

The authors declare that they have no known competing financial interests or personal relationships that could have appeared to influence the work reported in this paper.

## Acknowledgements

This work was supported by the National Natural Science Foundation of China (72474223, 72421002), the Science and Technology Innovation Program of Hunan Province (No. 2024RC3133), and the Innovation Research Foundation of National University of Defense Technology (No. JS24-04).

## Appendix

**Table S1. Summary of classical poverty indicators**

| Indicator | Proposer | Formula | Description | Highlights |
|---|---|---|---|---|
| ***Unidimensional poverty measurement*** | | | | |
| Gini Coefficient | Corrado (1912) | $G = 1 - 2\int_{0}^{1} L(X)dX$ | $L(X)$ is a Lorenz curve, and $X$ is the income or wealth. | Measures the inequality of income or wealth distribution |
| Headcount Index ($P_0$) | Foster et al. (1984) | $P_0 = \frac{1}{N}\sum_{i=1}^{N} I(y_i < z)$ | • $N$ is the total population (or sample).<br>• $I(\cdot)$ equals 1 if the income (or expenditure) $y_i$ is less than the poverty line $z$, and 0 otherwise.<br>• The poverty gap ($G_i$) represents the extent of the actual income (or expenditure) $y_i$, which lies below the poverty line $z$ for poor individuals and equals 0 for non-poor individuals.<br>• $G^P$ is the Gini coefficient of inequality among the poor.<br>• $\hat{G}_p$ is the Gini coefficient of the poverty gap ratios for the entire population. | Simply measures the proportion of the population living below the poor line |
| Poverty Gap Index ($P_1$) | Foster et al. (1984) | $\begin{cases} P_1 = \frac{1}{N}\sum_{i=1}^{N}\frac{G_i}{z} \\ G_i = (z - y_i)\times I(y_i < z) \end{cases}$ | | Further adds up the extent to individuals fall below the poverty line |
| Poverty Severity Index ($P_2$) | Foster et al. (1984) | $\begin{cases} P_2 = \frac{1}{N}\sum_{i=1}^{N}\left(\frac{G_i}{z}\right)^2 \\ G_i = (z - y_i)\times I(y_i < z) \end{cases}$ | | Further considers the inequality by weighting the poverty gaps |
| Sen Index | Sen (1976) | $P_s = P_0 G^P + P_1(1 - G^P)$ | | Combines measures of the proportion of poor people, depth of their poverty, and distribution of welfare among the poor |
| Sen-Shorrocks-Thon (SST) Index | Shorrocks (1995) | $P_{SST} = P_0 P_1^P (1 + \hat{G}_P)$ | | Considers the degree, incidence, and inequality of poverty jointly, and can evaluate the change of poverty sources over time |
| ***Multidimensional poverty measurement*** | | | | |
| Townsend | Townsend | - | A census-based index of material deprivation in four dimensions: | A measure of material deprivation within a population |
| Deprivation Index | (1988) | | • *Non-car ownership*<br>• *Non-home ownership*<br>• *Unemployment*<br>• *Household overcrowded* | |
| Human Poverty Index (HPI) | UNDP (1997) | - | HPI-1 measures the deficiencies in the three indices:<br>• *Longevity*<br>• *Knowledge*<br>• *Daily comforts*<br>HPI-2 additionally measures *social exclusion*. | Indicates the living standard in a country and uses deprivation as a means to measure poverty |
| Multidimensional Poverty Index (MPI) | Alkire and Santos (2010) | - | Selects three dimensions to measure poverty, including a total of ten dimensions:<br>• *Health: nutrition and child mortality*<br>• *Education: years of schooling and school attendance*<br>• *Living Standard: cooking fuel, sanitation, drinking water, electricity, floor, and assets* | Reflects the intensity and incidence of multidimensional poverty and the deprivation of individuals or families |
| PPI | Grameen Foundation (2005) | - | Has been developed with a specific aim to measure poverty at the household level in a particular country. | A country-specific poverty indicator to measure the distribution of poverty among the participants and track poverty levels over time |

3

4 **Table S2. Summary of novel data resources utilized to estimate poverty**

| Authors | Time | Region | Description | Source |
|---|---|---|---|---|
| ***Nighttime lights data*** | | | | |
| Elvidge et al. (1997) | 1994–1995 | 21 countries | The total land area covered in the study is nearly 26 million $km^2$, using a data volume of 3.1 gigabytes, which is less than 1% of the data volume required for a single complete coverage with Landsat TM data. The area lit is defined as regions with detectable lights in at least 10% of the cloud-free observations. | National Oceanic and Atmospheric Administration's (NOAA) National Geophysical Data Center (NGDC) |
| Ebener et al. (2005) | October 1994–March 1995 | 171 countries | This grid dataset resulted from a 6-month, 1-km resolution composite based on images by the U.S. Air Force DMSP-OLS. The key parameters extracted from the grid included the number of cells highlighted at night, the area lit (surface area being highlighted at night), the total frequency of light observation (indicating the total intensity of the highlighting), and the mean frequency of light observation (indicating the dispersion of this intensity). | NGDC |
| Sutton et al. (2007) | 1992–1993, 2000 | India, China, Turkey and U.S. | The data include global population databases known as LandScan and DMSP-OLS city light composites. | http://www.du.edu/sutton/GDPestimationData |
| Sutton and Costanza (2002) | 1995-1996 | Global | The dataset provides global coverage, capturing NTL for the entire Earth's surface at a 1-$km^2$ resolution. The NTL data are combined with a global land-cover dataset and ecosystem service values to estimate economic activity and ecosystem service product (ESP) at a high spatial resolution. | NGDC |
| Noor et al. (2008) | 2000 | 37 countries in Africa at the Administrative 1 level | The dataset uses a global human settlement NTL product at approximately 1×1 km spatial resolution. The brightness of the light pixels was measured on an arbitrary scale from 0 to 63 units. | NGDC |
| Elvidge et al. (2009) | 2003 | 233 countries (cover global extents between the 65° south and 65° north latitude) | The data are used to create cloud-free composites, screening out cloud-affected data and other anomalies. The NTLs are represented on a 30-arc-second grid. The brightness values range from 2 to 3 digital numbers (DN) to a saturation DN of 63, indicating the intensity of lighting. | The gray-scale image data are available at http://www.ngdc.noaa.gov/dmsp/download_poverty.html |
| Henderson et al. (2012) | 1992–2003 | 187 countries (cover the entire globe between the | This data series consist of 18 annual composites from four satellites, each annual composite being a raster dataset with values every 30 seconds of latitude and longitude (approximately 0.86 $km^2$ at the equator, decreasing with the | The Version 2 DMSP-OLS NTL Time Series data. Available at |
|  |  | 65° north and 65° south latitude) | cosine of latitude). The data are compiled and cleaned by the National Geophysical Data Center (NGDC) of NOAA, removing temporary features such as forest fires and excluding readings affected by auroral activity and cloud cover. Each grid value is an eight-bit integer (0–63) called a digital number (DN) averaged over all nights fitting certain criteria. | http://www.ngdc.noaa.gov/dmsp/global_composites_v2.html |
| Mellander et al. (2015) | 2006 | Sweden | Radiance data had pixel (light) values ranging from 0 to 846, while the saturated data have values ranging from 0 to 63. The NTL data are provided in 30-arc-second grids through geo-referenced TIFF images. The study used both radiance and saturated light emissions to correlate with economic activity, population, and establishment density. | The NTL data are made available by the U.S. National Oceanographic and Atmospheric Administration (NOAA).<br>The data are collected by the OLS flown on the DMSP satellites.<br>Data documentation and download can be found at the NOAA National Geophysical Data Center (NGDC), https://ngdc.noaa.gov/eog/ |
| McCallum et al. (2022) | 2015 | 49 countries spread across Africa, Asia, and the Americas | The VIIRS-DNB sensor images nearly the entire Earth nightly at an approximate resolution of 750 m. The WSF dataset, with an original spatial resolution of 10 m, represents the most detailed global inventory of human settlements. | The NTL data can be downloaded from the Earth Observation Group (EOG) of the Colorado School of Mines. Available at https://eogdata.mines.edu/download_dnb_composites.html |
| Wang and Li (2024) | 2015 | 33 countries in Africa | The dataset includes annual NTL images of Africa for the year 2015 from the VIIRS-DNB V2.1 dataset from the Earth Observation Group. The annual NTL image boasts a spatial resolution of 15 arc seconds (approximately 500 m at the equator). | https://eogdata.mines.edu/products/vnl/ |
| Tang et al. (2025) | 1992-2022 | Global | This dataset provides a global long-term nighttime light time series (1992–2022) named LRCC-DVNL, with a specific focus on low-light areas and dark sky areas. It achieves pixel-level alignment, restoration, and continuous calibration across DMSP-OLS (1992–2013) and NPP-VIIRS (2014–2022) data. With a spatial resolution of 1000 meters, the dataset effectively addresses data | https://doi.org/10.7910/DVN/15IKI5 |
| | | | gaps in high-latitude regions and resolves interannual continuity challenges between disparate sensors. | |
| ***Daytime satellite imagery data*** | | | | |
| Jean et al. (2016) | - | Nigeria, Tanzania, Uganda, Malawi, and Rwanda | The dataset includes daytime images from Google Static Maps at zoom level 16. It includes features such as urban areas, roads, bodies of water, and agricultural areas. | Google Static Maps API |
| Xie et al. (2016) | 2015 | Uganda | The dataset consists of more than 330,000 daytime satellite images, each with a resolution of 400 × 400 pixels at zoom level 16. The images are labeled with NTL intensity values ranging from 0 to 63. The dataset is balanced using a 3-component Gaussian mixture model to handle the imbalance in NTL intensity values. | Google Static Maps API |
| Head et al. (2017) | 2010<br>2013<br>2012<br>2011 | Rwanda<br>Nigeria<br>Haiti<br>Nepal | The dataset consists of millions of satellite images at zoom level 16, providing a pixel resolution of 2.5 m. Each image is set to 400 × 400 pixels, matching the land area covered by a single pixel of NTL data (1 $km^2$). The total area covered by these images included Rwanda, 26,338 $km^2$; Nigeria, 923,768 $km^2$; Haiti, 27,750 $km^2$; and Nepal, 147,181 $km^2$. The number of images downloaded for each country was as follows: Rwanda, 29,627 images; Nigeria, 1,036,956 images; Haiti, 33,490 images; and Nepal, 195,618 images. | Google Static Maps API |
| Yeh et al. (2020) | 2009–2017 | 23 countries in Sub-Saharan Africa | Each image has a spatial resolution of 30 m/pixel and included seven bands: RED, GREEN, BLUE, NIR (near infrared), SWIR1 (Shortwave Infrared 1), SWIR2 (Shortwave Infrared 2), and TEMP1 (thermal). | The data to replicate all findings in this paper are available at https://github.com/sustainlabgroup/africa_poverty |
| Ratledge et al. (2022) | 2006–2016 | Sub-Saharan Africa and Uganda | The dataset consists of multispectral satellite images, including RGB visual bands, near-infrared bands, and shortwave infrared bands, with a resolution of 30 m per pixel. It includes georeferenced asset wealth data from 27,174 enumeration areas across 25 countries in Sub-Saharan Africa, encompassing 641,621 household surveys conducted over a period of 13 years. | Data are available on GitHub at https://github.com/nwrat/RCQSB2022_public |
| ***Very high-resolution imagery data*** | | | | |
| Rogers et al. (2006) | 1992–1996 | Uganda | These data are processed into synoptic monthly (maximum value) composites to provide a record of monthly changes in an average year. The data include the middle infrared (MIR AVHRR channel 3), Land Surface Temperature (LST), Normalized Difference Vegetation Index (NDVI), air temperature (Tair), and vapor pressure deficit (VPD). | NASA Pathfinder program: http://terra.nasa.gov/. The Meteosat data are provided by the FAO ARTEMIS program: https://www.esa.int/ |
| Engstrom et al. (2022) | - | Sri Lanka | The imagery data covers approximately 3,500 km$^2$ in Sri Lanka, encompassing 47 divisional secretariat (DS) divisions and 1,291 villages (Grama Niladhari divisions). The satellite imagery consists of 55 unique “scenes” purchased from Digital Globe. | The satellite imagery consists of 55 unique “scenes” purchased from Digital Globe (https://www.digitalglobe.com/) |
| Watmough et al. (2019) | September 2004 | Sauri, Kenya | The dataset is a fine spatial resolution land use/land cover (LULC) map, pan-sharpened to a 0.6-m spatial resolution, covering an area of 10 × 16 km. It identifies eight land use classes: agriculture, building, grassland, non-vegetated (bare ground), road, shrub, water, and woodland. The dataset includes NDVI time series data from the 500-m resolution MODIS. | QuickBird |
| Kit et al. (2012) | 2003 | Hyderabad, India | The data covers 400 km$^2$ of urban territory within a rectangular bounding box of 78°22′–34′0″ east longitude and 17°18′–30′0″ north latitude. The imagery was radiometrically calibrated, pan-sharpened, corrected for sensor- and platform-induced distortions, and mapped to UTM projection zone 44 N. | QuickBird |
| Kit and Lüdeke (2013) | 2003–2010 | Hyderabad, India | The imagery is provided as a gridded dataset, radiometrically calibrated, pan-sharpened, corrected for sensor- and platform-induced distortions, and mapped to UTM projection zone 44 N. The image size is 38,144 × 68,287 pixels. | QuickBird (2003) and WorldView 2 (2010) |
| Kohli et al. (2012) | - | Kisumu, Kenya | The data include high-resolution images that allowed for a detailed analysis of physical characteristics such as building density, roof material, road layout, and vegetation. | GeoEye Foundation, http://www.geoeyefoundation.org/ |
| Kuffer et al. (2016) | 2009 | Mumbai | The dataset includes six WorldView-2 scenes (PAN: 0.5 m and MS: 2 m) covering a part of the city. It is supported by 80 ground-truth points collected between 2011 and 2013, and 180 sample points for accuracy assessment. | WorldView-2 |
| | 2004 | Ahmedabad | The dataset includes an OrbView image (available via USGS) and a Resourcesat image. Data on slum locations from the municipality (circa 2010) were used for training and assessment. | OrbView, Resourcesat |
| | 2004, 2005 | Kigali | The dataset features a QuickBird image from 2004 and municipal land use data from 2005, which were used for training and accuracy assessment. | QuickBird |
| Stark et al. (2024) | 2021 | Cape Town, Caracas, Lagos, Medellin, Mumbai, Nairobi, Rio de Janeiro, and Sao Paulo | The dataset comprises 8-bit RGB remote sensing images from PlanetScope satellites, covering eight cities in the Global South with a 3-m resolution, divided into patches of 88 × 88 pixels and annotated with three classes: background, formal built-up areas, and slums. It is supplemented with expert-mapped polygons and aerial imagery from Google Earth. | PlanetScope |
| ***Call Detail Records*** | | | | |
| Blumenstock and Eagle (2012) | January 2005–December 2008 | Rwanda | The dataset consists of phone call details from 1,529 users. It includes comprehensive data on the time and date of each call, and the proximate locations of the caller and receiver, which were determined by the cell towers through which the calls were routed. | Phone company |
| Blumenstock et al. (2015) | May 2008–May 2009 | Rwanda | The dataset encompasses information from 1.5 million unique individuals. It is organized into primary geographic units comprising of 30 districts and secondary geographic units represented by 300 cell towers. | Primary mobile phone operator in Rwanda |
| Aiken et al. (2022) | October–December, 2018, April–June, 2019, March-September 2020 | Togo | The dataset includes detailed records of various mobile activities. Calls contain the caller's phone number, the recipient's phone number, the date and time of the call, the call duration, and the ID of the cell tower through which the call was placed. SMS messages include the sender's phone number, the recipient's phone number, the date and time of the message, and the ID of the antenna through which the message was sent. The mobile data usage records consist of the phone number, the date and time of the transaction, and the amount of data consumed (combined upload and download). In addition, the dataset includes mobile money transactions detailing the sender's phone number, the recipient's | Togo's two mobile phone operators |
| | | | phone number (for peer-to-peer transactions), the date and time of the transaction, the amount of the transaction, and the broad category of the transaction type (cash in, cash out, peer-to-peer, or bill pay). | |
| Soto et al. (2011) | February–July 2010. | A main city in a Latin-American country | The dataset is from approximately 500,000 users, focusing on those with an average of two daily calls. The dataset covers the city, which is served by 920 base transceiver stations (BTS) towers. For each user, 279 features modeled from the CDR data were computed, including 69 behavioral variables, 192 social network features, and 18 mobility variables. | - |
| Gao et al. (2024) | November 2016 | Shenzhen, China | The dataset includes 11,812,888 origin-destination (OD) trips and 3,370,583 transit users, collected over six days from public transit (subway and bus) smart card records. | - |
| Frias-Martinez and Virseda (2012) | - | 12 cities in a Latin American country | The dataset includes five months of cell phone calls and SMS data from over 10,000,000 prepaid and contract subscribers. It contains encrypted originating and destination numbers, the time and date of each call, the call duration, and the BTS connected to the call. | A telecommunications company operating in the Latin American country |
| Pappalardo et al. (2016) | 2007 | France | The dataset records the geographic location of 87,000 phone towers and 5.7 billion calls made during 45 days by 20 million anonymized mobile phone users, resulting in a total size of 900 GB of information. | Orange mobile phone operator |
| Wang et al. (2024) | September 30, 2016–April 1, 2017 | Major American cities, specifically Boston, Chicago, and Miami | The dataset used in the study comprises mobility patterns from more than 560,000 anonymized users across Boston, Chicago, and Miami, tracking 687 types of activities to predict economic outcomes using machine learning. | Cuebiq, https://www.cuebiq.com/about/data-for-good/ |
| Rubio et al. (2010) | - | - | The dataset consists of CDRs for nearly 300,000 anonymized customers in an advanced economy and 700,000 anonymized customers in a developing economy. Each CDR includes information such as the originating encrypted number, the destination encrypted number, the time and date of the call, the duration of the call, and the BTS to which the cell phone was connected during the call. The dataset specifically focuses on subscribers with an average of at | Telefónica |
| | | | least two calls per day, encompassing data from 220,000 individuals in the advanced economy and 190,000 individuals in the developing economy. | |
| Xu et al. (2018) | 2011 (50 days) | Singapore | The dataset covers 4.4 million cellphone users and includes unique IDs, communication type (call/SMS), date, time, and user location. The locations are reported as the latitude and longitude of the cellphone tower. The network comprises approximately 5000 towers, with an average nearest distance of 100 meters between them. | - |
| | 2009 (4 months) | Boston Metropolitan Area | The dataset contains location estimations for approximately 1 million anonymous mobile phone users. These locations are generated through triangulation during calls, messaging, or web browsing activities. The uncertainty range for the location estimations has a mean of 320 m and a median of 200 m. | AirSage (http://www.airsage.com) |
| Blumenstock (2018) | July 2009 | Rwanda | The dataset includes 856 mobile phone subscribers. A 20-minute phone survey was conducted with a geographically stratified sample of subscribers. The survey data were merged with mobile phone records that described all transactions made by each subscriber since 2005. The mobile phone records included data such as phone calls, text messages, airtime purchases, and mobile money use. | - |
| | Nov 2015–Apr 2016 | Afghanistan | The dataset includes 1,234 respondents who participated in several rounds of face-to-face and phone-based interviews. The survey focused on male heads of households in the provinces of Kabul and Parwan. Each respondent's survey responses were matched with their mobile phone transaction records obtained from the Roshan mobile phone network, covering the period starting in January 2014. These data included phone calls, text messages, airtime purchases, and mobile money use. | - |
| Aiken et al. (2023) | November 2015–April 2016 | Six provinces in Afghanistan, with a specific focus | The dataset consists of detailed logs of 629,543 transactions, including 310,883 calls, 305,756 text messages, and 12,904 recharges. These data are matched to | Primary mobile phone operator in Afghanistan |
| | | on the Balkh province for the detailed analysis | survey respondents who consented to participate, covering phone numbers, call durations, text messages, and recharges. | |
| ***Mobile communications network*** | | | | |
| Eagle et al. (2010) | August 2005 | UK | The data consist of anonymized call logs recorded by the network operator as required by law and for billing purposes. The data encompass more than 90% of mobile phones and greater than 99% of residential and business landlines in the country, resulting in a network with 65 million nodes and 368 million reciprocated social ties. The mean geodesic distance in this network is 9.4, and the average degree of network neighbors is 10.1. The giant component contains 99.5% of all nodes. | - |
| Smith-Clarke et al. (2014) | December 1, 2011–April 28, 2012 | Côte d'Ivoire | The dataset contains aggregated CDRs of calls between 5 million Orange customers. It includes the total volume (471 million calls) and duration (960 million minutes) of calls routed through cell towers over a 20-week period. Côte d'Ivoire has a population of approximately 20 million and a total area of 33 million hectares. | Orange mobile phone operator |
| | A period of 6 weeks in early 2012 | An anonymous region | The dataset includes call records of approximately 928,000 customers over a 6-week period. It contains the total volume (40 million calls) and duration (170 million minutes) of the calls. Region B has a population of approximately 58 million and a total area of 19 million hectares. | - |
| Mao et al. (2015) | December 1, 2011–April 28, 2012 | Côte d'Ivoire | The dataset consists of anonymous metadata pertaining to 2.5 billion calls and SMS exchanges. It includes two types of mobile phone datasets: hourly antenna-to-antenna traffic and individual trajectories for 50,000 customers for two weeks, with a spatial resolution down to the individual antenna. The data pertain to calls between 1,231 cell towers distributed across 50 departments. The mobile penetration rate in Côte d'Ivoire is 83%. | The data were provided by France Telecom-Orange through the D4D Challenge in 2013 |
| ***Advertising data*** | | | | |
| Fatehkia et al. (2020) | March–April 2019 | Philippines | The dataset is derived from Facebook's Marketing API, which provides audience estimates typically used for advertising campaign planning. It includes information on the types of devices and network connections used by Facebook users in specific locations. | The data were collected from the Facebook Marketing API (https://developers.facebook.com/docs/marketing-api/audiences/reference/basic-targeting/#location) using the pySocialWatcher library (https://github.com/maraujo/pySocialWatcher) |
| | June–September 2019 | India | | |
| Fatehkia et al. (2022) | November 1, 2019–July 2020 | Lebanon | The dataset is derived from Facebook's Adverts Manager tool and marketing API. It includes estimates of monthly active users (MAU) who match specific targeting criteria such as geographic location, demographics (age and gender), self-declared education levels, inferred user behavior (devices and network connection types), language, and migrant status. | |
| Piaggesi et al. (2022) | January–September 2020 | Atlanta, Bogotá, Santiago, Casablanca | The dataset includes estimates of MAU based on various attributes such as demographics, technology use, and behaviors. These data were collected to provide a detailed mapping of socioeconomic status within city neighborhoods. | The datasets presented in this study can be found in the online repository: https://github.com/simonepiaggesi/predicting-city-poverty-facebook |
| ***Social media content*** | | | | |
| Quercia et al. (2012) | September 27 and December 10, 2010 | London | The dataset initially consisted of 250,000 profiles. Of these profiles, 157,000 specified geographic locations, and 1,323 specified London neighborhoods. After filtering for real profiles, the dataset was refined to include 573 profiles. | The datasets were sourced from Twitter profiles, with locations geo-referenced using the Yahoo! PlaceMaker API |
| Preoţiuc-Pietro et al. (2015) | Around August 5, 2014 | - | The dataset includes 5,191 users from 55 three-digit job groups across nine 1-digit Standard Occupational Classification (SOC) groups. It consists of 10,796,836 tweets. Each user is mapped to their income using their job title as a proxy. | Data are available for download from http://figshare.com/articles/Twitter_User_Income_Dataset/1515997 |
| Flekova et al. (2016) | - | - | The dataset comprises two large sets of tweets. Each dataset consists of users and their historical records of tweet content, profile information, and trait-level features extracted with high precision from their profile information. | - |
| Hasanuzzaman et al. (2017) | January 1, 2015–January 31, 2015 | - | The data consists of approximately 10 million tweets from 5,191 Twitter users. These tweets were used to analyze the temporal orientation (past, present, and future) and its correlation with the users' income. | Twitter streaming API, https://dev.twitter.com/streaming/overview |
| Lampos et al. (2016) | February 1, 2014–March 21, 2015 | UK | The dataset comprises two parts:<br>Data 1: A set of 1,342 Twitter user profiles in the UK and their corresponding 2,082,651 tweets<br>Data 2: An additional dataset obtained by randomly sampling all UK tweets posted in the same exact period as Data 1 (159,101,560 tweets) | https://figshare.com/articles/dataset/Socioeconomic_status_classification_of_social_media_users/1619703 |
| Hristova et al. (2018) | 2007–2015 | London, New York | The dataset comprises approximately 10 million geo-referenced pictures posted on Flickr from 2007 to 2015. These pictures were analyzed to quantify neighborhood cultural capital by categorizing cultural activities into nine distinct categories. | The data are made publicly available at http://goodcitylife.org/cultural-analytics/ |
| ***Social network relations*** | | | | |
| Vaca-Ruiz et al. (2014) | 2009–2012 | Brazil | The dataset includes detailed activity from 80,000 users, encompassing 13.1 million posts, 22 million reposts, and 4 million comments. It also contains data on 1.4 million repost cascades and 25,000 repost edges between cities. | Yahoo Meme |
| Norbutas and Corten (2018) | 2010 | The Netherlands | The dataset includes profile and relational data for more than 10 million users. The analysis focused on 6.27 million users, covering all 441 municipalities in the Netherlands and representing approximately 34.24% of the real population in each municipality. The network consists of 123.38 million undirected friendship ties among these users, with an average of 77 network neighbors per user. | Hyves, a Dutch online social network |
| Tóth et al. (2021) | 2011 | Hungarian | The dataset covers roughly 2 million individuals located in approximately 500 towns in Hungary. It comes from iWiW, a Hungarian online social network that was once popular and used by nearly 40% of the country's population. The dataset includes the location of the users at the town level and comprises more than 300 million friendship ties established by the end of 2011. | iWiW |
| ***Combination of social media content and social network relations*** | | | | |
| Lerman et al. (2016) | July–November 2014 | Los Angeles | The dataset consists of 6 million geo-tagged tweets made by 340,000 distinct users. These tweets were linked to US Census tracts through their locations, allowing the researchers to measure the emotions expressed in the tweets and the structure of social connections, and to use the area's socioeconomic characteristics in the analysis. | Twitter's location search API and Twitter4J API |
| Aletras and Chamberlain (2018) | - | - | The dataset contains Twitter users mapped to their occupational class and income. It originally included 5,191 users, but the study reported results on a subset of 4,625 users that are still publicly available. The dataset includes information on users' followers, which led to a set of 3,925,702 users, reduced to 53,199 unique accounts for analysis. | Twitter |
| Vaganov et al. (2020) | - | Russia | The dataset includes information about 18 million nodes (users) interconnected by 80 million undirected edges (friendships). Of these nodes, more than 400,000 bank clients were matched to their online accounts, with approximately 150,000 users' annual transaction activities matched. After filtering and extracting the largest connected component, the subset contained approximately 65,000 accounts. The full dataset includes the immediate neighbors of the client nodes, resulting in 1,658,492 nodes and 39,373,278 edges. | VK.com, a Russian online social media |
| Liao et al. (2024) | December 20, 2016 | Jilin Province, China | The dataset includes URL data provided by China Unicom. The process of generating URL data involves users accessing the Internet via signal communication with cellular towers. The dataset comprises 88,446 Internet users, 12,826 cellular towers, and 21,381,085 Internet records. | China Unicom |
| ***Multisource data*** | | | | |
| Steele et al. (2017) | November 2013–March 2014 | Bangladesh | The dataset comprises aggregated CDRs from Grameenphone subscribers, covering phone usage, top-up patterns, social network interactions, user mobility, and handset usage. The pre-aggregated datasets represent tower-level activity from 48,190,926 subscriber SIMs over a 4-month period. | Telenor/Grameenphone |
| | - | Bangladesh | The dataset includes 25 raster and vector datasets that encompass various environmental and physical metrics. These metrics include vegetation indices, nighttime lights, climatic conditions, and distances to roads and urban areas. Detailed information can be found in Table S1 of the source paper. | Details in Table S1 of the original paper |
| Pokhriyal and Jacques (2017) | January–December 2013. | Senegal | The dataset includes 11 billion mobile phone transactions (calls and texts) involving 9.54 million unique subscribers. It captures the spatiotemporal and individualistic patterns of mobile phone usage. | Orange mobile phone operator |
| | 1960–2014 | Senegal | The dataset includes various metrics such as NTL, vegetation cover (NDVI), urban areas, road density, proximity to services (schools and hospitals), and agrometeorological conditions (temperature, precipitation, and soil type). The data are available in both raster and vector formats at different spatial resolutions. Detailed information is provided in Table S1 of the original paper. | Publicly available datasets from various sources, including GIS and Earth Observation data |
| Njuguna and McSharry (2017) | July 2014–March 2015 | Rwanda | The dataset consists of 3.8 billion CDRs of calls and SMSs from 4.8 million subscribers across 580 towers. The data include fields such as timestamp, anonymized subscriber ID, start cell, and end cell. It is used to derive features such as mobile ownership per capita and average daily call volume per phone. | A leading mobile network operator in Rwanda |
| | December 2014 | Rwanda | The dataset consists of monthly composited cloud-free maps with a resolution of 15 arc seconds, capturing artificial light on the Earth's surface. These data are used to calculate nightlight intensity per capita. | NOAA |
| Niu et al. (2020) | 2015.1.3 | Guangzhou | The dataset includes remote sensing data from Landsat 8 images, which have a spatial resolution of 30 m. | Earth Explorer site, https://earthexplorer.usgs.gov/ |
| | January–December 2015 | Guangzhou | The dataset includes night-time light images from the VIIRS Day/Night Band, featuring monthly cloud-free composites. | NOAA/NGDC, http://www.ngdc.noaa.gov/eog/viirs/download_monthly.html |
| | - | Guangzhou | The dataset includes point-of-interest (POI) data comprising point locations with attributes such as site/facility names, addresses, and types. The types include shopping malls, office buildings, elementary education institutions, hospitals, and metro stations. | Baidu Maps, http://map.baidu.com/ |
| | - | Guangzhou | The dataset includes housing rent data, which has been transformed to estimate housing rents at the community level. | Online real estate platform Anjuke, https://guangzhou.anjuke.com/ |
| Kondmann et al. (2020) | 2013 | Sub-Saharan Africa | The dataset includes NTL data derived from the DMSP-OLS annual stable lights V4 composite of 2013 by NOAA. These data measure nightlight intensity on a scale from 0 to 63, with a spatial resolution of 1 × 1 km cell. | NOAA |
| | November 2018–February 2020 | Sub-Saharan Africa | The dataset includes more than 4 million geolocated tweets collected via the official Twitter API, covering the study period. It comprises tweet counts and 300-dimensional fastText embeddings of English language tweets. | Twitter API (https://developer.twitter.com/en/docs/twitter-api) |
| Zhao et al. (2019) | 2015 | Bangladesh | The dataset includes NTL data from the NPP-VIIRS Day/Night Band annual composite, with a spatial resolution of approximately 500 m. These data are used to reflect NTL intensity in Bangladesh. | Version 1 VIIRS Day/Night Band NTL. Available online: https://ngdc.noaa.gov/eog/viirs/download_dnb_composites.html |
| | 2015–2017 | Bangladesh | The dataset includes high-resolution Google Satellite Imagery, downloaded at zoom level 16. Each image has a size of 224 × 224 pixels, providing a spatial resolution of approximately 2.39 m. | Google Maps Platform-Maps Static API. Available online: https://developers.google.com/maps/documentation/maps-static/intro |
| | - | Bangladesh | The dataset includes road map data acquired from OSM, which is used to calculate road accessibility. | OSM, https://www.openstreetmap.org/ |
| | 2015 | Bangladesh | The dataset includes a land cover map acquired from the European Space Agency Climate Change Initiative project, with a spatial resolution of 300 m. These data are used to extract land cover features such as urban areas, cropland, tree cover, and water. | European Space Agency (ESA) Climate Change Initiative (CCI) project ESA CCI, https://www.esa-landcover-cci.org/ |
| Zheng et al. (2024) | 2022 | Fujian, China | County-level socioeconomic statistical data are used to construct the Multidimensional Poverty Index (MPI). | Statistics Bureau of Fujian Province |
| | 2022 | - | NPP/VIIRS monthly composite nighttime-light data are used to derive nighttime-light variables. | Colorado School of Mines Earth Observation Group |
| | 2022 | - | Geospatial variables include land use/land cover, average building height, road network density, monthly mean temperature, and POI data. | DAMO Academy; Wu et al. (2023); OpenStreetMap; National Tibetan Plateau Data Center; Amap |
| | 1999–2000 | - | DEM data are used to extract average altitude and flatland percentage. | SRTM |
| Tingzon et al. (2019) | 2016 | Philippines | The dataset includes night-time light images comprising 134,540 pixels, each approximately 0.25 $km^2$ in size. The NTL intensity values range from 0 to 122. | Visible Infrared Imaging Radiometer Suite Day/Night Band (VIIRS DNB) |
| | - | | The dataset includes daytime satellite imagery, consisting of 150,000 images in the training set and 13,454 images in the validation set. Each image covers approximately 0.25 $km^2$. | Google Static Maps API |
| | - | | The dataset includes high-resolution settlement data, providing detailed information on human settlements. It encompasses 16 road features, 28 building features, and 101 POI features. | DigitalGlobe |
| | - | | The dataset includes detailed information from OSM, contributed by a community of volunteer mappers. It covers roads, buildings, and POI. | OSM |
| Espín-Noboa et al. (2023) | - | Sierra Leone and Uganda | The dataset includes daylight satellite images recorded in RGB bands with a resolution of 640 × 640 pixels, corresponding to approximately 2.5 m per pixel. | Google Maps Static API https://developers.google.com/maps/documentation/maps-static/overview |
| | - | | The dataset includes mobility maps that aggregate movement counts per day, divided into three 8-hour windows. These data are used to provide quick and localized responses after natural disasters and health crises. | Movements between tiles are obtained through the Meta “Data for Good” platform |
| | | | The dataset includes infrastructure data comprising 54 features derived from OSM. These features include the number of buildings or ATMs and the distance to the closest road or POI. | OSM |
| | - | | The dataset includes nightlight intensity data, consisting of monthly average radiance composite images extracted from Google Earth Engine. | Earth Engine Data Catalog. 2022. VIIRS Stray Light Corrected Nighttime Day/Night Band Composites Version 1. https://developers.google.com/earthengine/d |
| | | | | atasets/catalog/NOAA_VIIRS_DNB_MONTHLY_V1_VCMSLCFG.<br>Google. 2022. Earth Engine.<br>https://earthengine.google.com/ |
| Chi et al. (2022) | - | 135 low- and middle-income countries (LMICs) worldwide. | The dataset includes data on mobile phone network infrastructure, encompassing the number of cell towers, Wi-Fi access points, and mobile devices in each area. This information is used to infer telecommunications connectivity, which correlates with regional wealth. | Facebook (https://research.fb.com/category/connectivity/) |
| | - | | The dataset includes high-resolution satellite imagery and nightlight data. The satellite images provide detailed visual information about land use, infrastructure, and economic activity, whereas the nightlight data are used to estimate economic activity based on the brightness of regions at night. | National Centers for Environmental Information Earth Observation Group (https://www.ngdc.noaa.gov/eog) |
| Marty and Duhaut (2024) | - | - | NTL data from DMSP-OLS (available from 1992 to 2013) are combined with those form VIIRS (available from 2012 to present) into a single dataset. | VIIRS and the DMSP-OLS harmonized dataset from Li et al. [79] |
| | 2017–2020 | - | Daytime imagery from Sentinel 2 and Landsat 7: Sentinel-2 provides 10-m resolution surface reflectance data, which have been available since 2017, capturing images approximately every week. Landsat 7 offers imagery from 1999 to 2022 at a 30-m resolution. For both the Sentinel and Landsat data, the authors used tiles of 224 × 224 pixels, matching the tile size used to train the VGG16 model. | Google Earth Engine |
| | - | - | MOSAIKS was developed to streamline the generation of informative features from high-resolution satellite imagery. The MOSAIKS platform facilitates feature extraction for any given location, enabling users to train machine learning models using these features. The MOSAIKS API provides 4000 features for each location, with each location corresponding to approximately a 1-km grid. | http://www.globalpolicy.science/mosaiks. |

| | | | |
|---|---|---|---|
| - | - | The study utilizes Facebook advertising data to measure the proportion of the population on Facebook and the proportion of Facebook users with specific attributes such as using an expensive phone or having interests in luxury goods. | https://developers.facebook.com/docs/marketing-apis/ |
| - | - | OSM: The authors extract the count of each type of POI within a 2.5-km radius of each survey location and determine the minimum distance to each POI. In addition, they extracted the length of each road type near each survey location and calculated the minimum distance to each road type. The road types include motorways, trunk roads, primary roads, secondary roads, tertiary roads, service roads, and pedestrian paths. | OSM |
| - | - | Land cover characteristics: The European Space Agency's Climate Change Initiative Land Cover dataset is provided annually from 1992 to 2018 at a 300-m resolution. This dataset categorizes each pixel into one of 36 land cover classes, as defined by the United Nations Land Cover Classification System (e.g., urban areas and croplands). | http://www.esa-landcover-cci.org/?q=node/197 |
| - | - | Long-term climate characteristics are derived from the WorldClim version 2 dataset. This dataset captures annual trends, seasonality, and extreme values of temperature and precipitation at an approximate resolution of 1 km, with values averaged from data collected between 1970 and 2000. | Long-term climate features from the WorldClim version 2 dataset |
| 2018–2020 | | The pollution measurement data are sourced from the Sentinel-5 Precursor (Sentinel-5P) satellite; and the aerosol optical depth (AOD) readings, from the Moderate Resolution Imaging Spectroradiometer (MODIS). Sentinel-5P, launched in 2017, is dedicated to atmospheric monitoring, while MODIS has captured images since 2000. | - |
| - | - | Synthetic aperture radar (SAR) data involves transmitting waves and measuring the strength and orientation of the waves reflected back to the satellite sensor. In this study, the researchers used the average and standard deviation of the VV and VH bands, and calculated the average and standard deviation of the VV/VH ratio. | - |

| | | | | |
|---|---|---|---|---|
| Tennant et al. (2025) | 2008–2020 | Ethiopia, Malawi, Tanzania, and Uganda | Household survey data are used to estimate structural consumption from productive assets and consumption expenditures, and to construct structural FGT poverty measures. | World Bank LSMS |
| | - | - | EO and geospatial covariates include population density, building footprints, remoteness, nighttime lights, elevation, slope, rainfall, temperature, and NDVI. | WorldPop; building-footprint data; travel-time/remoteness data; harmonized nighttime lights; CHIRPS; MERRA-2; AVHRR NDVI |
| Jung et al. (2026) | 2021 | Brazzaville, Congo | Restricted household-level social registry data are used to construct household-level MPI labels for targeting analysis. | Lisungi Social Safety Net Project; World Food Programme; Republic of Congo Ministry of Social Affairs |
| | 2023 | Brazzaville, Congo | Landsat-derived daytime imagery features are used as remote-sensing inputs. | Landsat 8/9 OLI-TIRS |
| | 2021-2022 | - | VIIRS nighttime-light statistics are used as remote-sensing inputs. | VIIRS nighttime lights |
| | - | - | Sentinel-2/MOSAIKS embeddings are used as image-based representation features. | Sentinel-2/MOSAIKS |
| | 2019–early September 2021 | - | Geotagged tweets are used to construct tweet counts and topic-weight features. | Twitter API V2 |
| | - | - | IGN administrative boundaries and POI data are used to construct geographic accessibility features. | Institut Géographique National (IGN) |
| | 2014 | - | OSM building footprints are used to calculate block-level built-environment features. | OpenStreetMap |
| | 2019–2023 | - | Internet-connectivity features include fixed and mobile download speed, upload speed, and latency. | Ookla Speedtest data |
| Jung et al. (2026) | 2018 | Zambia | DHS wealth-index data are used as ground-truth labels for village-cluster wealth prediction. | Zambia DHS |
| | 2015 | - | Household consumption expenditure data are used for validation and for mapping predicted wealth to consumption distributions. | LCMS |

| | | | |
|---|---|---|---|
| 2010/2015 | - | SAE poverty indices are used for additional ward-, district-, and province-level poverty analysis. | World Bank SAE based on 2010 Census and 2015 LCMS |
| 2022–2023 | - | Administrative caseload data record ward-level distribution of Food Security Packs. | MCDSS |
| - | - | Satellite-derived features include NTL, NDVI, daytime-image embeddings, and MOSAIKS representations. | - |
| 2019–2021 | - | Geotagged tweets are used to construct tweet counts and BERTopic-based topic features. | Twitter API V2 |
| - | - | OSM POI and building footprints are used to derive POI-distance and urban-morphology features. | OpenStreetMap |
| 2019–2023 | - | Fixed and mobile internet speed-test features are used to construct connectivity indicators. | Ookla |